\documentclass[reprint,aps, physrev]{revtex4-2}

\usepackage{physics}
\usepackage{graphicx}
\usepackage{dcolumn}
\usepackage{bm}

\usepackage{listings}
\usepackage[table,dvipsnames]{xcolor}
\definecolor{DarkGreen}{rgb}{0,0.5,0}
\definecolor{DarkPink}{RGB}{231,84,128}
\definecolor{maroon}{RGB}{128, 0, 0}

\usepackage{colortbl}
\usepackage{enumitem}
\usepackage{natbib}
\usepackage{amsmath, amssymb}
\usepackage{braket}

\usepackage{tikz}
\usetikzlibrary{quantikz}
\usepackage{graphicx}
\usepackage{dsfont}

\onecolumngrid 
\usepackage{nameref}
\usepackage{hyperref} 
\usepackage{cleveref} 
\usepackage{xparse}

\usepackage{verbatim}
\usepackage{balance}
\usepackage{ifthen}
\usepackage[caption=false]{subfig} 
\usepackage{ragged2e} 
\DeclareCaptionJustification{justified}{\justifying}

\makeatletter
\def\Ketbra#1{\@ifnextchar\bgroup{\KetbraTwo{#1}}{\KetbraOne{#1}}}
\def\KetbraOne#1{\mathinner{\ket{#1} \bra{#1}}}
\def\KetbraTwo#1#2{\mathinner{\ket{#1} \bra{#2}}}
\makeatother

\newcommand{\OC}[1]{\textcolor{red}}

\begin{document}


\title{Non-local resources for error correction in quantum LDPC codes}

\author{Omprakash Chandra}
\email{omprakash.chandra@hdr.mq.edu.au}
\author{Gopikrishnan Muraleedharan}
\author{Gavin K. Brennen}%
\affiliation{%
 Center for Engineered Quantum Systems, School of Mathematical and Physical Sciences, Macquarie University, 2109 NSW, Australia}%

\date{\today}

\begin{abstract}
Quantum low density parity check (qLDPC) codes are an attractive alternative to the surface code due to their relatively high code rate and distance. However, unlike the surface code which has simple, geometrically local, stabilizer checks, high performing qLDPC codes have non-local stabilizers that are challenging to measure.
Recent advancements have shown how to deterministically perform high-fidelity, cavity mediated many-body gates, enabling the encoding and decoding of non-local GHZ states. We integrate this non-local resource into the DiVincenzo-Aliferis method of fault-tolerant stabilizer measurement for quantum hypergraph product and lifted product codes. Using circuit-level noise simulations, including the noise optimized cavity mediated gate, we find promising thresholds of \(0.84 \%-0.60 \%\) for the hypergraph product code and psuedo-threshold of \(0.3\%-0.4\%\) for the lifted product codes, with cavity cooperativities in the range $C\sim 10^4-10^6$. We propose a compatible tri-layer architectural layout for scheduling stabilizer measurements, enhancing circuit parallelizability. 
\end{abstract}
\maketitle

\section{Introduction}

Scalable quantum computing requires quantum error correction (QEC) to tame errors accumulated from environmental noise and imperfect gate operations. Since their invention \cite{Shor:1995} a plethora of QEC code families have been developed characterized by a variety of features including: encoding rate, the code distance, the set of transversal logical gates supported by the code, fault tolerance relative to a given noise model, and the error threshold within that noise model. 
 The surface code is one of the most   extensively studied QEC codes \cite{Kitaev1997, bravyi1998quantum, kitaev2003fault} due to its shared advantages of: a high threshold for error rates, relatively straightforward decoding, and geometrically local stabilizer measurements which makes it compatible with nearest neighbor connected two-dimensional architectures \cite{dennis2002topological}. These features make it a popular choice for current experimental implementations of fault-tolerant quantum computers \cite{Google2021, IBM2021, Taminiau2018}. Despite its advantages, the surface code has significant limitations hindering scalability, primarily due to a poor encoding rate. Asymptotically, the encoding rate, defined as the ratio of the number of logical qubits $k$ to the number of physical qubits $n$, approaches zero as \(n \rightarrow \infty\) \cite{Bravyi2010}.

Surface codes are but one example of a larger class of codes known as quantum low density parity check (qLDPC) codes \cite{Breuckmann:2021}. Inspired by classical LDPC codes, their quantum cousins are stabilizer codes where the weight of any stabilizer is bounded by a constant, and any qubit has overlap with no more than a constant number of stabilizer checks. Hypergraph product codes (HGP) \cite{tillich2013quantum} and Lifted product codes (LP) \cite{panteleev2021quantum} are types of qLDPC codes that offer promising alternatives to the surface code. Keeping the physical qubits $n$ fixed, these codes offer higher encoding rate, $k/n$, than the surface code. Both their encoding rate and distance scales favorably with the block size of the code. For the codes considered in this work, $k/n \propto O(n^{\xi})$ and $d \propto O(n^{\lambda})$, for some $0.5 < \xi < 1$ and $0 < \lambda < 0.5$. Recent work demonstrates that Bivariate-Bicycle codes (BB codes), which is a type of HGP code exhibits high thresholds \cite{bravyi2024high}, which means they can tolerate higher error rates before failing. This robustness makes them particularly well-suited for near-term quantum processors, where error rates are still relatively high. The high thresholds of HGP codes are achieved by utilizing advanced decoding algorithms, such as BPOSD \cite{roffe2020decoding} and BPLSD \cite{hillmann2024localized}, in combination with overlapping window techniques \cite{skoric2023parallel}, which efficiently manage the complex error patterns in these codes. However, both HGP and LP codes encounter difficulties with stabilizer measurement because their stabilizers are non-local. This requires complex operations that current technology finds challenging to implement. 

Recent research has demonstrated the potential to implement HGP codes in real quantum systems. For instance, the experimental advancement of neutral atom quantum computing using Rydberg interactions has made rapid progress \cite{saffman_quantum_2010,Levine2018,Browaeys2020}. To enable non-local gates, several strategies are available. It has been experimentally demonstrated that distant qubits can be connected by physically shuttling the constituent atoms in the tweezer trap arrays \cite{xu2024constant}. However, shuttling is a complex and time-consuming process that reduces the number of stabilizers that can be measured simultaneously and also introduces errors during the process. There have also been proposals to perform long-range Rydberg gates with the atoms in place by targeting different Rydberg excitations according to inter-qubit spacing \cite{pecorari2025high, poole2025architecture}. However, because the strength of the van der Waals interaction decreases as \(1/r^6\), the range of the interaction is limited to $\lesssim 7$ lattice sites. Additionally, parallelization is restricted in these cases: when performing a coupling gate, no other coupling gates can be executed within the Rydberg blockade radius
\cite{morgado_quantum_2021}. The first issue inhibits implementing long-range HGP codes, which offer a very high encoding rate and fairly high distance. The second issue hampers circuit parallelizability, increasing the overall QEC time and introducing several other errors. Finally, it has been proposed to use neutral atoms coupled to a cavity array to non-deterministically create Bell states which can then enable pair-wise non-local gates in a register \cite{ramette2022any}. 

We present a scheme that provides a solution for implementing long-range multi-qubit gates in qLDPC codes. Recent advancements have demonstrated high-fidelity non-local many-body gates by coupling qubits to a common bosonic mode, which enables the preparation of non-local GHZ states and more general multi-qubit gates with the qubits in place \cite{jandura2023nonlocal}. This approach avoids the complexity associated with shuffling. As suggested in \cite{jandura2023nonlocal}, we can integrate these non-local gates into the DiVincenzo-Aliferis method for fault-tolerant stabilizer measurement \cite{aliferis_2007}. The advantage is that each stabilizer measurement requires only two or, at most, four rounds of non-local resources. The first set of non-local gates encodes the ancilla block into a GHZ state. The second set decodes all ancilla blocks, including the redundified ones, provided they are adjacent. Otherwise, if each ancilla block is decoded separately, the number of non-local resource rounds increases to four. We extend this technique to long-range HGP and LP codes and present numerical results demonstrating high thresholds. We propose a scheduling scheme for measuring stabilizers using a tri-layer architecture to enhance the circuit parallelizability. We also discuss some near term algorithms that are implementable using our scheme.

The rest of the article is organized as follows. In Sec.~\ref{Tools}, we review essential concepts including non-local many-body gates in Sec.~\ref{non-local many body gates}, syndrome extraction circuits in Sec.~\ref{syndrome extraction}, and stabilizer quantum codes in Sec.~\ref{stabilizer codes}. In Sec.~\ref{steane code}, we implement the DiVincenzo-Aliferis method of syndrome extraction, utilizing non-local many-body gates. In Sec.~\ref{applying to hypergraph product codes}, the method is applied to HGP and LP codes. Section~\ref{architecture for syndrome extraction} introduces a tri-layer architecture for efficient syndrome extraction. We discuss the strategic placement of cavities and explore the scheduling and parallelizability of syndrome measurements. Finally, in Sec.~\ref{conclusion and outlook}, we summarize our study, discussing the challenges and near-term algorithms applicable to Rydberg atom quantum computers, followed by a discussion of future directions.

\begin{figure}[hthb]
    \centering   \includegraphics[width=1\linewidth]{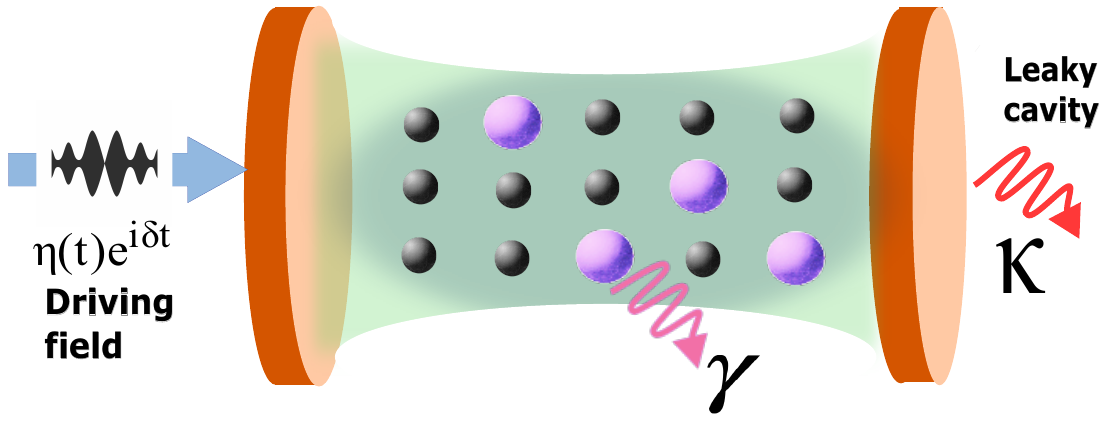}
    \caption{Illustration of the basic setup for the scheme with a cavity containing an array of three level spins spanned by a qubit and an excited state $\ket{e}$. The cavity mode is driven by an external classical field $\eta(t)e^{i\delta t}$ and decays at rate $\kappa$, while $\ket{e}$ leaks at a rate $\gamma$ to states outside the qubit space. The highlighted qubits are the ones involved in non-local interactions for preparing the GHZ state.}
    \label{Cavity setup}
\end{figure}

\section{Tools} \label{Tools}
\subsection{\label{non-local many body gates}Non-local many body gates}

We recapitulate the scheme introduced in \cite{jandura2023nonlocal} in order to motivate the form of the non-local gate, its associated error, and the embedding architecture described in Sec.~\ref{architecture for syndrome extraction}. The setup consists of $N$ three-level systems with computational basis states $\ket{0}$ and $\ket{1}$, and an excited state $\ket{e}$ with transition frequencies $\omega_0$ for $\ket{0}\leftrightarrow \ket{1}$ and $\omega_e$ for $\ket{1}\leftrightarrow\ket{e}$. A cavity mode, with annihilation (creation) operators $\hat{a} (\hat{a}^{\dagger})$ and frequency $\omega_c$, couples the transition $\ket{1}\leftrightarrow\ket{e}$ with coupling strength $g$. This cavity mode is driven by a complex classical field, $\eta(t)$, according to the Hamiltonian $H_{\rm drive} = 2|\eta| \sin(\omega_{L} t - \arg(\eta))(\hat{a}^{\dagger} + \hat{a})$. The classical field is detuned from the cavity by $\delta=\omega_c-\omega_L$ and from the $\ket{1}\leftrightarrow\ket{e}$ transition by $\Delta=\omega_e-\omega_L$.
The Hamiltonian in then transformed to the interaction picture (rotating frame) defined by the unitary, \(\hat {U}_r(t)=\exp\left[it(\omega_L (\hat{a}^\dag \hat{a} +\hat{n}_e) +\sum_j  \omega_0 \ket{0_j}\bra{0_j})\right]\). We assume that $\ket{e}$ decays at a rate $\gamma$ and is treated as leakage outside the qubit subspace by the introduction of a non-Hermitian term in the Hamiltonian. This will provide an expression for the fidelity, which is exact in the case of full leakage and serves as a lower bound otherwise.
After applying the rotating wave approximation one arrives at the effective Hamiltonian,
\begin{equation} 
    \hat{H}_{\text{eff}} = \delta \hat{a}^\dag \hat{a} + (\Delta-i\gamma/2)\hat{n}_e + [(g\hat{S}^{-} +i\eta)\hat{a}^\dag + hc],
    \label{eq:Hamiltonian}
\end{equation}
where, $\hat{n}_e = \sum_j \ket{e_j}\bra{e_j}$, $\hat S^+ = \sum_j \ket{e_j}\bra{1_j}$, $\hat S^- = (\hat S^+)^\dag$. The system evolves under the Lindblad master equation 
\[
\dot{\rho} = -i\hat{H}_{\text{eff}}\rho + i\rho \hat{H}_{\text{eff}}^\dag + L\rho L^\dag - \frac{1}{2}\{L^\dag L, \rho\}.
\]
Here, \( L = \sqrt{\kappa} \hat{a} \) is the jump operator, and \( 1/\kappa \) represents the lifetime of excitation in the cavity mode. We use a time-dependent pulse $\eta(t)$ over a duration $T$, with $\eta(0) = \eta(T) = 0$, while $g$, $\delta$, and $\Delta$ remain constant throughout the process. As described in~\cite{jandura2023nonlocal}, in the large detuning limit, keeping $T$ and $\eta / \Delta$ constant ($T\sim 20\times g^{-1}$ suffices), there exists a pulse profile that can generate a high-fidelity Mølmer-Sørensen type gate, $\hat{U} = e^{i \theta \hat{J}_z^2}$, where $\hat{J}_z = \frac{1}{2}\sum_{j=1}^N Z_j$, and $Z_j = \ket{0_j}\bra{0_j} -\ket{1_j}\bra{1_j} $.
To make the gate address only a subset of ancillary qubits needed for a cat state, one can adopt a variety of approaches: shelve the $\ket{1}$ state for qubits that should be spectators to an ancillary state $\ket{a}$ that doesn't couple to the cavity, or use addressable large AC-stark shifts applied only on the relevant ancilla to make them interact strongly with the cavity leaving the rest too far detuned to interact, or use a different species of spins for the ancilla with addressable transitions in frequency and space.

The evolution under the Hamiltonian $\hat{H}_{\text{eff}}$, in the absence of any losses ($\gamma, \kappa = 0$), is a unitary transformation given by $\hat{U} = e^{i \theta \hat{n}_1^2}$ where, $\hat{n}_1 = \sum_j \ket{1}_j\bra{1}_j$. This coupled with some single qubit rotations can generate entangling gates similar to Mølmer-Sørensen. In the presence of losses like cavity decay ($\kappa$) and excited state decay of the spins ($\gamma$), the evolution is no longer unitary and is described by a map given by,
\begin{equation}
     \mathcal{E_{\text{eff}}}(\rho)=\sum_{m,m'} \rho_{m,m'} e^{i\theta_{m,m'}} \ket{m}\bra{m'} \label{Eq:encoder}
\end{equation} where,
\begin{align}\label{Eq:phase}
    \theta_{m,m'} \approx & \  (m^2-m'^2)\theta+ N(m-m')\theta \\ \nonumber &+ i \frac{(m-m')^2 \theta}{2\sqrt{C} d_N} +i \frac{(m+m'+N)\theta d_N}{2\sqrt{C}},
\end{align}
and $\ket{m} = \ket{J=N/2,m_z=m}$. For the purposes of this paper we will always be using this map for encoding and decoding of GHZ state for which the parameter \(\theta=\pi/2\). It would be convenient for us to define the following parameters,

\begin{align}
    C =g^2/\kappa \gamma, d_N = [2(1+2^{-N})]^{-1/2}, \alpha = \frac{\pi}{4 \sqrt{C} d_N}, \label{Eq:parameters}
\end{align}
Where \(C\) is referred to as cavity cooperativity, \(d_N\) is weakly \(N\)-dependent parameter, and \(\alpha\)
quantifies the probability for an error on the spins induced by cavity as elaborated below in Sec. \ref{imperfect encoding and perfect decoding}. 
The map described in Eq.~\ref{Eq:encoder} is not trace-preserving which is justified for two reasons. First it provides a lower bound on the fidelity of the non-local gate \cite{jandura2023nonlocal}, and second, is compatible with many architectures, where decay from an excited state maps the spins to states outside the qubit subspace or can be driven to do so (see e.g. \cite{wu_erasure_2022}). 

Note that, while it may seem that the strength of the effective interaction between qubits is entirely independent of distance, this is not actually the case. Indeed, as required by causality, the strength of the coupling $g$ of the cavity mode to the qubits scales like $1/\sqrt{V}$ where $V$ is the quantization volume of the cavity. For the system sizes considered here, this is not an issue; however, it would ultimately impose a limitation on performing interactions between qubits in arbitrarily large arrays. 

\subsection{Syndrome extraction}\label{syndrome extraction}
Measuring stabilizers of a quantum error correcting code is the most vital step during quantum error correction. We need an efficient, and fault-tolerant syndrome extraction circuit so that errors arising during the syndrome extraction doesn't spread into data qubits. In addition to being fault-tolerant, the process should be highly parallelizable and fast to prevent backlog issues \cite{skoric2023parallel}.
Syndrome extraction can be performed using various methods, such as Shor's method, Steane's method, or the Flag qubit method, among others. Among these, Shor's method is particularly appealing due to its simplicity of implementation, as it requires only the preparation of a pure cat state. This simplicity makes it the most suitable choice for our purposes \cite{shor1995scheme,delfosse2020short,Chao_2020}. We briefly review Shor's method of syndrome extraction before discussing the DiVincenzo-Aliferis method.

\subsubsection{Shor's method}

\begin{figure}[htbp]
\centering
\hspace{-0.5cm}
\resizebox{0.5\textwidth}{!}{
\begin{quantikz}[row sep=0.1cm, column sep=0.1cm]
\lstick[4]{\text{ancilla}} \ket{+} & \ctrl{1} & \ctrl{2} & \qw & \ctrl{4} &\qw &\qw &\qw &\qw &\qw &\ctrl{5} & \qw & \qw &\qw & \rstick{X}\\
\ket{0} & \targ{}  & \qw & \qw & \qw & \qw &\qw & \qw &\qw &\qw & \qw&\ctrl{5} & \qw & \qw & \rstick{X} \\
\ket{0} & \qw & \targ{} & \ctrl{1} & \qw & \qw & \qw &\qw &\qw &\qw &\qw &\qw&\ctrl{5} & \qw & \rstick{X} \\
\ket{0} & \qw & \qw & \targ{} & \qw & \ctrl{1} & \qw & \qw &\qw &\qw&\qw &\qw &\qw &\ctrl{5} &\rstick{X} \\ [0.5cm]
\lstick{\shortstack{verification\\ qubit}}\ket{0} & \qw & \qw & \qw & \targ{} & \targ{} & \ghost{} & \rstick{Z}  \\ [0.5cm]
\lstick[4]{\text{data}} & \qw & \qw & \qw & \qw &\qw &\qw& \qw & \qw & \qw &\targ{} & \qw & \qw & \qw &  \\
 & \qw & \qw & \qw & \qw & \qw &\qw &\qw&\qw & \qw& \qw&\targ{} & \qw & \qw & \\
& \qw & \qw & \qw & \qw & \qw & \qw &\qw&\qw&\qw &\qw &\qw &\targ{} & \qw &  \\
& \qw & \qw & \qw & \qw & \qw & \qw & \qw &\qw &\qw&\qw &\qw &\qw &\targ{} &  
\end{quantikz}
}
\caption{Circuit representing Shor's style of fault-tolerant syndrome extraction \cite{548464} for $[[7,1,3]]$ Steane code where a verification qubit measures the parity $Z_1 Z_4$. An outcome of $+1$ indicates the ancilla is error-free, while $-1$ signals an error in ancilla preparation, requiring the batch to be discarded and preparation re-attempted.}
\label{fig:shor_syndrome_extraction}
\end{figure}
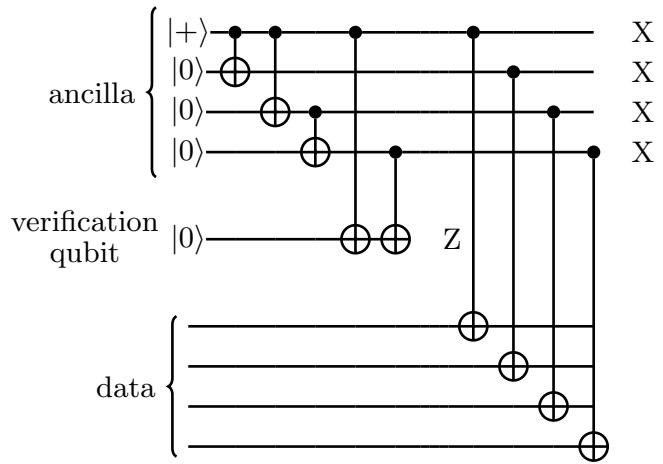

Shor's method is a syndrome extraction technique that uses ancilla qubits prepared in a `cat' state to perform stabilizer measurements. The circuit for stabilizer measurement using Shor's method \cite{548464} is shown in Fig.~\ref{fig:shor_syndrome_extraction}. The steps involved are,
\begin{enumerate}
    \item Prepare the ancilla qubits in a classical repetition code or ``cat" state.
    \item Verify the cat state.
    \item Apply CNOT gates coupling the ancilla and data qubits.
    \item Measure the ancilla qubits to obtain the stabilizer measurement outcome or syndromes.
\end{enumerate}

In the case of imperfect measurement, the syndrome extraction must be repeated until we get the same measurement outcome in succession. Generally, this can be achieved by repeating up to $O(d)$ times, where $d$ is the code distance \cite{aliferis2005quantum}. The above discussed methods ensure the fault-tolerance of Shor's method. These steps are repeated for all stabilizers to collect syndromes, which are then fed into a classical decoding algorithm. The decoder outputs a correction operator if the error is correctable or identifies a logical error if it is not. The verification step is time-intensive and introduces idling errors (or wait errors) on data qubits. This reduces the code's ability to protect the logical information.

\subsubsection{DiVincenzo-Aliferis method} \label{Divincezo}

In 2007, David P. DiVincenzo and Panos Aliferis~\cite{aliferis_2007} introduced a novel method of syndrome extraction in which ancilla verification can be bypassed and replaced with a decoding step. While skipping verification may result in the accumulation of errors, these errors can still be detected and corrected after decoding and measurement. The purpose of the decoder is to identify and invert any multi-qubit errors introduced by the encoder and propagated to the data. If the decoder itself is faulty, an additional decoder, referred to as `redundification of decoding', can be employed to distinguish between errors arising from the decoding process and those originating in other parts of the circuit. Although there are several ways to implement the procedure of skipping verification and post-processing after ancilla-data CNOT operations, DiVincenzo and Aliferis suggest that using a decoder is likely the most efficient approach \cite{aliferis_2007}. This method is particularly beneficial when measurement processes are slow, as it removes the need for immediate access to measurement outcomes.

This method is fault-tolerant as seen from the following arguments. Consider a case where an error occurs during CNOTs between the ancilla and data qubits. Since CNOT gates are implemented transversally, a single \(X\) error in the ancilla cannot cause more than one error in the data block. If a \(Z\) error propagates from the data to the ancilla during the CNOT operations, it will be manifested as a measurement error. This measurement error can be corrected by repeating the measurement \( d \) times, where \( d \) is the code distance, and then majority voting. Consider the case where a single error occurs during the ancilla encoding step. It's possible that more than one error can occur during this step, but still the procedure is fault tolerant. The encoder should be designed so that no logical errors arise at the output from a single error within it, and all first-order single-qubit errors can be detected during the decoding step. Finally, if a single fault occurs in the decoding step. We need to make sure that errors occurring during decoding are not mistaken for errors from other parts of the circuit. This confusion can arise if errors during encoding and ancilla-data coupling give the same syndrome as errors in decoding. In such cases, we have no way of distinguishing the location of actual error. Thus, the decoder must be carefully designed to avoid such confusion. Another way to solve this confusion is to redundify the decoder and measure both sets of ancilla separately. If the two measurement sets agree, we are assured that whatever error occurred was during encoding. If they don't agree, we will know that an error has occurred during decoding, and that the data qubits are unaffected. A detailed analysis of constructing an encoder and decoder for any stabilizer code, incorporating the effects of an erroneous cavity, is provided in Sec.~\ref{steane code}.

Verification involves measurements prior to the ancilla-data CNOTs, and these measurements take significantly longer compared to the single/two qubit gates or other non-local gates. The DiVincenzo-Aliferis method, even with the additional step of decoding that is not present in Shor's method, is much faster. Especially in our setup, we use cavities for decoding, which is much faster than the slow measurement process during verification. Another advantage of this method is that it allows us to manage slow measurements effectively, as all operations are Clifford, enabling efficient tracking and updating of the Pauli frame. Since we plan to work with qLDPC codes that have higher-weight stabilizers, Shor’s method would require multiple non-local CNOT gates during verification, and the number of non-local gates grows with the weight of the stabilizer. In contrast, the DiVincenzo-Aliferis method uses non-local resources four times: once for encoding and thrice for decoding as shown in Fig.~\ref{Fig:Divincezo-Aliferis circuit}. All three ancilla blocks can be decoded simultaneously using the same cavity, provided they are in the proximity. However, in crowded ancilla layers, this may lead to crosstalk, depending on the architecture. If crowding is not an issue, only two uses of non-local resources are needed. Otherwise, the ancilla blocks must be decoded separately, increasing the resource usage to four.

\vspace{0.5cm}

  \begin{figure*}[ht]
  \centering
  \begin{quantikz}[column sep=0.8cm, row sep=0.1cm, name=qcirc]
    \lstick{\text{Data block}} 
      & \qw 
      & \qwbundle{} 
      & \qw 
      & \gate{M} 
      & \qw       
      & \qw 
      & \qw           
      & \qw           
      & \qw 
      & \qw 
      \\[0.5cm]
    \lstick[1]{\text{Ancilla}}  
      & 
      & \lstick{$\ket{0}^{\otimes N}$}
      & \gate[1, style={fill=cyan, inner xsep=0.09cm, inner ysep=0.2cm}]{\mathcal{E}_{D^{-1}}}
      & \ctrl{-1}
      & \qw
      & \qw
      & \ctrl{1}
      & \ctrl{2}
      & \qw
      & \gate[style={fill=cyan, inner xsep=0.15cm, inner ysep=0.3cm}]{\mathcal{E}_{D}}
      & \meter{}
      \\[1cm]
      {} 
      & 
      & 
      & 
      & 
      & 
      & \lstick[1]{\(\ket{0}^{\otimes {N}}\)}
      & \targ{}
      & \qw
      & \qw
      & \gate[style={fill=cyan, inner xsep=0.15cm, inner ysep=0.4cm}]{\mathcal{E}_{D}}
      & \meter{}
      \\[1cm]
      & 
      & 
      & 
      & 
      & 
      & \lstick[1]{\(\ket{0}^{\otimes {N}}\)}
      & \qw
      & \targ{}
      & \qw
      & \gate[style={fill=cyan, inner xsep=0.15cm, inner ysep=0.4cm}]{\mathcal{E}_D}
      & \meter{} \\
  \end{quantikz}
  
  \begin{tikzpicture}[overlay, remember picture]
    \begin{pgfonlayer}{background}
      \draw[dashed, rounded corners, fill=blue!20, fill opacity=0.3] 
            (-4.8,2.9) rectangle (-1.3,5);
         \draw[dashed, rounded corners, fill=blue!20,fill opacity =0.3] (-0.3,-0.4) rectangle (7.7,5);
    \end{pgfonlayer}
    \node at (-2.6,2.6) {{\sc Encoding}};
    \node at (4,-0.6) {{\sc Decoding}};
  \end{tikzpicture}

  \begin{tikzpicture}[overlay, remember picture]
        \node at (-0.8,6.2) {\textbf{\textcolor{DarkGreen}{local}}};
        \node at (-2.3,3.8) {\textbf{\textcolor{red}{non-local}}};
        \node at (5.3,3.8) {\textbf{\textcolor{red}{non-local}}};
        \node at (5.3,1.95) {\textbf{\textcolor{red}{non-local}}};
        \node at (5.3,0.2) {\textbf{\textcolor{red}{non-local}}};
        \node at (3,0.8) {\textbf{\textcolor{DarkGreen}{local}}};
        \node at (2,2.5) {\textbf{\textcolor{DarkGreen}{local}}};
    \end{tikzpicture}

\caption{The circuit illustrates DiVincezo-Aliferis style of syndrome extraction. Due to the non-locality of the stabilizers, the targeted data qubits are spread out in space and the architecture is designed to support a neighborhood of ancillas local to each qubit. Hence, the encoding and the decoding steps are non-local while the gates coupling the ancilla to the data are local. Note that each non-local gate requires just one interaction with the cavity and two single qubit local transversal gates. After the ancilla interacts with data they are redundified to two blocks in order to distinguish encoding and decoding errors. The two levels of redundification are introduced to correct for leakage errors in the non-local map. The final measurements are three-outcome measurements in the \(Z\) basis: \(0\), \(1\), or neither—which help detect whether a qubit has leaked. This can be achieved by a \(Z\) measurement followed by a bit flip and another \(Z\) measurement. By comparing the measurement outcomes from the three sets of ancilla qubits, one can distinguish errors arising during encoding from those occurring during decoding.}
\label{Fig:Divincezo-Aliferis circuit}
\end{figure*}

\subsection{Stabilizer quantum codes} \label{stabilizer codes}
Stabilizer quantum codes were introduced by Daniel Gottesman \cite{gottesman1997stabilizer} in 1997. Given a stabilizer group \( S \) of order \(m\), we can define a subspace \( H \) on an \( N \)-qubit Hilbert space as the set of all states \( \ket{\psi} \) that satisfy \( S_i \ket{\psi} = (+1) \ket{\psi} \) for all \( S_i \in S \), where \( 1 \leq i \leq m \). This subspace is called codespace and it defines the quantum stabilizer code $C$. We use $[[n,k,d]]$ notation for quantum stabilizer codes where $n$ is the number of physical qubits, $k$ is the number of logical qubits and $d$ is the distance of the code. Here, $d(C)$ is defined as the minimum weight of a Pauli operator $P\in \Pi^N$ (where $\Pi=\{\pm I, \pm iI, \pm X, \pm iX, \pm Y, \pm iY, \pm Z, \pm iZ\}/\{\pm I, \pm iI\}$ is the equivalence class of Pauli group) that commutes with all the stabilizer generators $S_1,..S_m$ but $P\notin S$. Such Pauli operators are called logical operators. 

Stabilizer codes offer significant advantages. One of the key benefits is that, instead of specifying the basis states of the subspace, we can specify the generators of the stabilizer group associated with it. This provides a more compact and convenient representation of the quantum code. Additionally, detecting whether an error has occurred becomes much easier. A Pauli operator $P \in \Pi^N$, where $P \notin S$, is typically interpreted as an \textit{error} that changes the quantum state $\ket{\psi}$ into $P\ket{\psi}$. Such Pauli operators $P$ anti-commute with the stabilizer generators in $S$, provided they are not logical operators. This indicates that an error has occurred.

\subsubsection{Hypergraph product codes (HGP codes)}
HGP codes belong to the family of quantum low density parity check codes or qLDPC codes. Given two classical binary codes represented as $[n_1, k_1, d_1]$ and $[n_2, k_2, d_2]$, with their respective parity check matrices $H_1$ and $H_2$, we can use the hypergraph product method introduced by Tillich and Zémor \cite{tillich2013quantum} to construct a Calderbank-Shor-Steane (CSS) code \cite{CSS1,CSS2}. This method involves forming a generator matrix through the combination of two hypergraphs, each aligned with the classical code's parity check matrix. The \(X\) and \(Z\) stabilizer generators denoted by $G_X$ and $G_Z$ matrices in the symplectic form can be calculated using,

\begin{align}
\label{eq:HPG_CSS}
  \begin{split}
    G_X = & (H_1^T \otimes I_{r_2},I_{n_1}\otimes H_2),\\ 
    G_Z=&(I_{r_1}\otimes H_2^T, H_1 \otimes I_{n_2}).\\  
  \end{split}
\end{align}

 Each sublattice block is formed by taking the Kronecker product of two binary matrices, namely $H_1 \in F_2^{r_1 \times n_1}$ and $H_2 \in F_2^{r_2 \times n_2}$, and the Identity matrices denoted by $I_{r_i}$ and $I_{n_i}$, where $i=1,2$. The matrices $G_X$ and $G_Z$ have $r_1r_2$ and $n_1n_2$ rows, respectively (some of the rows can be linearly dependent). Both matrices have $N = r_2n_1 + r_1n_2$ columns, which determines the block length of the quantum code. The key aspect of this construction is the inherent fulfillment of the commutativity condition, specifically the symplectic product $G_XG_Z^T = 0$. This condition will ensure that all the stabilizers commute with each other.

There are four possible classical codes using the parameters we have defined. The first code, \( C_1 \), has parameters \([n_1, k_1, d_1]\) with its parity-check matrix \( H_1 \). The second code, \( C_2 \), has parameters \([n_2, k_2, d_2]\) and its parity-check matrix is \( H_2 \). Additionally, we can consider the transpose codes: \( C_1^T \), which has parameters \([n_1 - k_1, k_1^T, d_1^T]\) and the parity-check matrix \( H_1^T \), and \( C_2^T \), which has parameters \([n_2 - k_2, k_2^T, d_2^T]\) with the parity-check matrix \( H_2^T \). The resulting quantum code is: $[[{n_1 n_2} + (n_1 - k_1)(n_2 - k_2), k_1 k_2 + k_1^T k_2^T, \min(d_1, d_2, d^T_{1}, d^T_{2})]]$. We call qubits belonging to part $n_1 n_2$ as sector-1 qubits and, qubits belonging to part $(n_1-k_1)(n_2-k_2)$ as sector-2 qubits. $k_1 k_2$ number of logical qubits are entirely supported on sector-1 qubits, and the remaining $k^T_1 k^T_2$ logical qubits are entirely supported on sector-2 qubits. 

We can select two parity check matrices at random, each associated with a classical code, and use the hypergraph product shown in Eq.~\ref{eq:HPG_CSS} to create a quantum CSS code. If the original parity check matrices have low density or sparsity, the classical codes they represent are LDPC codes \cite{gallager1962low, sipser1996expander}. When we use sparse parity check matrices in the hypergraph product, the resulting parity check matrix for the quantum CSS code also remains sparse, leading to a quantum LDPC code or qLDPC code. For example, the parity check matrices corresponding to repetition code are sparse, and the hypergraph product of a repetition code with itself gives a qLDPC code, popularly known as Surface code. For more details on code construction, refer to Appendix~\ref{App:Details of Code Construction}.

\subsubsection{Lifted product codes (LP codes)}
Lifted product codes \cite{panteleev2021quantum} or LP codes are generalization of HGP codes where the elements of block matrices are replaced by elements from a commutative ring R, such as $R=F_2[x]/(x^l-1)$, or more generally, a group algebra $F_2G$ for a group $G$. The expression of parity check matrix of the resulting quantum code is, 
\begin{align}
\begin{split}
   H_X = &(A \otimes I_{m_B}, I_{m_A} \otimes B),\\
   H_Z= &(I_{n_A} \otimes B^T, A^T \otimes I_{n_B}). \\
\end{split}
\end{align}
Here, \( A \in M_{m_A \times n_A}(R) \) and \( B \in M_{m_B \times n_B}(R) \) are matrices over the ring \( R \). \( I_{m_A}, I_{m_B}, I_{n_A}, I_{n_B} \) are identity matrices of appropriate sizes. When \(R=F_2\), the LP code reduces to HGP codes. When \(R\) is a polynomial ring or a group algebra, the code takes on a quasi-cyclic structure. LP codes reduces the number of physical qubits as compared to HGP codes by symmetry reduction \cite{breuckmann2021quantum}, offering higher encoding rate. Block length, \(N\) of the resulting quantum code is given by \( \ell \cdot (n_A m_B + n_B m_A)
\), where \(l\) is the lift size, meaning each element of the ring is replaced by \(l\times l\) circulant matrices. The number of logical qubits, \( K\) is lower bounded by \(\ell \cdot (n_A - m_A) \cdot (n_B - m_B).\) The distance \(D\) scales as, \(\Theta( \frac{N}{\log(N)}) \). For more details look at \cite{panteleev2021quantum}.

\section{Using non-local resource for stabilizer measurement} \label{steane code}


We start with the $N$-qubit ancilla state initialized in the all-zero state (denoted as $\lvert 0\rangle^{\otimes N}$, which corresponds to all spins pointing down). This state resides in the Dicke space, which is the maximum angular momentum subspace of the \(2^N\)-dimensional Hilbert space. In the collective angular momentum basis, this state is expressed as \(\ket{J=\frac{N}{2}, m_z=-\frac{N}{2}}\), where \(N\) is the total number of spin-\(\frac{1}{2}\) particles, \(J\) is the total angular momentum, and \(m_z\) is the projection of angular momentum along the \(Z\) axis. Note that we use the convention that: $\ket{0}=\ket{\downarrow}=\ket{J=\frac{1}{2},m_z =\frac{-1}{2}}$ and $\ket{1}=\ket{\uparrow}=\ket{J=\frac{1}{2},m_z =\frac{1}{2}}$. 
Encoding of the ancilla state into the GHZ state is accomplished in three steps:
\begin{enumerate}
    \item We apply \( e^{-i\frac{\pi}{2}\hat{J}_y} \) to the initial state $\ket{0}^{\otimes N}$, which rotates each qubit by an angle of \(\frac{\pi}{2}\) about the \(y\)-axis, transforming the state to \(\ket{+}^{\otimes N}\).
    
    \item We then apply the map \(\mathcal{E}_{\mathrm{eff}}\) generated by \(\hat{H}_{\text{eff}}\) from cavity as defined in Eq.~\ref{Eq:encoder}. In the absence losses, the operation is unitary and is denoted by \(\hat{\mathcal{U}_c}\), where the subscript \(c\) means unitary generated by the cavity.
    
    \item We apply the inverse of the first operation which is \( e^{i \frac{\pi}{2}\hat{J}_y} \).
\end{enumerate}

The steps \(1\)-\(2\)-\(3\) collectively constitute the encoding operation, $\hat{E}_D^{-1}$, referred to as the `encoder'. In the absence of noise the encoder is a unitary map and is given by,
\begin{align}
\label{eq:encoding_unitary}
E_{D^{-1}}(\rho)&=e^{i\frac{\pi}{2}\hat{J}_y}\hat{\mathcal{U}_c}e^{-i\frac{\pi}{2}\hat{J}_{y}} \rho \ e^{i\frac{\pi}{2}\hat{J}_y}\hat{\mathcal{U}_c}^{\dagger}e^{-i\frac{\pi}{2}\hat{J}_{y}} \\ 
&=\hat{U}_E \ \rho \ \hat{U}^{\dagger}_E, \nonumber 
\end{align}
where, $\hat{U}_E=e^{i\frac{\pi}{2}\hat{J}_y}\hat{\mathcal{U}_c}e^{-i\frac{\pi}{2}\hat{J}_{y}}$. The next step involves ancilla-data controlled gates (C-M), where $M$ is the stabilizer to be measured, followed by the decoding map \(E_{D}(\rho)=\hat{U}^{\dagger}_E \ \rho \ \hat{U}_E\), and finally measurement of ancilla. 
In the presence of losses, the map is no longer unitary, introducing errors with some probability. In the following section, we will analyze the map in the presence of errors. 
  
\subsection{Cavity error analysis}
As mentioned above, losses in the cavity introduce errors that modify the cooperativity \( C \), as reflected in the second and third terms of the map \(\mathcal{E}_{\text{eff}}\) in Eq.~\ref{Eq:encoder}, rendering it non-unitary. Consequently, our encoding and decoding operations become faulty. Here, we will only consider the ideal scenario and the first-order failures in both encoding and decoding, which results in three possible scenarios: (1) perfect encoding $\&$ perfect decoding; (2) imperfect encoding $\&$ perfect decoding; (3) perfect encoding $\&$ imperfect decoding.

\subsubsection{Perfect encoding and perfect decoding}
\label{sec:perfect}

As discussed earlier, in the absence of any losses, both encoding and decoding operations are unitary. The cavity unitary \(\hat{\mathcal{U}}_c\) generated by $\hat{H}_{\text{eff}}$ takes the form \(e^{-i\theta \hat{n}^2}\), where \(\theta=\pi/2\) for GHZ state preparation, unitary becomes \(e^{-i\frac{\pi}{2}\hat{n}^2}\), where \(\hat{n}=\sum_j\ket{1_j}\bra{1_j}\). We can rewrite, 
\begin{align}
    \hat{n}=&\sum_j^N\frac{(I_j-Z_j)}{2}   \\  \nonumber
    = & \frac{N}{2}-J_z \quad \text{where } J_z=\sum_j \frac{Z_j}{2}\\ \nonumber
    \hat{n}^2=&J_z^2-NJ_z +\frac{N^2}{4}I. 
\end{align}
The cavity unitary becomes: \(\mathcal{U}_c=e^{-i\frac{\pi}{2}(J_z^2-NJ_z +\frac{N^2}{4}I)}\), from which we can seclude the constant term \(e^{-i\pi N^2/8} \). Note that, for simplicity, we omit the hats from operators. The encoding unitary as defined in Eq.~\ref{eq:encoding_unitary} then becomes: $\hat{U}_E=e^{i\frac{\pi}{2}\hat{J}_y} e^{-i\frac{\pi}{2}\left(\hat{J}_z^2 -N \hat{J}_z\right)}e^{-i\frac{\pi}{2}\hat{J}_{y}} = e^{-i\frac{\pi}{2}\left(\hat{J}_x^2-N \hat{J}_x\right)}$. We start with the initial state $\ket{0}^{\otimes N}=\ket{J=N/2,m_z=-N/2}$, in the combined angular momentum basis. For simplicity, we write \(\ket{J=N/2,m_z=-N/2} \rightarrow \ket{\Bar {0}}\) and \(\ket{J=N/2,m_z=N/2} \rightarrow \ket{\Bar {1}}\). The initial state in this notation is \( \ket{\Bar{0}} \). After the encoding operation, the initial state becomes,
\begin{align}
\rho_{_\mathrm{GHZ}} &= \hat{U}_E \ket{\bar{0}}\bra{\bar{0}} \hat{U}_E^{\dagger} \nonumber \\ 
&= \frac{1}{2} \Big( \ket{\bar{0}} + i\ket{\bar{1}} \Big) \Big( \bra{\bar{0}} - i \bra{\bar{1}} \Big). \label{Eq:rhoGHZ}
\end{align}

At this point, the combined (ancilla $+$ data) state is \( \upsilon_1 = \rho_{_\mathrm{GHZ}}\otimes \sigma\), where $\sigma$ denotes the data. The next step, as shown in Fig.~\ref{Fig:Divincezo-Aliferis circuit}, involves ancilla-data controlled operations for stabilizer measurement, $\hat{M}$, represented by \(\hat{V}\). Note that \(\hat{M}\) can represent any stabilizer of either \(X\) or \(Z\) type of any stabilizer code. For \(X\) type stabilizers we will use CNOTs and for \(Z\) type, the CNOTs will be replaced by CZs. The state after step 2 is,
\begin{align}
\upsilon_2 &= \hat{V} \Big( \rho_{_\mathrm{GHZ}} \otimes \sigma \Big) \hat{V}^{\dagger} \nonumber \\
&= \frac{1}{2} \Big( 
    \ket{\bar{1}}\bra{\bar{1}} \otimes M\sigma M 
  + \ket{\bar{0}}\bra{\bar{0}} \otimes \sigma  \nonumber \\
   &+ i \ket{\bar{1}}\bra{\bar{0}} \otimes M\sigma 
  - i \ket{\bar{0}}\bra{\bar{1}} \otimes \sigma M 
\Big). \label{Eq:upsilon2}
\end{align}

Next step is the redundification. Two sets of ancilla blocks are initialized in the state \(\ket{0}^{\otimes N} \), also represented as \(\ket{-N/2} \rightarrow \ket{\Bar{0}}\). During redundification, we copy the coherence with the help of CNOT gates. The resulting state is

\begin{align} 
    \upsilon_3 =&\frac{1}{2}\Big(\ket{\Bar{1}}_{a_1}\bra{\Bar{1}}\otimes \ket{\Bar{1}}_{a_2}\bra{\Bar{1}} \otimes \ket{\Bar{1}}_{a_3}\bra{\Bar{1}} \otimes M\sigma M \nonumber\\
    & +\ket{\Bar{0}}_{a_1}\bra{\Bar{0}}\otimes \ket{\Bar{0}}_{a_2}\bra{\Bar{0}} \otimes \ket{\Bar{0}}_{a_3}\bra{\Bar{0}}\otimes \sigma \nonumber \\
    &+i\ket{\Bar{1}}_{a_1}\bra{\Bar{0}}\otimes \ket{\Bar{1}}_{a_2}\bra{\Bar{0}} \otimes \ket{\Bar{1}}_{a_3}\bra{\Bar{0}}\otimes M\sigma \nonumber \\
    &-i\ket{\Bar{0}}_{a_1}\bra{\Bar{1}}\otimes \ket{\Bar{0}}_{a_2}\bra{\Bar{1}} \otimes\ket{\Bar{0}}_{a_3}\bra{\Bar{1}}\otimes\sigma M\Big). \label{Eq:upsilon_3}
\end{align}

All the ancilla blocks are decoded separately using decoding operation \(\hat{U}_E^{\dagger}\). Action of \(\hat{U}_E^{\dagger}\) is, 
\begin{align*}
    \hat{U}_E^{\dagger}\ket{\Bar{1}}=&-i\ket{\Bar{0}}+\ket{\Bar{1}} \\
    \hat{U}_E^{\dagger}\ket{\Bar{0}}=& \ket{\Bar{0}}-i\ket{\Bar{1}}.
\end{align*}

We use the relation above, expand the whole state and drop the cross terms since they don’t contribute to measurements in the \(Z\) basis. The expectation value of an operator \(\hat{O}\) in the state \(\rho\) is given by \(\text{Tr}(\rho \hat{O})\). When \(\rho\) is written in the eigenbasis of the measurement operator, only the diagonal terms affect the expectation value and cross terms vanish. This is because we're performing individual measurements in the \(Z\) basis. After rearranging, the state becomes:

\begin{align*}
\upsilon_4=\Big\{ &
  \ket{\bar{1}}_{a_1}\bra{\bar{1}} \otimes \ket{\bar{0}}_{a_2}\bra{\bar{0}} \otimes \ket{\bar{0}}_{a_3}\bra{\bar{0}} \\
+&  \ket{\bar{1}}_{a_1}\bra{\bar{1}} \otimes \ket{\bar{1}}_{a_2}\bra{\bar{1}} \otimes \ket{\bar{1}}_{a_3}\bra{\bar{1}} \\
+ & \ket{\bar{0}}_{a_1}\bra{\bar{0}} \otimes \ket{\bar{1}}_{a_2}\bra{\bar{1}} \otimes \ket{\bar{0}}_{a_3}\bra{\bar{0}} \\
+ &  \ket{\bar{0}}_{a_1}\bra{\bar{0}} \otimes \ket{\bar{0}}_{a_2}\bra{\bar{0}} \otimes \ket{\bar{1}}_{a_3}\bra{\bar{1}} 
\Big\} \otimes \frac{(I - M)}{2}\sigma_S \frac{(I - M)}{2} \\
+ \Big\{ &
  \ket{\bar{0}}_{a_1}\bra{\bar{0}} \otimes \ket{\bar{0}}_{a_2}\bra{\bar{0}} \otimes \ket{\bar{0}}_{a_3}\bra{\bar{0}} \\
+ &  \ket{\bar{1}}_{a_1}\bra{\bar{1}} \otimes \ket{\bar{1}}_{a_2}\bra{\bar{1}} \otimes \ket{\bar{0}}_{a_3}\bra{\bar{0}} \\
+&  \ket{\bar{0}}_{a_1}\bra{\bar{0}} \otimes \ket{\bar{1}}_{a_2}\bra{\bar{1}} \otimes \ket{\bar{1}}_{a_3}\bra{\bar{1}} \\
+&  \ket{\bar{1}}_{a_1}\bra{\bar{1}} \otimes \ket{\bar{0}}_{a_2}\bra{\bar{0}} \otimes \ket{\bar{1}}_{a_3}\bra{\bar{1}} 
\Big\} \otimes \frac{(I + M)}{2}\sigma_S \frac{(I + M)}{2}.
\end{align*}

The final step is to measure all the ancilla blocks. For each block, the measurement result should either be all \(1'\)s or all \(0'\)s. Any other result indicates an error occurred. We assign a bit value based on majority voting of the measurement outcomes. This gives us three bit values, one from each ancilla block. The stabilizer outcome is then determined by the Hamming weight of these bits. According to the rule above, if the Hamming weight is odd, we conclude that the state lies in the \(+1\) eigenspace of the stabilizer, else we are in the \(-1\) eigenspace.

\subsubsection{Imperfect encoding and perfect decoding} \label{imperfect encoding and perfect decoding}

Recall that the encoding operation \(\hat{U}_E\) consists of three independent steps: rotation \(e^{-i\frac{\pi}{2} \hat{J}_y}\), followed by the cavity unitary \(\hat{\mathcal{U}_c}\), and then  rotation \(e^{i\frac{\pi}{2} \hat{J}_y}\). We assume that each of these steps can fail independently. Considering only contributions first order in the error \(\alpha\) as defined in Eq.~\ref{Eq:parameters}, the action of the faulty encoding map on \(\rho=\ket{-N/2}\bra{-N/2}\), represented as \(\mathcal{E}_{D^{-1}}\) is:
\begin{align}
    \label{eq:depol}
    &\mathcal{E}_{D^{-1}}(\ket{-N/2}\bra{-N/2})  \approx (1 -( \underbrace{N \alpha d_N^2 + 5 Np_d/3}_{\tilde{p}_0} )) \rho_{GHZ} \nonumber \\
    &+  (\underbrace{2\alpha+ 8p_d/3}_{\tilde{p}_{\text{cavity}}}) J_x \rho_{GHZ} J_x +\alpha d_N^2 (J_x\rho_{GHZ} +\rho_{GHZ} J_x) \nonumber \\
    &-\alpha (J_x^2\rho_{GHZ} +\rho_{GHZ} J_x^2) +\frac{p_d}{3} \sum_{j=1}^N X_j \rho_{GHZ} X_j  \nonumber \\
    &+ Y_j \rho_{GHZ} Y_j+ Z_j \rho_{GHZ} Z_j.
\end{align}

The parameter \(p_d\) denotes the failure probability of \(Y\)-rotations. \newline
 The faulty encoding map \(\mathcal{E}_{D^{-1}}\), expanded to first order, can introduce a bit-flip error on any of the ancilla qubits with some probability, represented by the term \(J_x \rho J_x\). Additionally, it can introduce a measurement error simultaneously across all ancilla qubits via the terms \(Y \rho Y\) and \(Z \rho Z\). Note that this map is not trace-preserving, as it includes a leakage contribution given by the term \((1 - (N \alpha d_N^2 + \frac{5}{3} N p_d))\rho\). The terms \((J_x\eta+\eta J_x)\) and \((J^2_x\eta+\eta J^2_x)\) do not contribute to the final measurement done in the \(Z\) basis. For more details refer to Appendix~\ref{app:faulty encoding map}.

Since the interaction with the data is performed using transversal controlled gates, it cannot introduce more than one error in the data block. Furthermore, since bit-flips, \(X\), commute with the perfect decoding operation, this error syndrome can be detected, and the affected data qubits can be corrected accordingly. For more details look at Appendix \ref{app: state in faulty encoding and perfect decoding case}.

\subsubsection{Perfect encoding and imperfect decoding} \label{perfect encoding and imperfect decoding}

The remaining case is perfect encoding and imperfect decoding. We can follow the same arguments as for imperfect encoding. The faulty decoding map \(\mathcal{E}_D\) after taking into account the faulty unitary and faulty \(Y\)-rotation looks like,
\begin{align}
   &\mathcal{E}_D(\rho)  \approx (1 - \tilde{p}_0)  \hat{U}_E^{\dagger} \rho  \hat{U}_E  +  p_{\text{cavity}} J_x  \hat{U}_E^{\dagger} \rho  \hat{U}_E J_x+ \nonumber\\
   & \frac{p_d}{3} \left(\sum_{j=1}^N X_j  \hat{U}_E^{\dagger} \rho  \hat{U}_E X_j + Y_j  \hat{U}_E^{\dagger} \rho  \hat{U}_E Y_j + Z_j  \hat{U}_E^{\dagger} \rho  \hat{U}_E Z_j \right) \nonumber \\ \nonumber
    &+ \frac{p_d}{3}  \left(\sum_{j=1}^N X_j  \hat{U}_E^{\dagger} \rho  \hat{U}_E X_j + X_j \prod_k X_k  \hat{U}_E^{\dagger} \rho  \hat{U}_E \prod_k X_k X_j \right. \\ \nonumber
    & \left. + N \prod_k X_k  \hat{U}_E^{\dagger} \rho  \hat{U}_E \prod_k X_k \right). \\
\end{align}

Here, \(\tilde{p}_0 = \alpha N d_N^2 + 2N p_d\), \(p_{\text{cavity}} = 2\alpha\), and \(\rho\) denotes an arbitrary state. This map closely resembles the faulty encoding map defined in Eq.~\ref{eq:depol}, with the exception of the final term. The additional term appears because the map now acts on a general state. A similar term also arises in the encoding map, but in that case, the input state \(\ket{\frac{-N}{2}}\bra{\frac{-N}{2}}\) is invariant under \(Z\) errors, allowing those contributions to be absorbed into the existing error terms. \newline
Similar to the faulty encoding map, the error terms appear either as single-qubit measurement errors or as measurement errors affecting all ancilla qubits simultaneously. To mitigate these errors, we redundify the ancilla block as shown in Fig.~\ref{Fig:Divincezo-Aliferis circuit}, decode it, perform measurements, and then compare the measurement outcomes across all ancilla blocks. Since we focus only on first-order failures, we assume that the other two ancilla blocks remain unaffected. By comparing the measurement results, we can determine whether the error occurred during encoding and identify the affected data qubit. This allows us to apply the appropriate correction operator. If the error occurs during decoding, no correction is needed because the error arises after the ancilla qubits have interacted with the data qubits, leaving the data qubits unchanged. However, this error correction process relies on our ability to distinguish between encoding and decoding errors. Fortunately, in most cases, we can differentiate between the two. In the rare cases where we cannot, the error only introduces a single fault on a data qubit. Since we repeat the stabilizer measurement, this type of error can be addressed in the next round. A detailed analysis of the impact of each possible error term is provided in Appendix~\ref{app:redundification calculation}.

\subsection{Fault tolerance}

We will now present the following arguments, supported by calculations from previous sections, to demonstrate that our setup is fault-tolerant. We observe that the key element in proving the Fault-Tolerance (FT) for the DiVincenzo-Aliferis method lies in the careful design of the decoding circuit. Specifically, the decoder must be designed so that no single fault within it produces the same syndrome as any multi-qubit error caused by a single fault in the encoder or during ancilla-data interaction. When restricted to two-qubit gates, such a decoder can always be constructed for a distance-3 code \cite{aliferis_2007}. Alternatively, with the use of global gates, a decoder can be constructed for codes of any distance, as shown in the previous sections. However, in our case, since the encoding and decoding operations are performed collectively, the error propagation is more complex.

As discussed in Sec.~\ref{imperfect encoding and perfect decoding}, our encoder/decoder, to a first-order approximation, can introduce at most a single bit-flip error on any one of the ancilla qubits. Since separate ancilla qubits are used for each of the C-M gates, an error originating from an ancilla qubit can propagate to at most one data qubit and vice-versa. This approach prevents the occurrence of hook errors, which could otherwise be catastrophic \cite{dennis2002topological}. There is a small probability of a global bit-flip error arising from depolarization noise. For such an error to result from measurement faults, all ancilla measurements would need to fail simultaneously, a highly improbable event. While this may cause incorrect interpretation of the stabilizer outcome, since the data qubits remain unaffected, the error can be reliably detected and corrected by repeating the measurements.

An important aspect to ensure fault-tolerance of our protocol is the ability to distinguish errors originating from the encoding and from the decoding operations. To do that, we redundify the ancilla block as shown in Fig.~\ref{Fig:Divincezo-Aliferis circuit}, decode it, measure, and then compare the measurement outcomes across all ancilla blocks. Since we focus only on the first-order failures, we assume that the other two ancilla blocks remain unaffected. By comparing the measurement results, we can confidently determine whether the detected bit-flip errors, to first order, originated in the encoding or the decoding part. As mentioned earlier, in cases where the decoding is faulty and the syndromes for the bit-flip errors at the two decoding blocks disagree, we can conclude that the error, to first order, occurred in one of the decoding blocks, leaving the data block unaffected \cite{aliferis_2007}. In this scenario, no correction or recovery operations are needed for the data block. Conversely, if the syndromes agree, it indicates that the detected errors, if any, to first order, occurred either during the encoding or the ancilla-data controlled operations, both of which could affect the data.

The situation becomes complex when a fault occurs in any of the transversal C-M gates. The transversal operation can introduce at most one error in the data. The challenge arises when a fault in a C-M gate generates an error that, after propagating through the decoder, results in a syndrome overlapping with those caused by an error in the encoder. In this case, the corresponding correction or recovery operation might inadvertently introduce an additional error, resulting in two errors in the data and compromising fault tolerance. Fortunately, any bit-flip error (\(X\)) commutes with the decoding operation, allowing it to remain on the same qubit without causing ambiguity. The real issue arises when \(Z\) or \(Y\) errors occur during ancilla-data C-M operations, as these do not commute with the decoder. Such errors can propagate to other ancilla qubits, resulting in syndrome ambiguity. By carefully analyzing the commutation relationships between phase errors (\(Z\)), bit-and-phase errors (\(Y\)), and the decoding operation, we observed that single \(Z_i\) or \(Y_i\) errors occurring on any of the qubit indexed as \(i\) propagate through the decoder as follows: \(I_1 I_2..Z_i..I_N \leftrightarrow X_1X_2..Y_i..X_N\) and \(I_1 I_2..Y_i..I_N \leftrightarrow X_1X_2..Z_i..X_N\). These transformations produce distinct error syndromes, reducing the likelihood of ambiguity. This analysis can be easily extended to any stabilizer code by considering the Pauli string of corresponding weight of the stabilizers. In our case, in addition to considering the parity of the measurement outcomes, the outcome of each physical measurement is crucial to accurately locate the error, as explicitly demonstrated in Appendix.~\ref{app: state in faulty encoding and perfect decoding case}. Even when some errors produce overlapping syndromes after encoding and ancilla-data operations, the key point is that only a single data qubit is affected. Since we repeat the syndrome measurement multiple times, any such error can be corrected in the subsequent round.

It is worth noting that the whole noise process during encoding or decoding is not trace-preserving. Leakage errors can cause the ancilla atoms to move outside the qubit subspace. For ease of explanation, we denote this state as $\ket{a}$. Recognizing that our setup exhibits this type of noise, we propose to modify the measurement process to detect it. The $Z$ measurements performed at the end return $+1$ if the ancilla is in the $\ket{0}$ state and $-1$ otherwise. To verify that the ancilla remains within the qubit subspace, one can apply a bit-flip operation and repeat the measurement. This bit-flip operation is designed to affect only the qubit subspace. If the ancilla is outside the qubit subspace, the measurements will return $-1$ in both cases.

Next, we must ensure that we can distinguish leakage errors originating from encoding versus those from decoding. To achieve this, we require three sets of decoding. The ancilla qubits are redundified twice before decoding. We now examine all possible cases to determine whether they can be distinguished.

The first case occurs when encoding is faulty, causing one of the ancilla qubits to enter a state outside the qubit subspace. In this scenario, the controlled operation will not act on the corresponding data qubit, as the control qubit is in the $\ket{a}$ state. During decoding, the redundification process will also fail to copy information to the corresponding ancilla qubits, leaving them in the $\ket{0}$ state. This can be detected using the two-measurement scheme. Thus, if one of the ancilla qubits in the first decoder is in state $\ket{a}$ while the other two are in $\ket{0}$, we attribute the leakage to encoding. Notice that in this case the data qubits are corrupted, but still noise is limited to one data qubit.

The second case occurs when encoding is noiseless, but one of the decoders experiences a leakage error. In this scenario, all gates, including those involved in redundification, are noiseless. However, during decoding, with some probability, one of the ancilla qubits may transition to the $\ket{a}$ state. If this occurs in the second or third sets of ancilla qubits, the measurement outcome will clearly indicate that the noise happened during decoding. If one of the ancilla qubits in the first set is measured in state $\ket{a}$, we must check whether the other two sets of decoders are either all in $\ket{0}$ or all in $\ket{1}$. If this condition holds, the leakage likely occurred during the first decoder. Otherwise, the leakage originated during encoding. In this case data qubits are untouched.

In the rare case where the first decoder returns $\{a,1,1,\dots\}$ while all other measurements return $0$, we can assign the error to the first decoder with high probability. Although leakage errors during encoding can also produce this specific measurement outcome, they do so with significantly lower probability.

In summary, we have demonstrated that our design of encoder and decoder for the DiVincenzo-Aliferis method is fault-tolerant under three distinct scenarios: (i) bit-flip/phase errors from C-M or decoding operations, (ii) syndrome ambiguity, and (iii) leakage errors. The fault-tolerance argument developed here is true for any stabilizer quantum error correcting code.

\section{Numerical Results}\label{applying to hypergraph product codes}
We use the circuit in Fig.~\ref{Fig:Divincezo-Aliferis circuit} for stabilizer measurement. 
The fundamental building blocks of the circuit are physical qubits, which are susceptible to various errors depending on the chosen hardware. The types of errors that physical qubits can experience include:

\begin{enumerate}
    \item \textit{Imperfect reset}: Error occurred when a data/ancilla qubit is initialized in $\ket{1}$ instead of $\ket{0}$. This happens with some probability $p_{\mathrm{in}}$ (similarly for $\ket{+}/ \ket{-}$ state). 
    
    \item \textit{Single-qubit gate error}: while doing a single qubit gate, say $H$ we additionally do $\hat{X}, \hat{Z}$ or $\hat{Y}$ operations, each with probability $p_1/3$. This error model is also known as \textit{single-qubit depolarizing noise}. 

    \item \textit{Two-qubit gate error}: while doing the controlled operations between ancilla and data qubits, we additionally do $\hat{I} \otimes \hat{X}, \hat{I} \otimes \hat{Y}, \hat{I} \otimes \hat{Z}, \hat{X} \otimes \hat{I}, \hat{X} \otimes \hat{X}, \hat{X} \otimes \hat{Y}, \hat{X} \otimes \hat{Z}, \hat{Y} \otimes \hat{I}, \hat{Y} \otimes \hat{X}, \hat{Y} \otimes \hat{Y}, \hat{Y} \otimes \hat{Z}, \hat{Z} \otimes \hat{I}, \hat{Z} \otimes \hat{X}, \hat{Z} \otimes \hat{Y}$, or $\hat{Z} \otimes \hat{Z}$, each with probability $p_2/15$. This error model is also known as \textit{two-qubit depolarizing noise}. 

    \item \textit{Measurement error}: While measuring, say an ancilla qubit in $\hat{Z}$ basis, we incorrectly project into wrong state and report a wrong value with some probability $p_\mathrm{meas}$. 

    \item \textit{Wait or Idle error}: While we are doing error-correction rounds, our data qubits are waiting, represented as in identity gate $\hat{I}$. But instead, performing a single qubit gate $\hat{X}, \hat{Z}$ or $\hat{Y}$ with probability $p_{\mathrm{wait}}/3$. For all the numerical simulations we assume $p_{\mathrm{wait}}=0$. This is justified by the fact that Rydberg atoms have long coherence time compared to the idle time of data qubit \cite{morgado2021quantum}\cite{saffman2010quantum}.

    \item \textit{Cavity error}: When using cavity to prepare a GHZ state, we obtain the ideal state GHZ with a certain probability, while error terms appear with some probability as described in Eq.~\ref{eq:depol}. Approximately, the failure probability say \( p_{\text{cavity}} =2\alpha\) scales as \(1/\sqrt{C}\) \ref{eq:depol}. In both error models discussed below, we allow \( p_{\text{cavity}} \) to vary to account for imperfections in the cavity due to losses.
\end{enumerate}

The numerical simulations we perform involve a combination of these errors. The errors can occur randomly at any location in the circuit, with no correlations between them. Such an error model or noise is referred to as \textit{Markovian noise}. Depending on the hardware, each of the errors enumerated above can have different probabilities. In this work, we consider two different models:
\begin{itemize}
    \item \textbf{Hardware agnostic error model}: This model is independent of specific hardware implementations, assuming all errors except the cavity error occur with the same probability $p$ throughout the circuit. In our numerical simulations, we account for type 1,3,4 and 5 errors with equal probability. So, $p_1 = p_2 = p_{\mathrm{in}} = p_{\mathrm{meas}} = p$ and $p_{wait}=0$. Type 6 error which is error due to imperfect cavity, $p_{cavity}=x p_2$, where \(x\) is a real number. 

    \item \textbf{Custom error model}: In this model, type 3 errors (two-qubit depolarizing) have probability $p$ and type 2 errors (single-qubit depolarizing) have probability $p/10$. Type 4 errors (measurement) and type 1 errors (imperfect reset) each have a probability of $2p$. So, $p_2 = p$, $p_1 = p/10$, $p_{\mathrm{in}} = p_{\mathrm{meas}} = 2p$ and $p_{wait}=0$. The cavity error, $p_{cavity}=x p_2$, where \(x\) is a real number. 
\end{itemize}

We used \texttt{STIM} \cite{gidney2021stim} for our simulations. After we have collected sufficient number of sub-threshold data points, we fit all the codes in a code family to the equation \cite{quintavalle2022reshape},
\begin{align} 
\label{eq:fitting equation}
    P_L(p) & = A\left(\frac{p}{p_{th}}\right)^{ad},
\end{align}
where $P_{L}(p)$ represents the logical failure probability per syndrome extraction cycle, calculated as $P_L(p)=1-(1-P_L(p,d))^{1/d}$, with $P_L(p,d)$ being the total logical errors after $d$ rounds of syndrome extraction, and $d$ being the code distance. Here $A,a>0$ and $p_{th}$ is the threshold of a code family under the given error model and decoder. The logical failure probabilities for $p > 10^{-3}$ are determined numerically, after which the data points are fitted to the above equation and extended to $p < 10^{-3}$ to estimate the logical failure rates. Note that we do not consider leakage errors in any of our simulations. For details of numerical simulation refer to Appendix.~\ref{app:details of numerical simulation}. 

\subsection{ Results for HGP codes with \texorpdfstring{$h(x)=1+x+x^2$}{h(x)=1+x+x2}}

We used the code generated via check polynomial $h(x)=1+x+x^2$, with lift $=6,9,12$ under periodic boundary conditions. Details of code construction has been discussed in Appendix~\ref{App:Details of Code Construction}, and Table~\ref{tab:Codes h(x)=1+x+x^2} lists down the codes we get. Numerical performance of these codes is shown in Fig.~\ref{fig:HGP and LP peformance}. We obtain a high threshold, approximately \(8 \times 10^{-3}\), which is comparable to that of the surface code under an agnostic error model \cite{fowler_surface_2012}. The data points are extrapolated using the fitting function in Eq.~\ref{eq:fitting equation}, with parameter \(a \approx 3/4\) .

\begin{table}[ht] \label{tab:Codes h(x)=1+x+x^2}
\centering
\fbox{%
    \begin{tabular}{c|c}
    \textbf{Lift} & \textbf{Periodic} \\ \hline
    6             & $[[72,8,4]]$ \\ \hline
    9             & $[[162,8,6]]$ \\ \hline
    12            & $[[288,8,8]]$ \\ 
    \end{tabular}
}
\caption{Codes generated via check polynomial $h(x)=1+x+x^2$ under periodic boundary conditions.}
\label{tab:Codes_h(x)=1+x+x^2}
\end{table}

\begin{table*}[ht]
\centering
\begin{minipage}{0.48\textwidth}
    \centering
    \fbox{%
        \begin{tabular}{c|c|c}
        \textbf{$p_{cavity}/p$}  & \textbf{Threshold (\(p_{th}\))} & \textbf{\(C_{th}\)} \\ \hline
        0.5  & $7.99 \times 10^{-3}$ & $1.10 \times 10^{7}$\\ \hline 
        1  & $7.78 \times 10^{-3}$ & $2.89 \times 10^{6}$ \\ \hline
        2   & $6.94 \times 10^{-3}$ & $9.08 \times 10^{5}$ \\ \hline
        3  & $6.81 \times 10^{-3}$ & $4.19 \times 10^{5}$ \\ \hline
        4  & $5.78 \times 10^{-3}$ & $3.27 \times 10^5$  \\ \hline
        5  & $5.48 \times 10^{-3}$ & $2.33 \times 10^5$  \\ \hline
        10& $5.38 \times 10^{-3}$  & $6.04 \times 10^4$     
        \end{tabular}}
\end{minipage}
\hfill
\begin{minipage}{0.48\textwidth}
    \centering
    \fbox{%
        \begin{tabular}{c|c|c}
            \textbf{$p_{cavity}/p$} &\textbf{Threshold (\(p_{th}\))}  & \textbf{\(C_{th}\)} \\ \hline
        0.5 &  $8.43 \times 10^{-3}$ & $9.85 \times 10^6$ \\ \hline
        1 &  $8.12 \times 10^{-3}$ & $2.65 \times 10^6$\\ \hline
        2 &  $7.90 \times 10^{-3}$ & $7.01 \times 10^5$\\ \hline
        3 & $7.68 \times 10^{-3}$ & $3.30 \times 10^5$ \\ \hline
        4  & $7.08 \times 10^{-3}$ & $2.18 \times 10^5$  \\ \hline
        5   & $6.59 \times 10^{-3}$ & $1.61 \times 10^{5}$\\ \hline 
        10  & $6.09 \times 10^{-3}$ & $4.72 \times 10^4$ 
        \end{tabular}}
\end{minipage}
\caption{Error thresholds obtained from simulation results for the  \([[72,8,4]], [[162,8,6]]\), and \([[288,8,8]]\) codes generated via check polynomial \(h(x)=1+x+x^2\) with periodic boundary conditions under noise channels with different strengths of non-local gate noise. The left table presents results under the custom error model, while the right table shows results under the hardware agnostic error model. In both cases, $p_{cavity}$ represents the error from the imperfect cavity, and $p$ denotes the two-qubit depolarizing error after controlled operations between ancilla and data qubits. Data points are fitted using fitting function \ref{eq:fitting equation}, with \(a=1/2\). See main text for details about the cooperativity \(C_{th}\) calculation. We varied the ratio $p_{cavity}/p$ across the listed (randomly chosen) values to observe threshold variations. For each ratio of $p_{cavity}/p$, scalable quantum computing is possible if the physical error rate is below the corresponding threshold and the cooperativity is higher than the \(C_{th}\) values given. } 
\label{tab:threshold_agnostic_custom_periodic}
\end{table*}

Right Table~\ref{tab:threshold_agnostic_custom_periodic} lists down the \(p_{th}\) values for codes listed in Table.~\ref{tab:Codes h(x)=1+x+x^2} for a given ratio of \(p_{cavity}/p_2\) and their corresponding \(C_{th}\) under hardware agnostic error model and left table lists down the results for custom error model. As the quality of the cavity decreases, the corresponding threshold also reduces, which aligns with our expectations. For example, when the cavity error rate is 10 times the two-qubit depolarizing error rate in case of hardware agnostic error model, \( p_{cavity}/p_2 = 10 \), we get a threshold of approximately \( 6.09 \times 10^{-3} \), or 0.609\% and the corresponding cooperativity of around \( 4.72 \times 10^4 \). \newline

\begin{figure*}[htbp]
    \centering
    \subfloat[]{
        \includegraphics[width=0.45\textwidth]{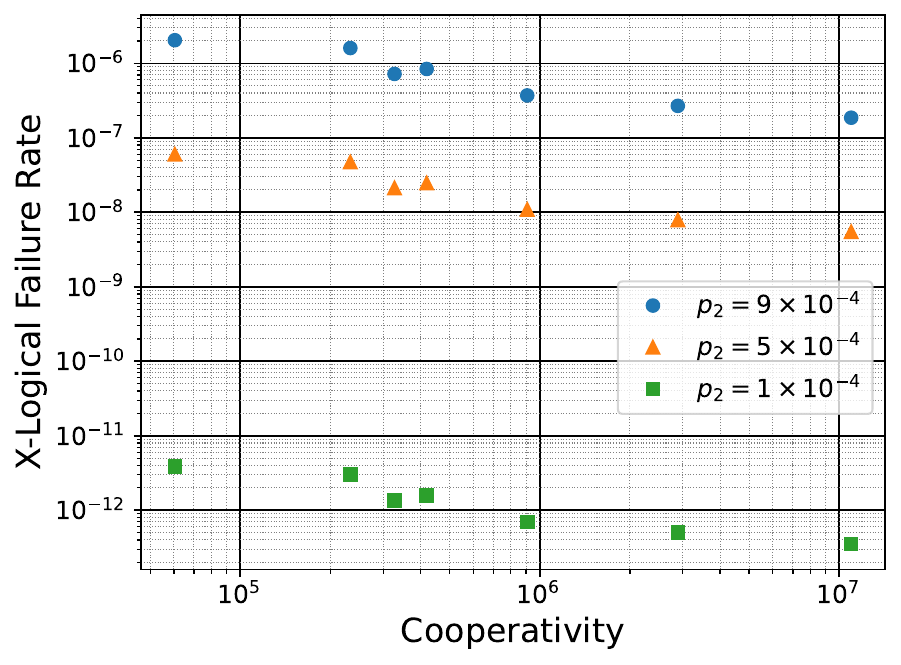}
        \label{fig:first}
    }
    \hspace{0.05\textwidth} 
    \subfloat[]{
        \includegraphics[width=0.45\textwidth]{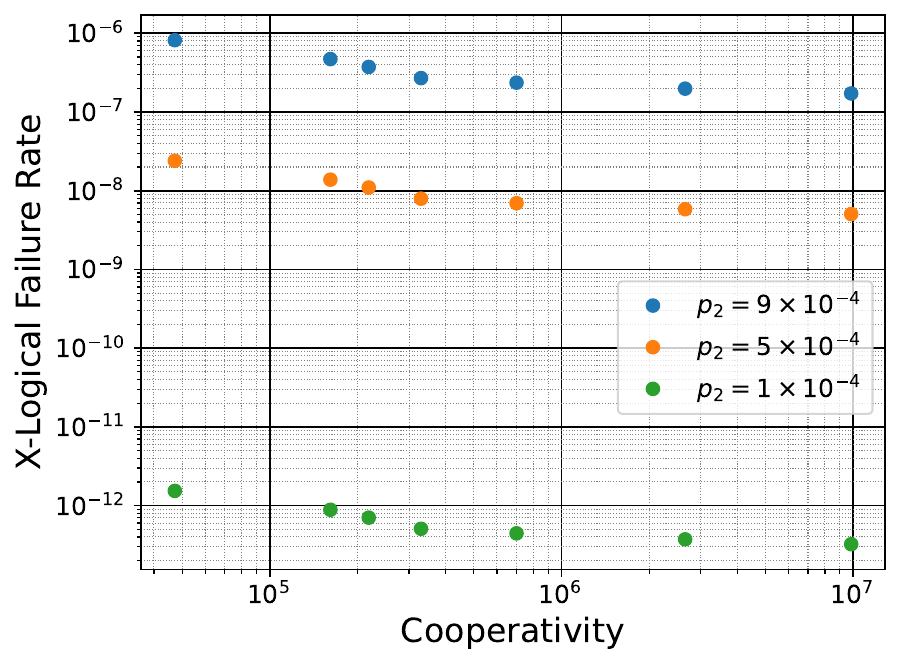}
        \label{fig:second}
    }
    \caption{ Plots showing failure rate of logical-X observables vs Cooperativity for $[[288,8,8]]$ code under agnostic error model. A set of data points is obtained by keeping the two-qubit gate error \(p_2\) fixed and varying the cavity error \(p_{\text{cavity}}\). (a) Figure shows the behavior of logical-\(X\) failure rate with cooperativity under Custom error model, b) and this under the Hardware-agnostic error model.}
    \label{fig:LFR_C}
\end{figure*}

We can also use this simulation data to determine the cooperativity required to achieve a specific logical error rate. Figure~\ref{fig:LFR_C} shows the plot for various fixed two-qubit error rates. For example, a fidelity of \(99.91\%\) has been demonstrated for two-qubit entangling gates \cite{IQM2024}. Using this value as \(p_2\), we can plot the logical failure rate (LFR) as a function of cooperativity. As shown in Fig.~\ref{fig:LFR_C}, a LFR of \(10^{-6}\) can be achieved with a cooperativity of approximately \(10^6\). The results improve exponentially with higher gate fidelities.


Experimentally, implementing codes with periodic boundaries is more challenging than those with open boundaries \cite{poole2024architecture} because interactions which are local on a periodic lattice are highly non-local with open boundaries. However, a key advantage of our cavity setup is that it enables the implementation of periodic boundary codes, doubling the number of logical qubits with only a few additional physical qubits while maintaining the same code distance.

\subsection{ Results for HGP codes with \texorpdfstring{\(h(x)=1+x+x^3+x^7\)}{h(x)=1+x+x3+x7}}

The true potential of our non-local resource lies in its ability to execute long-range non-local gates with minimal constraints, as discussed previously. This capability enables us to extend our proposal to high-rate codes generated by higher-degree polynomials, which produce highly non-local stabilizers. In this work, we specifically examine codes generated by the check polynomial $h(x) = 1 + x + x^3 + x^7$, which was previously explored in \cite{kovalev1212quantum}. We provide the code specifications, with construction details and the 2D layout in Appendix~\ref{App:Details of Code Construction}.

\begin{table}[ht]
\centering
\begin{tabular}{|c|c|}
\hline
\textbf{lift} & \textbf{Codes:Periodic Boundaries} \\ \hline
$15$ & $[[450,98,5]]$ \\ \hline
$30$ & $[[1800,98,10]]$ \\ \hline
\end{tabular}
\caption{Codes generated via check polynomial \(h(x) = 1 +
x + x^3 + x^7\) under open and periodic boundary condition.}
\label{tab:big codes}
\end{table}

\begin{figure*}
    \centering
    \includegraphics[width=\linewidth]{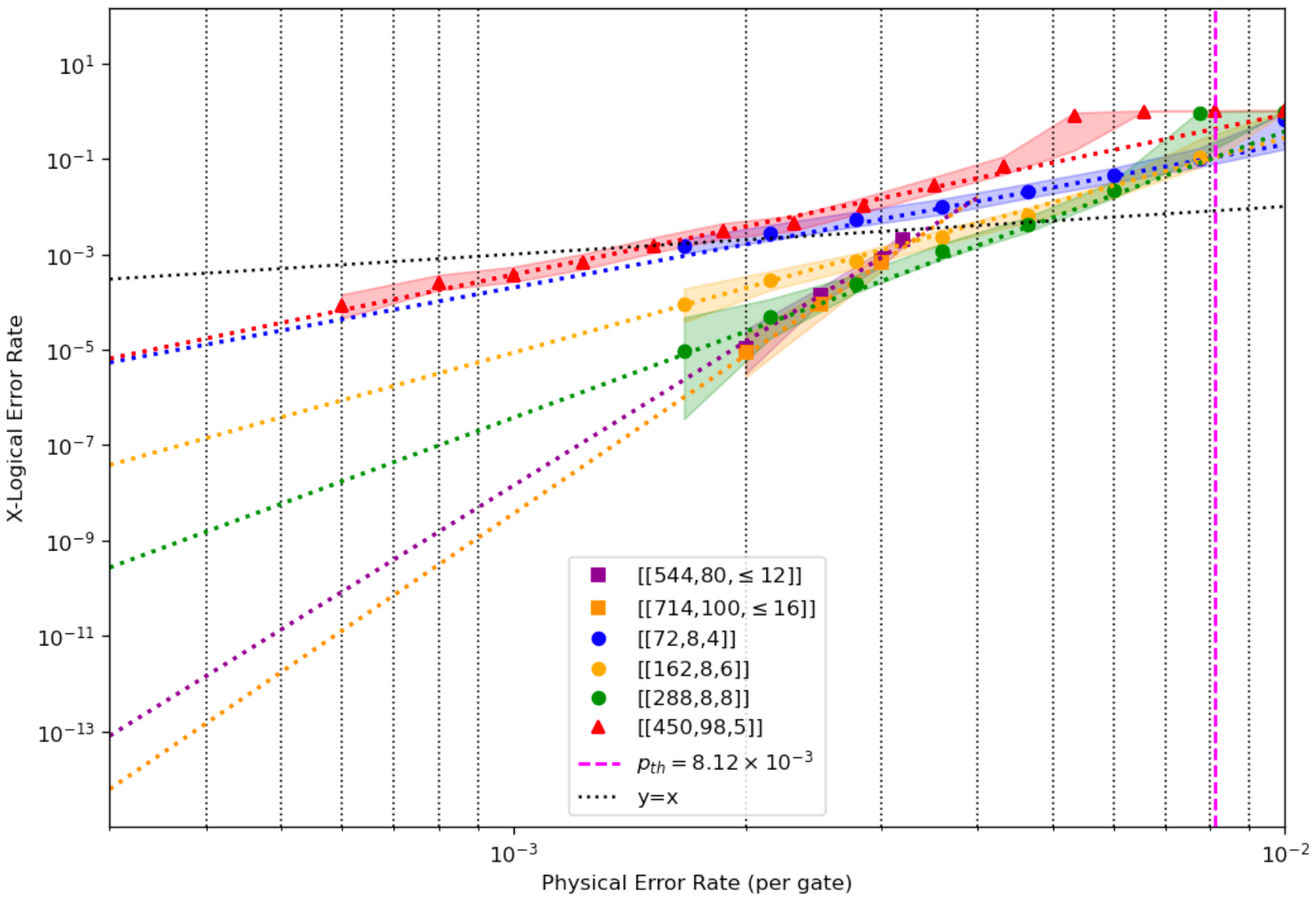}
    \caption{Plot showing logical-X error rate per syndrome extraction cycle vs. physical error rate (per gate). The plot is presented on a log-log scale. The simulations were carried out using \texttt{STIM}, with each data point based on $10^5$ Monte Carlo samplings. HGP codes \([[72,8,4]]\),  
\([[162,8,6]]\) 
    and \([[288,8,8]]\) 
    are constructed using check polynomial \(h(x)=1+x+x^2\) with different lifts. Their simulation is carried out under agnostic error model with the ratio \(p_\mathrm{cavity}/p = 1\), without accounting for erasure errors. Since they belong to the same code family, as evidenced by their convergence at a specific physical error rate value \(p_{th} \approx 8.12 \times 10^{-3}\), known as the threshold which is shown by vertical magenta line. 
    Their data points are modeled using the fitting equation~\ref{eq:fitting equation}, with \(a = 1/2\). In contrast, the HGP code \([[450,98,5]]\) 
  constructed using check polynomial \(h(x)=1+x+x^3+x^7\) represents a different family. The extended plot for this code is obtained using the same fitting function, but with \(a = 3/4\). Lifted Product (LP) codes like \([[544,80,\leq 12]]\) 
  and 
  \([[714,100,\leq 16]]\) 
  taken from \cite{xu2024constant} 
  are bench marked under a custom error model. Data points were extrapolated using the fitting function \(P_L(p) = \alpha p^{\beta},\) where \(P_L(p)\) is the normalized logical failure rate, and \(\alpha\), \(\beta\) are constants. For the \([[544,80,\leq 12]]\) code, the fitted line is \(P_L(p)=1.64 \times 10^{22}p^{10}\). For the \([[714,100,\leq 16]]\) code, the fitted line is \(P_L(p)=4.5\times 10^{24}p^{11}\).}

\label{fig:HGP and LP peformance}
\end{figure*}

Since our cavity setup allows for implementing periodic boundary codes, which doubles the number of logical qubits compared to open boundary codes, we will focus on studying periodic boundary codes. Table~\ref{tab:big codes} shows the different codes that could be generated using this check polynomial. Due to limitations of computational resources, we could only do the simulation for the $[[450,98,5]]$ code. See Fig.~\ref{fig:HGP and LP peformance} for the performance of this code. It is not possible to calculate the threshold with simulation from just one member of the code family. However, a pseudo-threshold can be estimated from the plot, which is determined when the logical failure rate intersects with the physical error rate at the \( y = x \) line. This point appears to be around \( p = 0.0015 \), or 0.15\%. The data points are fitted using fitting equation \ref{eq:fitting equation}, with \(a=3/4\).

\subsection{Results for LP codes}
LP codes, known for their higher encoding rate and distance, are an attractive option for implementation. We performed simulations with the LP codes from \cite{xu2024constant} and observed a logical error suppression of \(\approx 10^{-12}\) at a physical error rate of \(\approx 4 \times 10^{-4}\). While the exact code distance is difficult to compute, we used \(d\) rounds of syndrome extraction, where \(d\) serves as the upper bound.

\begin{table}[ht]
\centering
\begin{tabular}{|c|c|}
\hline
\textbf{lift} & \textbf{Codes:Periodic Boundaries} \\ \hline
$16$ & $[[544,80,\leq 12]]$ \\ \hline
$21$ & $[[714,100,\leq 16]]$ \\ \hline
\end{tabular}
\caption{Codes generated via check polynomial \(h(x) = 1 +
x + x^3 + x^7\) under open and periodic boundary condition.}
\label{tab:LP codes}
\end{table}

Compared to the results in \cite{xu2024constant}, our approach achieves several order of improvement in logical error suppression. This enhancement is primarily due to replacing shuffling with cavities for non-local gates during stabilizer measurements, which significantly reduces shuffle-induced errors and wait errors on data qubits. Moreover, the cavity error model mitigates errors by confining imperfections to single bit-flips on ancilla qubits, rather than causing more severe single- and two-qubit gate faults. Also, using dedicated ancilla qubits for each ancilla-data controlled operation effectively eliminates Hook errors \cite{dennis2002topological}.

Figure~\ref{fig:HGP and LP peformance} shows the numerical performance of LP codes. The data points were extrapolated using the fitting function \(P_l(p) = \alpha p^{\beta}\), where \(P_l(p)\) is the normalized logical failure rate, and \(\alpha, \beta\) are constants. Since the studied LP codes belong to different families, estimating a threshold was not feasible. However, the extrapolated line, derived from the numerical data, is shown in Figure~\ref{fig:HGP and LP peformance}.

\section{Architecture for Syndrome extraction circuit}\label{architecture for syndrome extraction}

\begin{figure}[htbp!]
    \centering
    \includegraphics[width=1\linewidth]{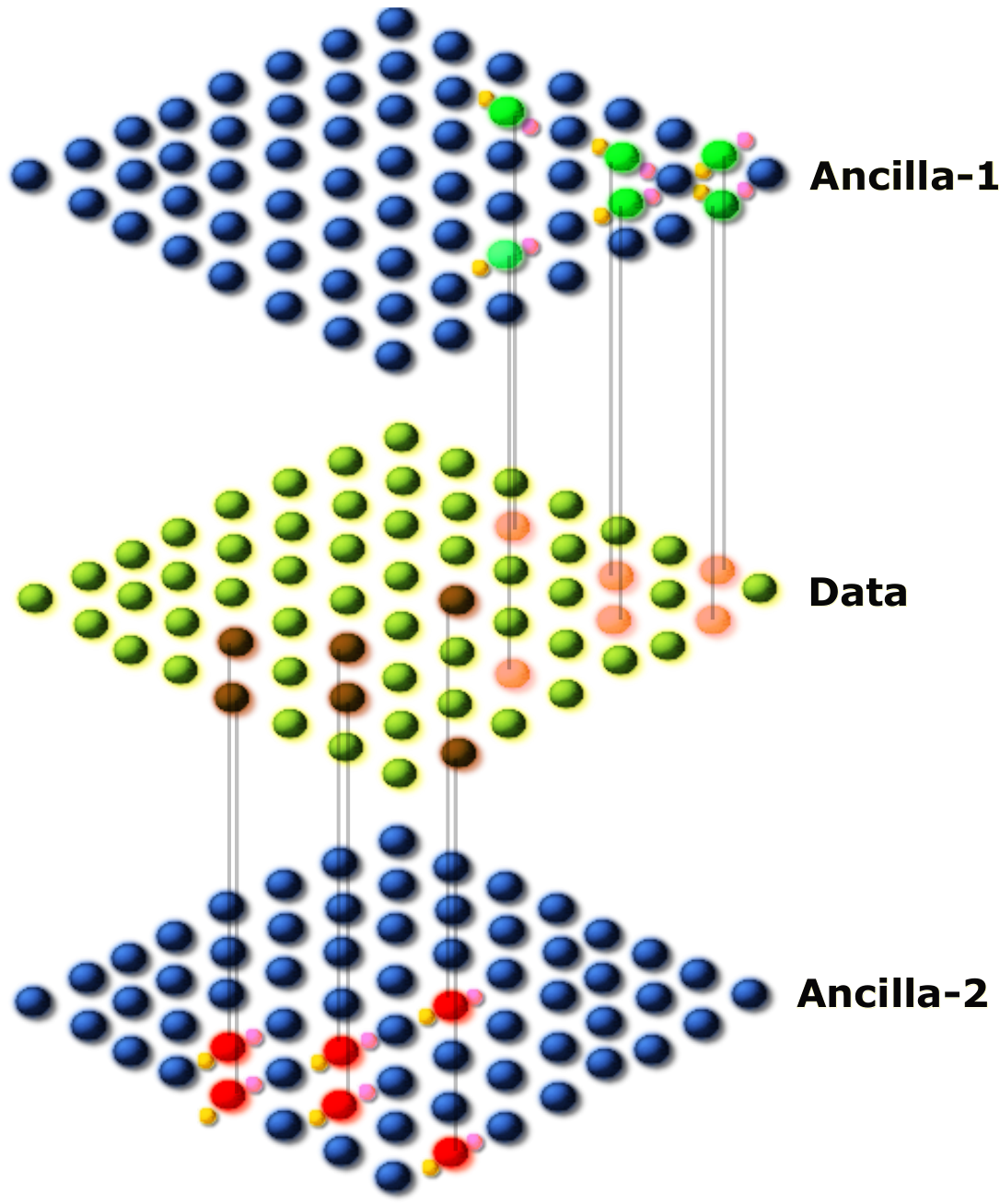}
    \caption{ An illustration of tri-layer architecture for stabilizer measurement. The top and bottom layers contain ancilla qubits, labeled as ancilla-1 and ancilla-2, respectively. The middle layer contains data qubits, which are encoded in the logical space of a LDPC code. Qubits shown in crimson in the middle layer represent the support of a \(Z\) stabilizer. Green-colored qubits in ancilla-1 are targeted by the cavity for GHZ state preparation, while yellow and pink colored qubits are used for redundification of decoding. The same method is applied to the \(X\) stabilizers, but this time using ancilla-2. The vertical lines indicate the controlled-\(M\) gates, where \(M\) is \(X\) or \(Z\), targeting the data. }
    \label{Fig:tril-layer stack}
\end{figure}

We propose a 3-dimensional tri-layer architecture for scheduling stabilizer measurements. The top and bottom layers represent ancilla qubits, labeled as `ancilla-1' and `ancilla-2', respectively. The middle layer contains data qubits encoded in the logical space of a qLDPC code. Ancilla-1 is specifically used to measure \(Z\) stabilizers, while ancilla-2 for \(X\) stabilizers. The data qubits of both HGP and LP codes can be arranged in a 2D layout such that the support of both types of stabilizers is limited to a single row and a single column, as illustrated in Fig.~\ref{app/fig/[[450,98,5]] layout}. This structure simplifies the implementation of cavity-based GHZ state preparation. 

We begin by identifying the support of a \(Z\) (or \(X\)) stabilizer on the data qubits and then select the neighboring ancilla qubits from ancilla-1 (or ancilla-2) accordingly. These selected ancilla qubits participate in a cavity-mediated interaction to prepare a GHZ state. Since some of these qubits may be located several lattice sites apart, this constitutes a non-local operation. Following GHZ state preparation, local two qubit gates (C-M) are applied between ancilla-1 (or ancilla-2) and the corresponding data qubits. The information is then transferred to an adjacent set of ancilla qubits for redundancy in the decoding process. All ancilla sets are subsequently decoded using the cavity interaction. Importantly, a single round of non-local operations is sufficient to decode all ancilla blocks. However, if spatial constraints (i.e., crowding) prevent decoding all blocks simultaneously via a single cavity interaction, the decoding must be performed separately. This can increase the number of non-local operations to four rounds. Finally, all ancilla blocks are measured in the \(Z\) basis.

\begin{figure}[htbp]
    \centering
    \includegraphics[width=1\linewidth]{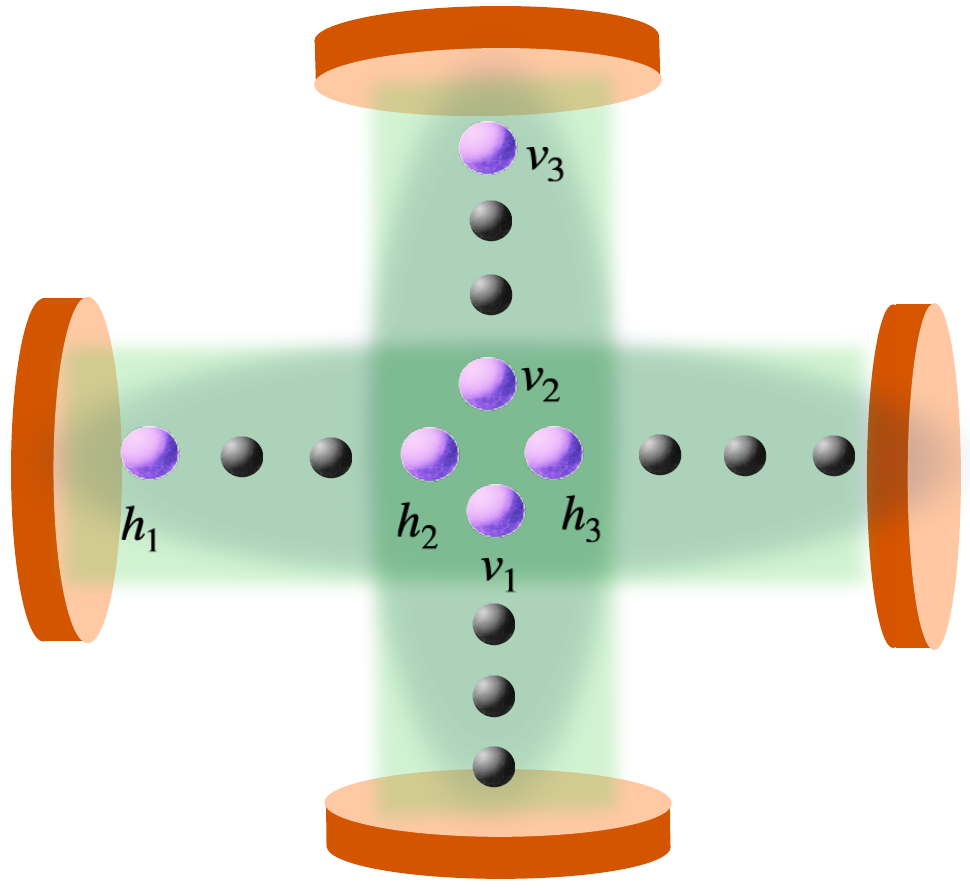}
    \caption{An illustration of GHZ state preparation using microwave cavities. We use two cavities designated as cavity-1 and cavity-2 to prepare a $\ket{GHZ}_w$ state. Cavity-1 prepares GHZ state on qubits placed horizontally, while cavity-2 prepares GHZ on qubits placed vertically. We can measure parity of any two qubits to project the combined state into $\ket{GHZ}_w$ state.}
    \label{fig:pair of cavity coupling atoms to prepare GHZ state}
\end{figure}

The motivation for proposing a tri-layer architecture is threefold. First, placing ancilla and data qubits in the same layer—as done in the surface code—leads to spatial congestion, especially when using a \(w\)-qubit GHZ state, \(\ket{\text{GHZ}}_w = (\ket{0}^{\otimes w} + \ket{1}^{\otimes w})/\sqrt{2}\), to measure a \(w\)-weight stabilizer. For instance, the codes listed in Table~\ref{tab:big codes} include stabilizers of weight 8, requiring 8 ancilla qubits for each \(X\) and \(Z\) stabilizer. Hosting all of these in the same layer as the data qubits would cause crowding, increase the likelihood of crosstalk, and restrict the number of stabilizers that can be measured in parallel—ultimately limiting parallelization.
Parallelization and spatial locality are critical for the speed and efficiency of quantum error correction protocols. A tri-layer architecture addresses these issues by separating data and ancilla qubits across layers. As noted earlier, stabilizers in hypergraph product (HGP) and lifted product (LP) codes have support localized along specific rows and columns. This structure enables a strategic layout of cavities aligned along rows and columns, as illustrated in Fig.~\ref{Fig:tril-layer stack}, following ideas also explored in~\cite{ramette2022any}. The tri-layer setup streamlines cavity placement. For each stabilizer measurement, only two to four pairs of cavities need to be activated: one pair for encoding and one (or up to three) for decoding, depending on whether all ancilla blocks can be decoded simultaneously. Finally, this architecture helps mitigate crosstalk between data and ancilla qubits—an inherent problem when all qubits reside in a single plane. \newline
To create $\ket{GHZ}_w$ state, we employ two distinct cavities, referred to as cavity-1 and cavity-2, as depicted in Fig.~\ref{fig:pair of cavity coupling atoms to prepare GHZ state}. Let's label the horizontal qubits as $h_1, h_2, h_3$ and the vertical qubits as $v_1, v_2, v_3$. We first use cavity-1 to prepare the GHZ state on the horizontal qubits: $\ket{GHZ}_{h_1 h_2 h_3}$. Then, similarly we use cavity-2 to prepare the GHZ state on the vertical qubits: $\ket{GHZ}_{v_1 v_2 v_3}$. We can then measure the parity between any two horizontal and vertical qubits, such as $Z_{h_3} Z_{v_1}$, to project the system into a combined GHZ state,

\begin{align}
\ket{GHZ}_{h_1 h_2 h_3}=&\left(\ket{000}_{h_1 h_2 h_3}+\ket{111}_{h_1 h_2 h_3}\right) \\ \nonumber
\ket{GHZ}_{v_1 v_2 v_3}=&\left(\ket{000}_{v_1 v_2 v_3}+\ket{111}_{v_1 v_2 v_3}\right).
\end{align}

Upon measuring $Z_{h_3} Z_{v_1}$, with the measurement outcome $m$, the combined state $\ket{GHZ}_{h_1 h_2 h_3} \otimes \ket{GHZ}_{v_1 v_2 v_3}$ is projected into the $\ket{GHZ}_6 = \left(\ket{000,000} + \ket{111,111}\right)/\sqrt{2}$ state if $m=0$. If $m=1$, apply the correction $X_{h_1} X_{h_2} X_{h_3}$ or $X_{v_1} X_{v_2} X_{v_3}$ to prepare the $\ket{GHZ}_6$ state. The correction term after measurement can be written as $\left(X_1 X_2 X_3\right)^m$.

\subsection{Scheduling stabilizer measurement}

\begin{figure*}[htbp]
    \centering   
    \includegraphics[width=0.9\linewidth]{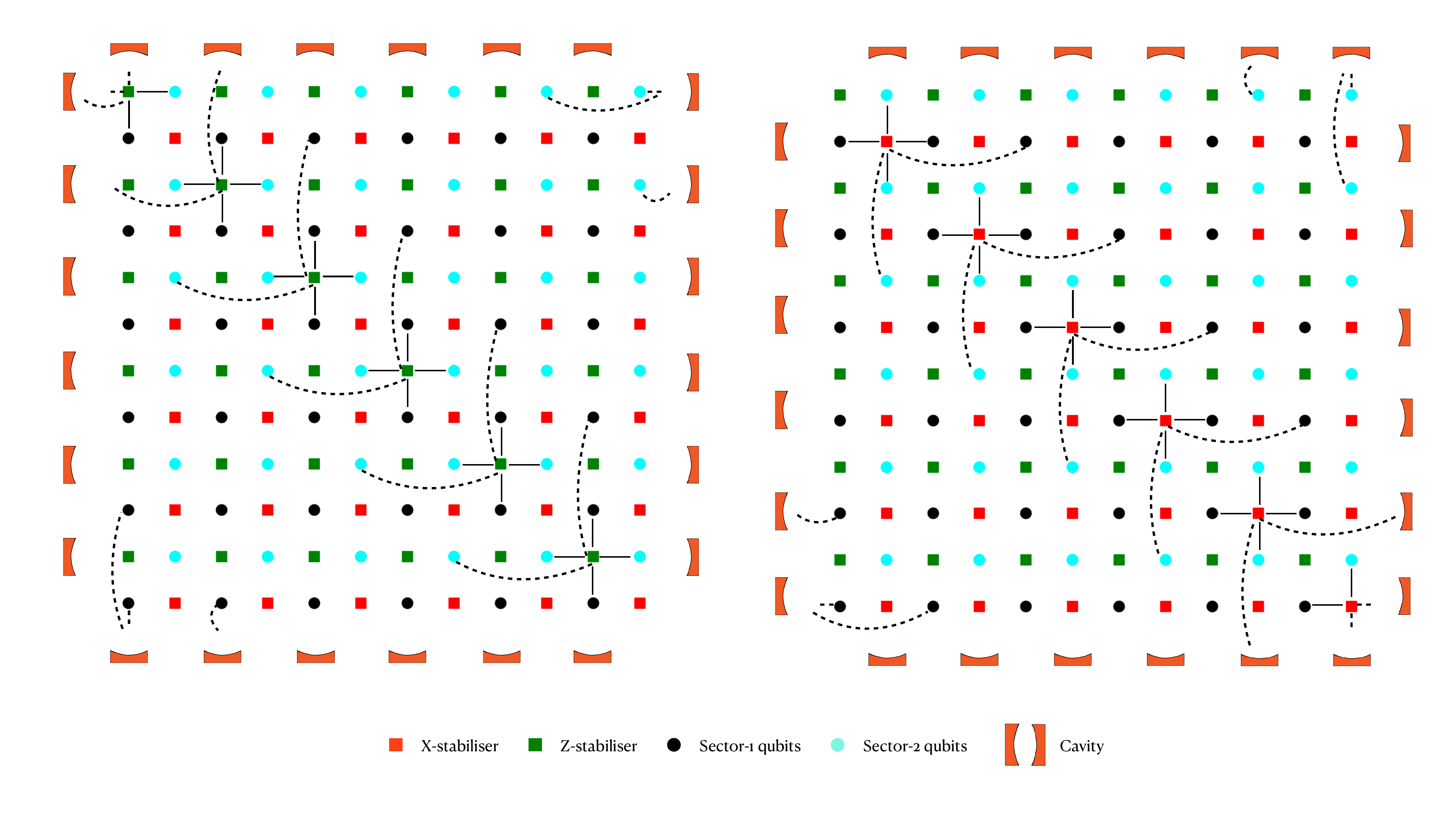}
    \caption{2-D layout of $[[72,8,4]]$ code generated via check polynomial $h(x)=1+x+x^2$, lift=6 with periodic boundary conditions. (a) Left figure shows GHZ state preparation in ancilla-1 using the non-local resources. Consequently, no other stabilizer measurement can be performed along the same row and column, as the cavities are occupied. However, we can measure another Z-check two unit cells below, as its support does not overlap with the same row or column, thus avoiding any cavity conflict. Similarly, we can proceed downwards along the diagonal. (b) we can proceed in the same fashion in ancilla-2 which we explicitly have for measuring $X$-checks. }
    \label{fig:scheduling stab measurement}
\end{figure*}

As discussed above, we can arrange the physical qubits of a HGP and LP code such that each \(X\) (or Z) stabilizer has support only along a single row and column. As illustrated in Fig.~\ref{fig:scheduling stab measurement}, cavities can be positioned along these rows and columns allowing targeted qubit operations for gate implementation. During a stabilizer measurement, such as a \(Z\) type as shown in left hand side of Fig.~\ref{fig:scheduling stab measurement}, cavities along its row and column are required. Consequently, no other stabilizer measurement can be performed along the same row and column, as those cavities are already in use. However, we can measure another Z-stabilizer two rows down along the diagonal, as its support does not overlap with the previous stabilizer, thus avoiding any cavity conflict. Similarly, we can proceed downwards along the diagonal. Since we have a separate ancilla layer, ancilla-2, for measuring X-stabilizers, we can proceed in the same fashion. This is shown in the right side of Fig.~\ref{fig:scheduling stab measurement}. 

Now we use cavities to prepare a GHZ state and proceed with controlled operations between ancilla and data. However, the C-M gate for \(X\) and \(Z\) stabilizers between ancilla and data must be applied carefully. There are two ways we could think of doing this: i) We can start with an \(X\) or \(Z\) stabilizer and apply the C-M gate between ancilla and data in an alternating fashion, i.e., first perform C-M gate for \(X\) (or \(Z\)), wait until all the gates are done, then perform C-M gates for \(Z\) (or \(X\)). Continue alternating in this manner. Or ii) We can do all the gates for \(Z\)-checks in diagonal in one time-step, and then do all the gates for \(X\)-checks in diagonal in another time-step.

Mixing the controlled gates of \(X\) and \(Z\) stabilizers may result in measuring the wrong operator with a global phase of \(-1\). To avoid this issue, it is best to wait until all the ancilla-data gates of \(Z\) (or \(X\)) have been applied before proceeding to those of \(X\) (or \(Z\)). In a single time step, all stabilizers along a diagonal can be measured. In subsequent time steps, we can move up or down along the anti-diagonal to perform the rest of the stabilizer measurements. This method ensures no cavity overlap and effectively parallelizes the syndrome extraction. For a code \([[2n^2, 2k^2]]\) obtained through the hypergraph product of a classical code \([n, k]\) with itself, all stabilizers can be measured in \(2(2n - 1)\) time steps. The number of stabilizers measured per time step varies with the number of checks along a diagonal.

\section{Conclusion and outlook} \label{conclusion and outlook}

We have developed an approach for performing QEC for a large class of codes by integrating non-local gates with the DiVincenzo-Aliferis method for stabilizer measurements. By coupling qubits to a cavity, high-quality cat states can be deterministically encoded and decoded in one step each. Then we incorporated the effect of losses in the cavity, and studied the stabilizer measurement circuit, and found the method is fault-tolerant for any stabilizer code. Next, we incorporated the cavity error model into circuit-level noise simulations of the hypergraph product and lifted product codes, achieving a promising threshold, thereby advancing towards large-scale fault-tolerant quantum computing. The \texttt{STIM} circuit we developed for syndrome extraction can be generalized to any generating polynomial \(h(x)\) with an arbitrary degree. 

We numerically tested the effect of circuit-level noise for codes generated by \(h(x) = 1 + x + x^2\). For a hardware-agnostic error model, we achieved a threshold ranging between $0.84\%-0.60\%$ for corresponding values of cooperativity in the range $9.85 \times 10^6-4.72 \times 10^4$. For the custom error model, the threshold values were between 0.8\% - 0.53\%, and the cooperativity ranged from $1.1\times 10^7-6.2 \times 10^4$. A logical failure rate of \(10^{-9}\) can be achieved with a cooperativity of approximately \(10^6\), by increasing the two-qubit gate fidelity to 99.96\%. Recent experimental work coupling Rydberg atoms to microwave cavities \cite{Schuster:2020, kumar2023quantum} has achieved single atom cooperativities of $C\approx 2.2 \times 10^4$, which is not far from the requirements found in our work. For lifted product codes, we observe a pseudo-threshold in the range of \( 0.3\% - 0.4\% \). Remarkably, logical errors are suppressed to approximately \( 10^{-12} \) when the two-qubit gate error is around \( 4 \times 10^{-4} \), a value that is within reach of current technological capabilities \cite{chang2023high}.
However, we struggled to perform simulations for larger codes, involving a lot of physical qubits, listed in Table~\ref{tab:big codes}. We observed that the \texttt{Sinter} package used to perform Monte Carlo sampling possibly causes memory leakage, because memory usage per core starts to increase over time and the code eventually crashes due to memory error. We tried to sample without using the \texttt{Sinter} package and found that the memory usage per core remains constant over time but ends up taking longer than using the \texttt{Sinter} package.

We also proposed a tri-layer architecture to efficiently parallelize stabilizer measurements, enabling a complete error correction cycle to be performed in \(2(2n-1)\) time steps, where \(n\) represents the \([n,k]\) classical code underlying the hypergraph product and the lifted product code. It makes efficient use of the cavity arrangements and also makes parallelization of stabilizer measurements possible. We became aware of a recent work where a four-qubit logical GHZ state was prepared with each logical qubit encoded in a non-local HGP code on a trapped-ion platform \cite{hong2024entangling}.

In our setup, performing computations with a $[[N,K,D]]$ code requires $3N$ physical qubits (roughly). Recent experimental advancements have demonstrated the ability to control up to $10,000$ Rydberg atoms \cite{Bohorquez_2023}. For \(N \sim 3000\), utilizing HGP codes, we can achieve \(K \sim 100-120\) and \(D \sim 10-15\). In contrast, using LP codes, it is possible to achieve \(K \sim 100\) with only \(N=714\) and \(D \leq 16\). Moreover, the LP codes can potentially support even higher numbers of logical qubits if the number of physical qubits is increased to \(3000\). Note that if we deploy redundification we will need more physical qubits than estimated here.\newline 

The ground state of an \(n_x \times n_y\) instance of the Fermi-Hubbard model can be estimated using \(2N\) logical qubits, where \(N = n_x \times n_y\). With a Rydberg atom quantum computer capable of handling around 100 logical qubits, it is possible to solve at least a \(7 \times 7\) instance, requiring up to 98 logical qubits \cite{cade2020strategies}. Even though high cooperativity is not currently available in the case of Rydberg atoms in optical cavities, it is potentially possible with microwave cavities. To implement the above-mentioned or other algorithms on a Rydberg atom quantum computer using HGP/LP codes, a fault-tolerant implementation of logical Clifford and non-Clifford gates is required. It is therefore essential to implement a fault-tolerant method for executing logical gates to realize a quantum processor. Moving forward, we plan to investigate the implementation of logical operators using our proposed setup. \newline

\section{Acknowledgments}
We thank Yumang Jing for insightful discussions on cavity error analysis.
GKB thanks Guido Pupillo, Sven Jandura, and Laura Pecorari for many helpful discussions. We also extend our gratitude to Craig Gidney for assistance with \texttt{STIM}-related questions, and to Joschka Roffe and Timo Hillmann for their help with the BPOSD package. OC thanks Pablo Poggi and Vineesha Shrivastava for their help with the figures. OC is supported by Sydney Quantum Academy, Sydney, Australia. We acknowledge support from the Australian Research Council Centre of Excellence for Engineered Quantum Systems (Grant No. CE 170100009). GM and GKB acknowledge funding from BTQ Technologies Corp.

\balance
\newpage
\appendix

\onecolumngrid
\section{Detailed calculation of state and map} \label{app:detailed calculation of state under faulty encoding and decoding maps}
\subsection{Imperfect encoding and perfect decoding} 

\subsubsection{The faulty encoding map} \label{app:faulty encoding map}

We approximate the effect of the faulty encoding map to first order. Here, $\mathcal{E}_{\text{eff}}$, given in Eq.~\ref{Eq:encoder} represents the mapping of the state $\rho = \sum\limits_{m,m'}\rho_{m,m'}\ket{\frac{N}{2},m}\bra{\frac{N}{2},m'}$ under the evolution of $H_{\text{eff}}$, given in Eq.~\ref{eq:Hamiltonian}. Since we have an extra $y$ rotation, we represent the state in $x$-basis instead, $\rho = \sum\limits_{m,m'} \rho_{m,m'}\ket{\frac{N}{2},m_x = m}\bra{\frac{N}{2},m_x = m'}$.
\begin{align}
    \mathcal{E}_{D^{-1}}(\rho) &= e^{i\frac{\pi}{2} \hat{J}_y} \mathcal{E}_{\text{eff}}\left(e^{-i\frac{\pi}{2} \hat{J}_y} \rho e^{i\frac{\pi}{2} \hat{J}_y}  \right) e^{-i\frac{\pi}{2} \hat{J}_y} \nonumber\\
    &= \sum_{m,m'} \rho_{m,m'} e^{i \theta_{\text{-}m,\text{-}m'}}\ket{\frac{N}{2},m_x = m}\bra{\frac{N}{2},m_x = m'}.
\end{align}

Here, $\theta_{m,m'}$ is defined in Eq.~\ref{Eq:phase}. The map can be divided into the ideal unitary (which prepares the perfect GHZ state) and a non-unitary part. In the following calculation, we assume that the $Y$ rotation is error free. A detailed analysis of the effect of depolarizing noise from the $Y$ rotation is given later. Note that the unitary and non-unitary parts commute. After Taylor expanding the non-unitary part we obtain,

\begin{align}
    \mathcal{E}_{D^{-1}}(\rho) & = e^{i\frac{\pi}{2} \hat{J}_y} \mathcal{E}_{\text{eff}}\left(e^{-i\frac{\pi}{2} \hat{J}_y} \rho e^{i\frac{\pi}{2} \hat{J}_y}  \right) e^{-i\frac{\pi}{2} \hat{J}_y} \\
    &= e^{i\frac{\pi}{2} \hat{J}_y} \mathcal{E}_{\text{eff}}\left(e^{-i\frac{\pi}{2} \hat{J}_y} \sum_{m,m'} \rho_{m,m'} \ket{J=\frac{N}{2},M_x = m}\bra{J=\frac{N}{2},M_x = m'} e^{i\frac{\pi}{2} \hat{J}_y}\right)e^{-i\frac{\pi}{2} \hat{J}_y} \\
    &=e^{i\frac{\pi}{2} \hat{J}_y} \mathcal{E}_{\text{eff}}\left(\sum_{m,m'} \rho_{m,m'} \ket{J=\frac{N}{2},M_z = -m}\bra{J=\frac{N}{2},M_z = -m'} \right)e^{-i\frac{\pi}{2} \hat{J}_y} \\
    &= e^{i\frac{\pi}{2} \hat{J}_y} \left(\sum_{m,m'} \rho_{m,m'} e^{i \theta_{-m,-m'}}\ket{J=\frac{N}{2},M_z = -m}\bra{J=\frac{N}{2},M_z = -m'} \right)e^{-i\frac{\pi}{2} \hat{J}_y} \\
    &= \sum_{m,m'} \rho_{m,m'} e^{i \theta_{-m,-m'}}\ket{J=\frac{N}{2},M_x = m}\bra{J=\frac{N}{2},M_x = m'}
\end{align}

From Eq.~\ref{Eq:phase} we have 
\begin{align}
    \theta_{-m,-m'} &= \left[ \left(m^2 - m'^2\right) - \left(m-m'\right)N\right]\theta + \frac{i \theta}{\sqrt{C} d_N} \left(m^2 + m'^2 - 2 mm'\right) \\
    &= m^2 \theta \left(1+\frac{i}{\sqrt{C}d_N}\right) - m'^2 \theta \left(1-\frac{i}{\sqrt{C}d_N}\right) -mN\theta + m' N\theta - \frac{2i}{\sqrt{C}d_N} \theta mm'
\end{align}

So we can write the map as,

\begin{align}
    \mathcal{E}_{D^{-1}}(\rho) & = e^{i \left(1 + \frac{i}{\sqrt{C}d_N}\right)\theta \hat{J}_x^2} e^{-iN\theta \hat{J}_x} \mathcal{E}'(\rho)e^{iN\theta \hat{J}_x}  e^{-i \left(1 - \frac{i}{\sqrt{C}d_N}\right)\theta \hat{J}_x^2} \\
    & = e^{i \theta \hat{J}_x^2} e^{-iN \theta \hat{J}_x} e^{\frac{-1}{\sqrt{C}d_N} \theta \hat{J}_x^2} \mathcal{E}'(\rho)e^{\frac{-1}{\sqrt{C}d_N}\theta \hat{J}_x^2} e^{iN \theta \hat{J}_x}e^{-i \theta \hat{J}_x^2}\\
    & = \hat{U} \left[e^{\frac{-1}{\sqrt{C}d_N} \theta \hat{J}_x^2} \mathcal{E}'(\rho)e^{\frac{-1}{\sqrt{C}d_N}\theta \hat{J}_x^2}\right] \hat{U}^{\dagger} \\
    \mathcal{E}'(\rho) & = \sum_{s=0}^{\infty} \left(\frac{2\theta}{\sqrt{C}d_N}\right)^s \frac{1}{s!} \hat{J}_x^s \rho \hat{J}_x^s \\
    e^{\frac{-1}{\sqrt{C}d_N} \theta \hat{J}_x^2} \mathcal{E}'(\rho)e^{\frac{-1}{\sqrt{C}d_N}\theta \hat{J}_x^2} &= \sum_{q,r,s=0}^{\infty}  (-1)^{q+r} 2^s\left(\frac{\theta}{\sqrt{C}d_N}\right)^{(q+r+s)} \frac{1}{q!r!s!} \hat{J}_x^{(s+2q)} \rho \hat{J}_x^{(s+2r)}.
\end{align}
Finally the map becomes,
\begin{align} 
    \mathcal{E}_D^{-1}(\rho) &= e^{-N\alpha d_N^2} \sum_{\substack{q_1,q_2, \\ r_1,r_2,s=0}}^{\infty} d_N^{2r} 2^s\alpha^{(q+r+s)} \frac{1}{q_1!q_2!r_1r_2!s!} \nonumber \\&
    \times \hat{J}_x^{(2q_1 + r_1 +s)} \hat{E}_D^{-1}(\rho) \hat{J}_x^{(2q_2 + r_2 + s)}.
\end{align}
with $q = q_1 + q_2$, $r = r_1 + r_2$, and \(\alpha = \theta/({2\sqrt{C}d_N})\). For GHZ state preparation, \(\theta=\pi/2\) and \(\alpha=\pi/(4d_N\sqrt{C})\). Note that $\hat{E}_D^{-1}$ in the equation above is the unitary part.  For \(\alpha \ll 1\), which is compatible with QLDPC codes where \(N\) is constant and where we have large \(C\) for good non-local gates, we only keep the first order terms with $q+r+s \leq 1$,

\begin{align} \label{app/final cavity error term}
\mathcal{E}_D^{-1}&(\rho) \approx e^{-N\alpha d_N^2}\left[\tau  + 2\alpha\hat{J}_x \tau \hat{J}_x \right.\nonumber \\ 
&\left. + \alpha d_N^2\left(\hat{J}_x \tau + \tau \hat{J}_x \right)  - \alpha \left(\hat{J}_x^2 \tau + \tau \hat{J}_x^2 \right) \right] \\ \nonumber
&\approx \tau \left(1- N\alpha d_N^2 \right) + 2\alpha\hat{J}_x \tau \hat{J}_x \nonumber \\ &+ \alpha d_N^2 \left(\hat{J}_x \tau + \tau \hat{J}_x \right)  - \alpha \left(\hat{J}_x^2 \tau + \tau \hat{J}_x^2 \right),
\end{align}

where $\tau = \hat{E}_D^{-1}(\rho)$ is the output of perfect encoding. We have the following relations from angular-momentum algebra, 
\begin{align}
    \hat{J}_x & =\frac{\hat{J}_+ + \hat{J}_-}{2} \\
    \hat{J}_x^2 & = \hat{J}_+^2 + \hat{J}_-^2 + \hat{J}_+\hat{J}_- + \hat{J}_-\hat{J}_+ .
\end{align}
The terms $\left(\hat{J}_x \tau + \tau \hat{J}_x\right)$, and $\left(\hat{J}_x^2 \tau + \tau \hat{J}_x^2\right)$ will not contribute to the final measurement as they are not diagonal in the $Z$ basis. We can see this by expanding the terms in angular momentum basis as,
\begin{align}
\bra{m} \hat{J}_x^2 \tau \ket{m}=\bra{m}(\hat{J}_+^2 + \hat{J}_-^2 + \hat{J}_+\hat{J}_- + \hat{J}_-\hat{J}_+)\tau \ket{m}.
\end{align}
The terms $\bra{m}(\hat{J}_+^2 + \hat{J}_-^2)\tau\ket{m}$ will not contribute to the diagonals. The leftover term: $(\hat{J}_+\hat{J}_- + \hat{J}_-\hat{J}_+)$ can be rewritten as $(\hat{J}^2-\hat{J}_z^2)$ and this term keeps the GHZ state invariant as shown below, 

\begin{align}
    \bra{m}(\hat{J}^2-\hat{J}^2_z)\tau\ket{m}=(J(J+1)-m^2) \bra{m}\tau \ket{m}.
\end{align}

Since $\hat{J}^2$ preserves the angular momentum basis and the $\hat{J}_z^2$ terms remain undetectable in the final Z-basis measurement, we omitted these terms in our simulations. Despite their complexity, they are inconsequential and do not affect the results. So the map after faulty encoding is
\begin{align}
\label{eq:cavity_state}
    \tilde{\tau} \approx (1-p_0) \tau + p_{\text{cavity}} \hat{J}_x \tau \hat{J}_x ,
\end{align} where \( p_0 = N\alpha d_N^2 \) and \( p_{\text{cavity}} = 2\alpha \). \newline

Now, let's consider the case when the \( Y \) rotations are also faulty. We model a faulty \(Y\)-rotation using a depolarization error model. Let $p_d$ be the probability of a depolarizing noise after the rotation. At every step, we only keep errors up to the first order. The effect of depolarizing noise is as follows,
\begin{align}
\mathcal{E}_y(\rho) = (1-N p_d) \tau + \frac{p_d}{3} \sum_{j=1}^N X_j \tau X_j + Y_j \tau Y_j + Z_j \tau Z_j,
\end{align}
where $\tau = e^{-i \frac{\pi}{2} J_y} \rho e^{i \frac{\pi}{2} J_y}$. Let $\rho_0$ be the initial state, $\rho_1$ be the state after the first $Y$ rotation, $\rho_2$ be the state after the cavity map, and finally $\tau$ be the state after second $Y$ rotation. We put tildes on each of them to denote the output of noisy map.
\onecolumngrid
\begin{align}
    \tilde{\rho}_1 &= (1-N p_d) \rho_1 + \frac{p_d}{3} \sum_{j=1}^N X_j \rho_1 X_j + Y_j \rho_1 Y_j + Z_j \rho_1 Z_j \\
    \tilde{\rho}_2 &= (1 - \frac{N \theta d_N}{2\sqrt{C}}) (1-Np_d) \rho_2 + 2 \alpha (1-Np_d) J_z \rho_2 J_z + \alpha(1-Np_d) d_N^2\left( J_z \rho_2 + \rho_2 J_z\right) \nonumber \\
    &+ \alpha (1-Np_d) \left(J_z^2 \rho_2 + \rho_2 J_z^2\right) + \frac{p_d}{3} E_D^{-1}\left(\sum_{j=1}^N X_j \rho_1 X_j + Y_j \rho_1 Y_j + Z_j \rho_1 Z_j\right) \nonumber \\
    &\approx (1 - \frac{N \theta d_N}{2\sqrt{C}} - Np_d ) \rho_2 + 2 \alpha J_z \rho_2 J_z + \alpha d_N^2\left( J_z \rho_2 + \rho_2 J_z\right) + \alpha \left(J_z^2 \rho_2 + \rho_2 J_z^2\right) \nonumber \\
    &+ \frac{p_d}{3}  E_D^{-1}\left(\sum_{j=1}^N X_j \rho_1 X_j + Y_j \rho_1 Y_j + Z_j \rho_1 Z_j\right) \\
   \tilde{\tau} & \approx (1 - \frac{N \theta d_N}{2\sqrt{C}} - 2 Np_d ) \tau + \frac{p_d}{3} \sum_{j=1}^N X_j \tau X_j + Y_j \tau Y_j + Z_j \tau Z_j + 2 \alpha J_x \tau J_x + \alpha d_N^2\left( J_x \tau + \tau J_x\right) \nonumber \\
    &+ \alpha \left(J_x^2 \tau + \tau J_x^2\right) + \underbrace{\frac{p_d}{3}  e^{i \frac{\pi}{2} J_y}  E_D^{-1}\left(\sum_{j=1}^N X_j \rho_1 X_j + Y_j \rho_1 Y_j + Z_j \rho_1 Z_j\right) e^{-i \frac{\pi}{2} J_y}}_\sigma \\
    \sigma &= \frac{p_d}{3} e^{-i\frac{\pi}{2}\left(\hat{J}_x^2-N \hat{J}_x\right)} \left(\sum_{j=1}^N X_j \rho_0 X_j + X_j Z_j \rho_0 Z_j X_j + Z_j \rho_0 Z_j \right) e^{i\frac{\pi}{2}\left(\hat{J}_x^2-N \hat{J}_x\right)} \nonumber \\
    &=  \frac{p_d}{3}  \sum_{j=1}^N \left(2 X_j \tau X_j + \tau \right) \nonumber \\
    &\approx \frac{p_d}{3} \left( 8 J_x \tau J_x + N \tau\right).
\end{align}

Here we have used that \(\rho_0 = \ket{0}\bra{0}^{\otimes N}\). In the last step, we approximated the individual bit-flip errors by a global \(J_x\) error. To see this, consider the expansion:
\begin{align}
    J_x \tau J_x &= \left( \frac{1}{2} \sum_i X_i \right) \tau \left( \frac{1}{2} \sum_j X_j \right) \\
    &= \frac{1}{4} \sum_{i,j} X_i \tau X_j \\
    &\approx \frac{1}{4} \sum_i X_i \tau X_i.
\end{align}
 We ignore the cross terms with \(i \ne j\) because we trace out those contributions when measuring in the \(Z\)-basis. These terms typically affect only off-diagonal components of the density matrix in that basis, which do not contribute to our final measurement outcomes. There are additional off-diagonal terms that similarly have no effect, and we omit those as well. Combining all of this, we obtain

\begin{align}
\label{eq:depol_encoding}
    \tilde{\tau} & \approx (1 - \frac{N \theta d_N}{2\sqrt{C}} - \frac{5}{3} Np_d ) \tau  +  (\frac{\theta}{\sqrt{C}d_N}+ \frac{8 p_d}{3}) J_x \tau J_x \nonumber \\ 
    &+ \frac{p_d}{3} \sum_{j=1}^N X_j \tau X_j + Y_j \tau Y_j + Z_j \tau Z_j
\end{align}

For simplicity, we write the above expression in terms of parameter \(\alpha=\theta/(2 \sqrt{C}d_N)\). For GHZ state preparation \(\theta=\pi/2\), which implies \(\alpha=\pi/(4d_N \sqrt{C})\). Expression becomes, 
\begin{align} \label{eq:appendix, faulty encoding map}
    \tilde{\tau}=(1-N\alpha d_N -5Np_d/3)\tau +(2\alpha+8p_d/3)J_x\tau J_x + \frac{p_d}{3}  \sum_{j=1}^N \left(X_j \tau X_j + Y_j \tau Y_j + Z_j \tau Z_j \right)
\end{align}

This states looks similar to Eq.~\ref{eq:cavity_state} where the probabilities have been modified. The extra depolarization noise can be considered as a noise after the encoding step.

\subsubsection{Computing the state} \label{app: state in faulty encoding and perfect decoding case}
Now let's compute the state in each step of stabilizer measurement circuit and see how the final state before measurement looks like. In the calculations below, $\upsilon$ denotes the combined ancilla \(+\) data state without noise, and \(\tilde{\upsilon}\) denotes the state with noise. Now, we will go through each steps of the circuit shown in Fig.~\ref{Fig:Divincezo-Aliferis circuit}:

We start step-1 with the ancilla in all-zero state. After encoding the ancilla is in the state \( (1-\tilde{p}_0)\rho_{_\mathrm{GHZ}} + \tilde{p}_{\text{cavity}}\hat{J}_x \rho_{_\mathrm{GHZ}} \hat{J}_x\), where \(\rho_{_\mathrm{GHZ}}\) is the perfect GHZ state as defined in Eq.~\ref{Eq:rhoGHZ}. We have kept aside the depolarizing term. We will consider this later. The combined state after step 1 is, 
\begin{equation}
\tilde{\upsilon}_1 \approx \left((1-\tilde{p}_0)\rho_{_\mathrm{GHZ}} + \tilde{p}_{\text{cavity}}\hat{J}_x \rho_{_\mathrm{GHZ}} \hat{J}_x\right) \otimes \sigma.
\end{equation}
Where \(\sigma = \ket{\psi}\bra{\psi}\) is the state of the data qubits. 

Step 2 is ancilla-data C-M gates. From the perfect encoding and perfect decoding case \ref{sec:perfect}, we know how $\rho_{_\mathrm{GHZ}}$ transforms after this step. Now we need to determine the transformation of $\hat{J}_x\rho_{_\mathrm{GHZ}} \hat{J}_x$. Instead of directly calculating the transformation of $\hat{J}_x\rho_{_\mathrm{GHZ}} \hat{J}_x$ , we can track the transformation of the operator $\hat{J}_x$ and then apply to $\upsilon_2$. Data qubits in the support of \(X\) stabilizers is denoted as a \(\{q_j\}_{\mu}\) where \(\mu\) denotes a stabilizer and \(j\) denotes the index of the data qubits in support of that stabilizer. We also define the set of indices \(\{j\}\) which stores the information about the support of a stabilizer. Say \(\mu=M\), representing an \(X\) stabilizer then \(\{q_j\}_{\mu=M}\) represents the data qubits in support of stabilizer \(M\) such that, \(M=\prod_{i\in \{q_j\}_{\mu=M}} X_i\). The set of ancilla neighboring the set of data qubits \(\{q_j\}_{\mu=M}\) is \(\{a_j\}_{\mu=M}\). We can write the collective \(X\)-rotation operator acting on \(\{a_j\}_{\mu=M}\) as \(\hat{J}_x^{[a]} = \sum\limits_{i \in \{a_j\}_{\mu=M}} \hat{X}_i/2\).  
We know CNOTs spread an \(X\) error acting on control to target \ref{App:spread of Pauli errors}, data qubits will get inflicted by an \(X\) error if the corresponding ancilla qubit gets an \(X\) error.  
We will use $X_i$ for Pauli-X acting on ancilla qubits \(\{a_j\}\) and $X'_i$ for Pauli-X acting on data qubits \(\{q_j\}\). Note that we skipped the index \(\mu\) for convenience. \newline 
The \(\hat{J}^{[a]}_x\) operator spreads the \(X\) error bit-wise to the data qubits according to the circuit shown in Fig.~\ref{Fig:Divincezo-Aliferis circuit}. If we denote the combined controlled operations as \(\hat{V}\), we get, 
\begin{align}
    \hat{\mathcal{X}} = \hat{V} \left(\hat{J}^{[a]}_x \otimes \mathds{1}\right)\hat{V}^{\dagger} &= \frac{1}{2} \sum_{i \in \{j\}} \hat{X}^{[a]}_i\otimes \hat{X}'^{[q]}_i,
\end{align}
where the superscript \([a]\) and \([q]\) denotes the operations acting on ancilla and data, respectively. 
The state after this step becomes:
 \begin{align}
    \tilde{\upsilon}_2 &\approx \hat{V} \tilde{\upsilon}_1\hat{V}^{\dagger} \nonumber \\
    &=\hat{V} \left(((1-\tilde{p}_0)\rho_{_\mathrm{GHZ}} + \tilde{p}_{\text{cavity}} \hat{J}_x\rho_{_\mathrm{GHZ}} \hat{J}_x)\otimes \sigma \right)\hat{V}^{\dagger} \nonumber \\
    &= (1-\tilde{p}_0) \upsilon_2 + \tilde{p}_{\text{cavity}} \hat{V} \left(\hat{J}_x \otimes \mathds{1}\right)\hat{V}^{\dagger} \ \upsilon_2 \ \hat{V} \left(\hat{J}_x \otimes \mathds{1}\right)\hat{V}^{\dagger} \nonumber \\
    &= (1-\tilde{p}_0) \upsilon_2 + \tilde{p}_{\text{cavity}} \hat{\mathcal{X}} \upsilon_2 \hat{\mathcal{X}}^{\dagger}
 \end{align}

Step 3 is the decoding step. Since we are considering the case of perfect decoder, $\hat{E}_D$ transforms the perfect state $\upsilon_2$ to $\upsilon_3$. So the imperfect state becomes,

\begin{align}
     &\tilde{\upsilon}_3 \approx (\hat{U}_E\otimes \hat{I})\left((1-\tilde{p}_0) \upsilon_2 + \tilde{p}_{\text{cavity}} \hat{\mathcal{X}} \upsilon_2 \hat{\mathcal{X}}^{\dagger}\right)(\hat{U}_E^{\dagger}\otimes \hat{I}) \nonumber \\  
     &=(\hat{U}_E \otimes \hat{I})(1-\tilde{p}_0)\upsilon_2(\hat{U}_E^{\dagger}\otimes I) \nonumber \\
     &+ \tilde{p}_{\text{cavity}} \Big[(\hat{U}_E\otimes \hat{I}) \hat{\mathcal{X}} (\hat{U}_E^{\dagger}\otimes \hat{I})\Big] \upsilon_3 \Big[ (\hat{U}_E\otimes \hat{I}) \hat{\mathcal{X}}^{\dagger} (\hat{U}_E^{\dagger}\otimes \hat{I})\Big] \nonumber \\ 
     &=(1-\tilde{p}_0) \upsilon_3 + \tilde{p}_{\text{cavity}} \hat{\mathcal{X}} \upsilon_3 \hat{\mathcal{X}}^{\dagger}. 
\end{align}

Here, \(\tilde{\upsilon}_3\) represents the imperfect state after step 3, while \(\upsilon_3\) denotes the final perfect state from case~\ref{sec:perfect} as defined in Eq.~\ref{Eq:upsilon_3}. The last step follows from the fact that \(\hat{U}_E\) commutes with \(\hat{\mathcal{X}}\), i.e., \([\hat{U}_E, \hat{\mathcal{X}}] = 0\). We substitute the terms of \(\upsilon_3\) into the above equation and obtain error terms like,

\begin{align}\label{Eq:steane_meas}
&\sum_{i \in \{j\}} X^{[a]}_i \ket{\frac{N}{2}}\bra{\frac{N}{2}}X^{[a]}_{i}\otimes X_i'^{[q]}(1+M)\sigma(1+M)X_i'^{[q]}  \nonumber \\
&+X_i^{[a]}\ket{-\frac{N}{2}}\bra{-\frac{N}{2}}X_i^{[a]}\otimes X_i'^{[q]}(1-M)\sigma(1-M)X_i'^{[q]}.
\end{align}

Here, \(M\) refers to a general stabilizer. 

Now we analyze the effect of the depolarizing noise term as stated by map in Eq.~\ref{eq:appendix, faulty encoding map}. The $X_i$ term proceed exactly like the $J_x$ noise we discussed and, $Y_i$ can be rewritten as $Z_iX_i$. Therefore, we only need to analyze how the $Z_i$ type error propagates throughout the circuit. These errors commute with the ancilla-data controlled gates. After passing through the decoder, we get 
\begin{align}
     (\hat{U}_E\otimes \hat{I})(Z_i) \upsilon_2 (Z_i)(\hat{U}_E^{\dagger}\otimes \hat{I}) = \prod_j X_j  \  \upsilon_3 \  \prod_j X_j \\ 
\end{align}
Note that, \([\hat{U}_E,Z_i]=X_1...Y_i..X_N\), but \(Y\) error in \(Z\) basis measurement is effectively an \(X\) error and we can effectively say that all the qubits have flipped. The errors act on the ancilla qubits after interaction with data. So this is basically a measurement error and the effect of this is to modify strength of the measurement error. Since all qubits flipping due to single qubit measurement error is of higher order and therefore very rare, an event like this can be detected.
Final step involves measuring all the ancilla qubits and depending upon the measurement outcome we can tell which data qubits have been affected.


\subsection{Perfect encoding and imperfect decoding}

\subsubsection{Imperfect decoding map}
The imperfect decoding map in the absence of \(Y\)-rotation errors is, on an initial state, $\rho_0$ is
\begin{align}
    \tilde{\tau} &=\mathcal{E}_D(\rho_0)= e^{-i\frac{\pi}{2}\hat{J}_y} \mathcal{E}_{\text{eff}}^{-1}\left(e^{i\frac{\pi}{2}\hat{J}_y}\rho_0 e^{-i\frac{\pi}{2}\hat{J}_y}\right) e^{i\frac{\pi}{2}\hat{J}_y} \\
    &= \sum_{m,m'} \rho_{m,m'} e^{-i \theta^*_{m,m'}}\ket{\frac{N}{2},m_x = m}\bra{\frac{N}{2},m_x = m'} \\
    &\approx \tau - \frac{N\theta d_N}{2\sqrt{C}} \tau + \frac{\theta}{\sqrt{C}d_N}\hat{J}_x \tau \hat{J}_x \nonumber \\ &- \frac{\theta d_N}{2\sqrt{C}} \left(\hat{J}_x \tau + \tau \hat{J}_x \right)  - \frac{\theta}{2\sqrt{C}d_N} \left(\hat{J}_x^2 \tau + \tau \hat{J}_x^2 \right).  
\end{align}
Assuming that no other errors occurred before the decoding, the first noise that it encounters is the depolarizing noise due to imperfect \(Y\)-rotation. We analyze the state in each operation,   
\begin{align}
    \tilde{\rho}_1 &= (1-N p_d) \rho_1 + \frac{p_d}{3} \sum_{j=1}^N X_j \rho_1 X_j + Y_j \rho_1 Y_j + Z_j \rho_1 Z_j \\
    \tilde{\rho}_2 &= (1 - \frac{N \theta d_N}{2\sqrt{C}}) (1-Np_d) \rho_2 + 2 \alpha (1-Np_d) J_z \rho_2 J_z - \alpha(1-Np_d) d_N^2\left( J_z \rho_2 + \rho_2 J_z\right) \nonumber \\
    &- \alpha (1-Np_d) \left(J_z^2 \rho_2 + \rho_2 J_z^2\right) + \frac{p_d}{3} E_D\left(\sum_{j=1}^N X_j \rho_1 X_j + Y_j \rho_1 Y_j + Z_j \rho_1 Z_j\right) \nonumber \\
    &\approx (1 - \frac{N \theta d_N}{2\sqrt{C}} - Np_d ) \rho_2 + 2 \alpha J_z \rho_2 J_z  - \alpha d_N^2\left( J_z \rho_2 + \rho_2 J_z\right) - \alpha \left(J_z^2 \rho_2 + \rho_2 J_z^2\right) \nonumber \\
    &+ \frac{p_d}{3}  E_D\left(\sum_{j=1}^N X_j \rho_1 X_j + Y_j \rho_1 Y_j + Z_j \rho_1 Z_j\right) \\
   \tilde{\tau} & \approx (1 - \frac{N \theta d_N}{2\sqrt{C}} - 2 Np_d ) \tau + \frac{p_d}{3} \sum_{j=1}^N X_j \tau X_j + Y_j \tau Y_j + Z_j \tau Z_j + 2 \alpha J_x \tau J_x - \alpha d_N^2\left( J_x \tau + \tau J_x\right) \nonumber \\
    & - \alpha \left(J_x^2 \tau + \tau J_x^2\right) + \underbrace{\frac{p_d}{3}  e^{-i \frac{\pi}{2} J_y}  E_D\left(\sum_{j=1}^N X_j \rho_1 X_j + Y_j \rho_1 Y_j + Z_j \rho_1 Z_j\right) e^{i \frac{\pi}{2} J_y}}_\sigma \\
    \sigma &= \frac{p_d}{3} e^{i\frac{\pi}{2}\left(\hat{J}_x^2-N \hat{J}_x\right)} \left(\sum_{j=1}^N X_j \rho_0 X_j + X_j Z_j \rho_0 Z_j X_j + Z_j \rho_0 Z_j \right) e^{-i\frac{\pi}{2}\left(\hat{J}_x^2-N \hat{J}_x\right)} \nonumber \\
    &= \frac{p_d}{3}  \left(\sum_{j=1}^N X_j \tau X_j + X_j \prod_k X_k \tau \prod_k X_k X_j + N \prod_k X_k \tau \prod_k X_k \right)  \nonumber
\end{align}

Combining everything we get,
\begin{align}
\label{eq:depol_decoding}
    \tilde{\tau} & \approx (1 - \frac{N \theta d_N}{2\sqrt{C}} - 2 Np_d ) \tau  +  (\frac{\theta}{\sqrt{C}d_N}) J_x \tau J_x+ \frac{p_d}{3} \left(\sum_{j=1}^N X_j \tau X_j + Y_j \tau Y_j + Z_j \tau Z_j \right) \nonumber \\
    &+ \frac{p_d}{3}  \left(\sum_{j=1}^N X_j \tau X_j + X_j \prod_k X_k \tau \prod_k X_k X_j + N \prod_k X_k \tau \prod_k X_k \right),
\end{align}

which can be rewritten as, 
\begin{align}
    \tilde{\tau} & \approx (1 - \tilde{p_0}) \tau  +  p_{\text{cavity}} J_x \tau J_x+ \frac{p_d}{3} \left(\sum_{j=1}^N X_j \tau X_j + Y_j \tau Y_j + Z_j \tau Z_j \right) \nonumber \\
    &+ \frac{p_d}{3}  \left(\sum_{j=1}^N X_j \tau X_j + X_j \prod_k X_k \tau \prod_k X_k X_j + N \prod_k X_k \tau \prod_k X_k \right),
\end{align}

where \(\tilde{p}_0= \alpha N d_N^2+2Np_d\), \(p_{\text{cavity}}=2\alpha\) and \(\alpha=\theta/(2d_N \sqrt{C})\).

\subsubsection{Computing the state}
Since the error is only in the decoding step, in the absence of depolarizing noise, we get the final state to be,
\begin{align}
    \tilde{\upsilon}_3 & \approx (1-p_0)\upsilon_3 + p_{\text{cavity}} \left(\hat{J}_x \otimes \mathds{1}\right) \upsilon_3 \left(\hat{J}_x \otimes \mathds{1}\right).
\end{align}
The state after taking into account the imperfect \(Y\)-rotation is

\begin{align} \label{eq:state after imperfect decoding}
\tilde{\upsilon}_3 & \approx (1 - \tilde{p}_0 ) \upsilon_3 + p_{\text{cavity}} J_x \upsilon_3 J_x + \frac{p_d}{3} \Bigg(\sum_{j=1}^N X_j \upsilon_3 X_j + Y_j \upsilon_3 Y_j  \nonumber \\
& + Z_j \upsilon_3 Z_j \Bigg) +  \\
& \frac{p_d}{3}\left( \sum_{j=1}^N X_j \prod_k X_k \upsilon_3 \prod_k X_k X_j + N \prod_k X_k \upsilon_3 \prod_k X_k \right).
\end{align}
where $\tilde{p}_0 = \alpha Nd_N^2 + 2 Np_d$. Note that in this case, there is no modification to $p_{\text{cavity}}$. Also, the last term we got in this case was not there in the case of imperfect encoding and we get that because of the commutation relation between the \(Y\) and \(Z\) errors from the depolarizing error model and the decoding operation.

\subsection{Leakage error analysis}

We note that the noisy encoding or decoding process is not trace preserving. This is shown by the fact that the trace of the final state in Eq.~\ref{eq:depol_encoding} or Eq.~\ref{eq:depol_decoding} is less than 1. This is due to the fact that the $\ket{1}$ state can decay into a state outside the qubit subspace, which we can denote by $\ket{a}$. The Kraus operators corresponding to this decay process are given by
\begin{align}
\mathcal{K}_1 &= \sqrt{p} \ket{a}\bra{1}    \\
\mathcal{K}_2 &= \ket{0}\bra{0} + \ket{a}\bra{a} + \sqrt{1-p} \ket{1}\bra{1}.
\end{align}

Let us look at different cases how this leakage error can affect.
\subsubsection{Leakage during encoding}
The state after perfect encoding is 

\begin{align}
    \upsilon_1 &= \left[\ket{0^{\otimes N}} + i \ket{1^{\otimes N}}\right]\left(\bra{0^{\otimes N}} - i \bra{1^{\otimes N}}\right) \otimes \ket{\psi}\bra{\psi} \nonumber \\
    &= \left[\Ketbra{0^{\otimes N}}  + i \Ketbra{{1^{\otimes N}}}{{0^{\otimes N}}}
 -i \Ketbra{0^{\otimes N}}{1^{\otimes N}} + \Ketbra{1^{\otimes N}}\right] \otimes \Ketbra{\psi}
\end{align}

We calculate what happens when Krauss operators act on it. Here we assume that first qubit decays. But analysis is similar for any qubit. The resulting state is 

\begin{align}
    \tilde{\upsilon_1} &=  \left[ p\Ketbra{a, 1^{\otimes {(N-1)}}} + \Ketbra{0^{\otimes N}}  + i \sqrt{1-p}  \Ketbra{{1^{\otimes N}}}{{0^{\otimes N}}}
 \right. \nonumber \\
    &\quad \left.-i \sqrt{1-p}  \Ketbra{0^{\otimes N}}{1^{\otimes N}} + (1-p) \Ketbra{1^{\otimes N}}\right] \otimes \Ketbra{\psi}
\end{align}

When this state passes through the controlled stabilizer gates, we get 

\begin{align}
     \tilde{\upsilon_2} &=p\Ketbra{a, 1^{\otimes {(N-1)}}} \otimes \tilde{M} \Ketbra{\psi}\tilde{M}  + \Ketbra{0^{\otimes N}} \otimes \Ketbra{\psi} + i \sqrt{1-p}  \Ketbra{{1^{\otimes N}}}{{0^{\otimes N}}} \otimes M \Ketbra{\psi} \nonumber\\
    & -i \sqrt{1-p}  \Ketbra{0^{\otimes N}}{1^{\otimes N}}  \otimes  \Ketbra{\psi} M  + (1-p) \Ketbra{1^{\otimes N}} \otimes M \Ketbra{\psi}M 
\end{align}

Next step is the redundification. Here we consider 2 sets of ancilla for redundification. So after this step we get.

\begin{align}
    \tilde{\upsilon_3} &= p\Ketbra{a, 1^{\otimes {(N-1)}}} \otimes \Ketbra{0, 1^{\otimes {(N-1)}}} \otimes \Ketbra{0, 1^{\otimes {(N-1)}}}\otimes  \tilde{M} \Ketbra{\psi}\tilde{M} \nonumber \\
    & + \Ketbra{0^{\otimes N}} \otimes \Ketbra{0^{\otimes N}} \otimes\Ketbra{0^{\otimes N}} \otimes \Ketbra{\psi} \nonumber \\
    & + i \sqrt{1-p}  \Ketbra{{1^{\otimes N}}}{{0^{\otimes N}}} \otimes \Ketbra{{1^{\otimes N}}}{{0^{\otimes N}}} \otimes \Ketbra{{1^{\otimes N}}}{{0^{\otimes N}}} \otimes M \Ketbra{\psi} \nonumber\\
    & -i \sqrt{1-p}  \Ketbra{0^{\otimes N}}{1^{\otimes N}}  \otimes  \Ketbra{0^{\otimes N}}{1^{\otimes N}}  \otimes \Ketbra{0^{\otimes N}}{1^{\otimes N}}  \otimes \Ketbra{\psi} M \nonumber\\
    & + (1-p) \Ketbra{1^{\otimes N}} \otimes \Ketbra{1^{\otimes N}} \otimes \Ketbra{1^{\otimes N}} \otimes M \Ketbra{\psi}M 
\end{align}

Now the decoding part does not affect the $\ket{a}$ state. So it just acts as if there was one less ancilla qubit. The state after decoding is

\section{Analysis of errors under two level redundification} \label{app:redundification calculation} 

The error-free cavity unitary is \(\hat{U}_E=e^{i \frac{\pi}{2} J_y}e^{\frac{-i\pi}{2} (J_z^2 - N J_z)}e^{-i\frac{\pi}{2}J_y}=e^{\frac{-i\pi}{2} (J_x^2 - N J_x)}\), where \(J_x\) is the collective angular momentum operator. \(\hat{U}_E\) acts on the computational basis states as follows,  

\begin{align}
    \hat{U}_E\ket{0}^{\otimes N}= & \ket{0}^{\otimes N} + i\ket{1}^{\otimes N}, \\
    \hat{U}_E\ket{1}^{\otimes N}= & i\ket{0}^{\otimes N} + \ket{1}^{\otimes N}.
\end{align}

Decoding unitary \(\hat{U}_E ^{\dagger}\) acts as:

\begin{align}
    \hat{U}_E ^{\dagger}\ket{0}^{\otimes N}= & \ket{0}^{\otimes N} - i\ket{1}^{\otimes N}\\
    \hat{U}_E ^{\dagger} \ket{1}^{\otimes N}= & -i\ket{0}^{\otimes N} + \ket{1}^{\otimes N}.
\end{align}

Note that we use the representation of state \(\ket{0}^{\otimes N}\) and \(\ket{\frac{-N}{2}}\) interchangeably. 
Step-by-step circuit analysis:

\begin{itemize}
    \item Start with ancilla state \(\ket{0}_{a_1}^{\otimes N}\) and data state \(\ket{\Psi}\).

    \item Apply encoding unitary \(\hat{U}_E\otimes I\):
    \begin{align}
        (\hat{U}_E\otimes I)\ket{0}_{a_1}^{\otimes N}\otimes\ket{\Psi}= (\ket{0}_{a_1}^{\otimes N}+i\ket{1}_{a_1}^{\otimes N})\otimes\ket{\Psi}.
    \end{align}

    \item C-M gates between ancilla-data to implement stabilizer \(M\):
    \begin{align}
        \ket{0}_{a_1}^{\otimes N}\otimes\ket{\Psi}+ i\ket{1}_{a_1}^{\otimes N}\otimes M\ket{\Psi}.
    \end{align}

    \item Introduce two additional ancillas \(\ket{0}_{a_2}^{\otimes N}, \ket{0}_{a_3}^{\otimes N}\) and apply transversal C-M:
    \begin{align}
        &\ket{0}_{a_1}^{\otimes N}\ket{0}_{a_2}^{\otimes N}\ket{0}_{a_3}^{\otimes N}\ket{\Psi}+i\ket{1}_{a_1}^{\otimes N}\ket{1}_{a_2}^{\otimes N}\ket{1}_{a_3}^{\otimes N}M\ket{\Psi}.
    \end{align}

    \item Decode all ancilla blocks separately with \(\hat{U}_E^{\dagger} \otimes \hat{U}_E^{\dagger} \otimes \hat{U}_E^{\dagger}\):
    \begin{align}
        & (\ket{0}^{\otimes N}-i\ket{1}^{\otimes N})_{a_1}(\ket{0}^{\otimes N}-i\ket{1}^{\otimes N})_{a_2}(\ket{0}^{\otimes N}-i\ket{1}^{\otimes N})_{a_3}\nonumber\\
        &+ i(-i\ket{0}^{\otimes N}+\ket{1}^{\otimes N})_{a_1}(-i\ket{0}^{\otimes N}+\ket{1}^{\otimes N})_{a_2}(-i\ket{0}^{\otimes N}+\ket{1}^{\otimes N})_{a_3}.
    \end{align}

    Upon expanding, we have:

    \begin{equation} \label{eq: three level redundification}
    \begin{aligned}
        &(\ket{0}^{\otimes N}_{a_1}\ket{0}^{\otimes N}_{a_2}\ket{0}^{\otimes N}_{a_3}
        -\ket{0}^{\otimes N}_{a_1}\ket{1}^{\otimes N}_{a_2}\ket{1}^{\otimes N}_{a_3}
        -\ket{1}^{\otimes N}_{a_1}\ket{0}^{\otimes N}_{a_2}\ket{1}^{\otimes N}_{a_3}
        -\ket{1}^{\otimes N}_{a_1}\ket{1}^{\otimes N}_{a_2}\ket{0}^{\otimes N}_{a_3})(I-\hat{M})\ket{\Psi}\\[0.3cm]
        &-i(\ket{0}^{\otimes N}_{a_1}\ket{0}^{\otimes N}_{a_2}\ket{1}^{\otimes N}_{a_3}
        +\ket{0}^{\otimes N}_{a_1}\ket{1}^{\otimes N}_{a_2}\ket{0}^{\otimes N}_{a_3}
        +\ket{1}^{\otimes N}_{a_1}\ket{0}^{\otimes N}_{a_2}\ket{0}^{\otimes N}_{a_3}
        +\ket{1}^{\otimes N}_{a_1}\ket{1}^{\otimes N}_{a_2}\ket{1}^{\otimes N}_{a_3})(I+\hat{M})\ket{\Psi}.
    \end{aligned}
    \end{equation}

    \item Final step is the measurement of all ancilla blocks in the \(Z\) basis. In each ancilla block, the measurement outcome is assigned a bit value after majority voting. We obtain three bit values corresponding to each of the ancilla blocks. The Hamming weight of these bit values determines the stabilizer outcome. \newline
    \textit{Example}: If the weight of the stabilizer is \(N=4\) (as in Steane's code) and after measurement and majority voting we obtain 
\begin{align*}
    (0,0,0,0)_{a_1} &\rightarrow 0_{a_1}, \\ 
    (1,1,1,1)_{a_2} &\rightarrow 1_{a_2}, \\ 
    (1,1,1,1)_{a_3} &\rightarrow 1_{a_3}, 
\end{align*}
the final outcome is \((0_{a_1},1_{a_2},1_{a_3})\), which corresponds to the \(+1\) eigenspace of the stabilizer \(M\). 

Upon careful observation, we find that the Hamming weight of the bit values from all three ancilla blocks is sufficient to determine the stabilizer measurement outcome: if the Hamming weight is even, we are in the \(+1\) eigenspace; if it is odd, we are in the \(-1\) eigenspace.
\end{itemize}

In the following section, we will go through each of the error-prone parts of the circuit~\ref{Fig:Divincezo-Aliferis circuit} and analyze their effect.

\subsection{Error in Encoding}

The encoding operation involves first applying the \(\pi/2\) angle \(Y\)-rotation \(e^{-i\frac{\pi}{2}\hat{J}_y}\) followed by cavity unitary \(\hat{\mathcal{U}}_c = e^{-i\frac{\pi}{2}(\hat{J}_z^2 - N\hat{J}_z)}\) and finally \(-\pi/2\) angle \(Y\)-rotation \(e^{i\frac{\pi}{2}\hat{J}_y}\). All three of these operations can potentially introduce errors. Let's analyze them step-by-step. 

\subsubsection{Error in cavity unitary \(\hat{\mathcal{U}}_c\)}

If the unitary \(\hat{\mathcal{U}}_c\) is faulty but the \(Y\)-rotations are perfect, then with some probability, bit-flip errors (\(\hat{J}_x\) errors) can occur, as described in Eq.~\ref{app/final cavity error term}. These errors flip an individual ancilla qubit in the first ancilla block \(a_1\). Note that we are considering all other operations in the circuit as perfect. Due to the transversal C-M operations, this error propagates from the ancilla in \(a_1\) to a data qubit interacting with it via C-M gates, and subsequently propagates further, affecting one ancilla each in the other two blocks \(a_2\) and \(a_3\). Since, bit-flips commute with \(\hat{U}_E\) the error stay on the same qubit after decoding operation. However, at the measurement stage, these propagated errors do not impact the final outcome, because we employ a majority voting procedure that effectively suppresses such errors as long as majority of the ancillas remain unaffected. \newline
\textit{Example}: To illustrate, consider a scenario where the stabilizer is product of \(X\)s and an error \(X^1_{{a_1}}\) occurs after encoding, flipping the first ancilla qubit in block \(a_1\). Assuming the first ancilla in each block \(a_1\), \(a_2\), and \(a_3\) interact via CNOTs, this error propagates, causing additional flips \(X^1_{{a_2}}\) and \(X^1_{{a_3}}\). Using the same example as above, in the ideal scenario as given in Eq.~\ref{eq: three level redundification}, we might measure the ancilla blocks as \((0,0,0,0)_{a_1}\), \((1,1,1,1)_{a_2}\), and \((1,1,1,1)_{a_3}\) with some probability. But due to the introduced errors, we instead measure \((1,0,0,0)_{a_1}\), \((0,1,1,1)_{a_2}\), and \((0,1,1,1)_{a_3}\) and after majority vote get \(0_{a_1}, 1_{a_2}\) and \(0_{a_3}\). This reasoning similarly applies if any single ancilla undergoes a bit-flip after the encoding stage. Therefore, our protocol successfully protects against \(\hat{J}_x\) errors arising from the faulty encoding unitary. \newline

\subsubsection{Error in both cavity unitary \(\hat{\mathcal{U}}_c\) and \(Y\)-rotations}

Now let's consider errors in both the cavity unitary \(\hat{\mathcal{U}_c}\) and the \(Y\)-rotations. As given in Eq.~\ref{eq:depol}, with probability \(p_d/3\), we obtain an additional depolarizing term:

\[
\sum_{j=1}^N ( X_j\tau X_j+Y_j\tau Y_j+Z_j \tau Z_j ).
\]

\begin{itemize}
    \item This means that with some probability, we can get an \(X_j \tau X_j\) term, which introduces a bit-flip on any one of the ancilla qubits in \(a_1\). The analysis in the previous section has shown that our protocol is robust against such errors.

    \item Now, consider the \(Z_j \tau Z_j\) term, which applies a phase flip to any one of the ancilla qubits in \(a_1\). This error propagates through both the ancilla-data C-M gates and the CNOT gates used for redundification. However, after decoding, it manifests as a measurement error. 

    \textit{Example:} Suppose that after imperfect encoding, we obtain a \(Z^1_{a_1}\) error on the first qubit in ancilla block \(a_1\). This error remains unchanged through both the ancilla-data C-M gates and the redundification CNOT gates. After the decoding operation \(U_E\), it transforms as:\[
    U_E Z^1_{a_1}U_E^{\dagger}=(Y^1X^2X^3X^4)_{a_1}.
    \]
    In the ideal scenario, as given in Eq.~\ref{eq: three level redundification}, we would measure the ancilla blocks in the states:
    \[
    (0,0,0,0)_{a_1}, \quad (1,1,1,1)_{a_2}, \quad (1,1,1,1)_{a_3}
    \]
    with some probability. However, after the \((Y^1X^2X^3X^4)_{a_1}\) error, we perform measurements, majority voting, and assign bit values, resulting in:
    \[
    (1,1,1,1)_{a_1} \rightarrow 1_{a_1}, \quad
    (1,1,1,1)_{a_2} \rightarrow 1_{a_2}, \quad
    (1,1,1,1)_{a_3} \rightarrow 1_{a_3}.
    \]
    The Hamming weight of this outcome is odd, indicating projection into the ``+1" sector of the stabilizer. However, in reality, the system was in the ``-1" sector, signifying a measurement error. Since the stabilizer measurement is repeated \(d\) times, where \(d\) is the distance of the QLDPC code, we are protected from these errors.

    \item The final term, \(Y_j \tau Y_j\), applies a \(Y_j=Z_jX_j\) error to one of the qubits in ancilla block \(a_1\). Since
    \[
    \text{CNOT}(Y)\text{CNOT}^{\dagger} = Y \otimes X,
    \]
    a bit-flip is introduced on the data qubit as well as on the qubits in ancilla blocks \(a_2\) and \(a_3\) that interact via CNOT gates. After decoding, similar to the previous case, this also results in a measurement error.

    \textit{Example}: Extending the previous example, suppose a \(Y^1_{a_1}\) error occurs on the first qubit in ancilla block \(a_1\). After the CNOT interactions, it transforms into:
    \[
    Y^1_{a_1}X^1_{a_2}X^1_{a_3}.
    \]
    After the decoding operation, it further transforms into:
    \[
    (Z^1X^2X^3X^4)_{a_1} X^1_{a_2} X^1_{a_3}.
    \]
    In the ideal scenario, we would measure:
    \[
    (0,0,0,0)_{a_1}, \quad (1,1,1,1)_{a_2}, \quad (1,1,1,1)_{a_3}
    \]
    with some probability. However, after the \((Z^1X^2X^3X^4)_{a_1} X^1_{a_2} X^1_{a_3}\) error, measurement and majority voting yield:
    \[
    (0,1,1,1)_{a_1} \rightarrow 1_{a_1}, \quad
    (0,1,1,1)_{a_2} \rightarrow 1_{a_2}, \quad
    (0,1,1,1)_{a_3} \rightarrow 1_{a_3}.
    \]
    The Hamming weight is again odd, indicating the ``+1" sector of the stabilizer when, in reality, it was in the ``-1" sector. As in the previous case, since we repeat the stabilizer measurement multiple times, we remain protected from these errors.
\end{itemize}

\subsection{Error in ancilla-data C-M gates}

As shown in Fig.~\ref{Fig:Divincezo-Aliferis circuit}, encoding is followed by transversal C-M gates between the ancilla and data. Here we discussed the case when C-M gates are all CNOTs. This can easily be generalized for other stabilizer measurements. We model a faulty CNOT using two-qubit depolarizing noise. This means that when we apply a CNOT, an error term is introduced with probability \(p\). The operation can be written as:

\begin{align}
    \widetilde{CNOT} = (1-p)\, CNOT + p\, E,
\end{align}

where \(\widetilde{CNOT}\) is the faulty operation, CNOT is the ideal operation, and  \(E=\{\hat{I} \otimes \hat{X}, \hat{I} \otimes \hat{Y}, \hat{I} \otimes \hat{Z}, \hat{X} \otimes \hat{I}, \hat{X} \otimes \hat{X}, \hat{X} \otimes \hat{Y}, \hat{X} \otimes \hat{Z}, \hat{Y} \otimes \hat{I}, \hat{Y} \otimes \hat{X}, \hat{Y} \otimes \hat{Y}, \hat{Y} \otimes \hat{Z}, \hat{Z} \otimes \hat{I}, \hat{Z} \otimes \hat{X}, \hat{Z} \otimes \hat{Y}, \hat{Z} \otimes \hat{Z} \}\), with each error occurring with probability \(p/15\). Note that the control qubits are in the ancilla block \(a_1\), and the target qubits are in the data. The analysis of error terms from the set \(E\) follows a similar argument as in the previous section. This means that, in the end, we will have either measurement errors or local bit-flips in the ancilla blocks, which can be resolved using majority voting.

\subsection{Error in Decoding}

As highlighted in Fig.~\ref{Fig:Divincezo-Aliferis circuit}, ancilla redundification through CNOTs, followed by the cavity unitary and measurement, constitutes the decoding operation. Note that both the CNOTs and the cavity unitary can be faulty. However, we do not consider cases where both events occur simultaneously, as such an occurrence represents a \(p^2\) event, where \(p\) is the probability of each event occurring independently. Let us analyze each case separately. \newline

\subsubsection{Error in CNOTs for redundification}

As shown in Fig.~\ref{Fig:Divincezo-Aliferis circuit}, the information from ancilla block \(a_1\) is copied to ancilla blocks \(a_2\) and \(a_3\) through transversal CNOTs. We will only consider the case when one of the CNOTs fails, as having more than one CNOT fail is a higher-order event in \(p\), where \(p\) is the probability of any one CNOT failing. 

If terms like \(I \otimes X, I \otimes Y, I \otimes Z, X \otimes I, Y \otimes I, Z \otimes I\) are picked, then only a single qubit in an ancilla block will be flipped. Similarly, if weight-\(2\) terms like \(X\otimes X, X\otimes Y\), and similar terms are picked, two qubits in two separate ancilla blocks will be flipped. However, if the error originates from encoding, it flips one qubit in each of the ancilla blocks, resulting in three bit-flips in total. 

In this way, we can distinguish between errors caused by encoding failures and those due to CNOT failures. If an error is identified as originating from one of the CNOTs during redundification, no correction is needed since the error occurred after the ancilla-data interaction, leaving the data qubit unaffected.

\subsubsection{Error in decoding operation \(\hat{U}_E^{\dagger}\) and CNOTs for redundification}

All three ancilla blocks \(a_1, a_2\), and \(a_3\) are decoded separately, and with probability \(p\), any one of the decoding operations can fail. Two or all three operations being faulty are \(p^2\) and \(p^3\) events, respectively, which we do not consider. Hence, we will only analyze the case where any one of the decoding operations fails.

Consider the first decoding operation consisting of ancilla block \(a_1\) failing while the other two remain perfect. From Eq.~\ref{eq:state after imperfect decoding}, we know how a faulty decoder acts. With probability \(p_{\text{cavity}}\), we get the \(J_x \upsilon_3 J_x\) term, which introduces a bit-flip on any one of the qubits. With probability \(p_d/3\), we obtain  
\[
\sum_{j=1}^N \left(X_j \upsilon_3 X_j + Y_j \upsilon_3 Y_j + Z_j\upsilon_3 Z_j\right),
\]
which results in a bit-flip, bit-phase-flip, or phase-flip on any one of the qubits. After measurement, these errors can be detected, and since we have redundification, we can compare the measurements of all the ancilla blocks. This allows us to determine whether the error occurred during decoding or encoding. If we find that the error occurred during decoding, no correction is needed, as the error took place after the ancilla-data interaction, leaving the data qubits unaffected.  

However, the third term  
\[
\sum_{j=1}^N \left(X_j \prod_k X_k \upsilon_3 \prod_k X_k X_j + N \prod_k X_k \upsilon_3 \prod_k X_k \right),
\]
which occurs with probability \(p_d/3\), indicates that all qubits will be flipped, ruining the majority voting process and eliminating our ability to determine whether the error arose from the encoding or decoding operation. Fortunately, the probability of this error occurring is several orders of magnitude lower than the probability of the first two error types combined.  

\textit{Example:} In the perfect case, suppose we measure the terms  
\[
(0,0,0,0)_{a_1}, \quad (1,1,1,1)_{a_2}, \quad (1,1,1,1)_{a_3}
\]
from Eq.~\ref{eq: three level redundification} with some probability. If one of the error terms listed above occurs, it introduces \(X^1_{a_1}\) on the first qubit in ancilla block \(a_1\), changing the measurement outcome to  
\[
(1,0,0,0)_{a_1}, \quad (1,1,1,1)_{a_2}, \quad (1,1,1,1)_{a_3}.
\]
After majority voting and assigning bit values, we still obtain \(0_{a_1},1_{a_2},1_{a_3}\), indicating the ``-1" sector of the stabilizer. However, note that only one qubit in \(a_1\) has flipped, while the qubits in the other two ancilla blocks maintain the same measurement outcome. This scenario is only possible if the decoding operation acting on \(a_1\) has failed. If the error had originated from encoding, then the bit-flips would have propagated to the respective interacting qubits in the other two ancilla blocks through CNOTs. This distinction allows us to differentiate between encoding and decoding errors. As mentioned earlier, no correction is needed in this case since the data qubit remains unaffected.  

However, if the third term occurs, it introduces  
\[
(X^1X^2X^3X^4)_{a_1},
\]
changing the measurement outcome to  
\[
(1,1,1,1)_{a_1}, \quad (1,1,1,1)_{a_2}, \quad (1,1,1,1)_{a_3},
\]
which results in \(1_{a_1},1_{a_2},1_{a_3}\), indicating the ``+1" sector instead of the expected ``-1" sector. This effectively leads to a measurement error. In this case, there is no way to detect the error.

\section{Details of numerical simulation} \label{app:details of numerical simulation}

We used \texttt{STIM} \cite{gidney2021stim} for our simulations. We performed both the hardware-agnostic and custom error model simulations. To infer how the threshold changes with respect to the cooperativity, $C$, of the cavity we varied the ratio of cavity error to two-qubit depolarizing error $\left(p_{cavity}/p_2\right)$. By varying the cavity error, we can study the relationship between the failure probability, \( p_{\mathrm{cavity}} \), and the threshold of a code.

We refer back to the circuit shown in Fig.~\ref{Fig:Divincezo-Aliferis circuit} for stabilizer measurement throughout our simulations. Stabilizers and logical operators were generated using the \texttt{bposd} package \cite{Roffe_LDPC_Python_tools_2022}. We designed the syndrome extraction circuit in \texttt{STIM} to be adaptable to any check polynomial \( h(x) \), enabling the generation of the corresponding syndrome extraction circuit for both HGP and LP codes with \( d \) rounds. Our numerical simulations support polynomials of arbitrarily high degree, given sufficient computational resources.

We created two ancilla layers that replicate the arrangement of the data qubits: one for \(Z\) stabilizers (top layer) and another for \(X\) stabilizers (bottom layer), as illustrated in Fig.~\ref{Fig:tril-layer stack}. To prepare a GHZ state for the measurement of \(X\) stabilizers, we use the cavity arrangements in the bottom layer and apply the gate described in Eq.~\ref{eq:encoding_unitary}. But due to limitations of \texttt{STIM} package, we used H and CNOT gates and approximated the errors. The dominant error is $\hat{J}_x = \sum_i ^N X_i/2$, where \( N \) denotes the number of ancilla qubits involved. We approximate this error by neglecting the cross terms, which we anyway cannot catch while measurement, resulting in the error model where a single bit-flip can happen at random on anyone of the ancilla. To model this error in \texttt{STIM}, we used \texttt{CORRELATED} and \texttt{ELSE$\_$CORRELATED} functions as explained in Appendix ~\ref{App:correlated and else correlated function}. We then applied CNOTs between ancilla and data layers, followed by 2-qubit depolarizing errors. We can use qubits adjacent to the normal ancilla qubits for redundification, as shown in Fig.~\ref{Fig:tril-layer stack}. We use the same error model for decoding as the one described for encoding. We measured all ancilla qubits in the Z-basis followed by appropriate measurement errors and repeated the syndrome extraction round \( d \) times, where \( d \) is the code distance.

In \texttt{STIM}, stabilizer generators cannot be directly declared. Instead, the set of deterministic measurements that determine the stabilizer outcome (either `0' or `1') are passed into a function called \texttt{DETECTORS}. We initialize all data and ancilla qubits in the zero state, and for each \(X\) and \(Z\) stabilizer, we begin by declaring the corresponding \texttt{DETECTORS}. Two types of detectors were used in our simulations: (i) tracking errors across time by XORing the current and previous round of ancilla measurements, and (ii) local detectors XORing measurements of different ancilla blocks to distinguish between encoding and decoding errors. As mentioned above, we perform $d$ rounds of error correction. Following the declaration of stabilizers, we specify the set of logical observables of interest using the \texttt{OBSERVABLE\_INCLUDE} function. Since we initialize the data qubits in the all-zero state, after the projective stabilizer measurements, the system is projected into the logical zero state, \(\ket{0}_L\). At this point, the logical-Z observables are deterministic. We then measure all logical-Z observables, and if any of them have flipped, it indicates the occurrence of a logical-\(X\) error. \texttt{STIM} then generates space-time graph of the entire circuit, where nodes represent detectors and edges correspond to error mechanisms that can trigger these detectors. The space-time graph of $d$ rounds of syndrome extractions is decoded using the sinter integration of the \texttt{BP+OSD} decoder \cite{Roffe_LDPC_Python_tools_2022, roffe2020decoding}. We used the min-sum algorithm for belief propagation, with a maximum of 30 iterations and a scaling factor of 0.625, utilizing a parallel update schedule. If belief propagation fails to converge, its output is sent to \texttt{OSD-0} for post-processing. For further details, refer to Appendix ~\ref{App:Decoding}.

The decoder outputs a correction operator, denoted as \(c\). If \(c \notin \text{rowspace}(H)\), meaning the correction operator does not belong to the stabilizer group, it indicates that a logical error has occurred and the decoding attempt has failed. \texttt{Sinter} package then perform sampling multiple times to estimate the logical error rate per round for various physical error rates. This procedure is repeated for all the codes in the same code family. We estimate the threshold of the code family by plotting logical error rate vs physical error rate in log-log scale. The threshold is given by the point where all the codes intersect, and below this point, we observe a sudden change in the slope of all the curves. It means that logical failure rate can be exponentially suppressed once the physical error is below the threshold. \newline
After we have collected sufficient number of sub-threshold data points, we fit all the codes in a code family to the equation \cite{quintavalle2022reshape},
\begin{align} 
    P_L(p) & = A\left(\frac{p}{p_{th}}\right)^{ad},
\end{align}
where $P_{L}(p)$ represents the logical failure probability per syndrome extraction cycle, calculated as $P_L(p)=1-(1-P_L(p,d))^{1/d}$, with $P_L(p,d)$ being the total logical errors after $d$ rounds of syndrome extraction, and $d$ being the code distance. Here $A,a>0$ and $p_{th}$ is the threshold of a code family under the given error model and decoder. The logical failure probabilities for $p > 10^{-3}$ are determined numerically, after which the data points are fitted to the above equation and extended to $p < 10^{-3}$ to estimate the logical failure rates. 


After obtaining the threshold values for different ratios of \(p_{\mathrm{cavity}}/p_2\), as shown in Tab. \ref{tab:threshold_agnostic_custom_periodic}, we compute the corresponding cooperativity \(C_{th}\) when \(p_2=p_{th}\). \(J_x\) errors are coming both from cavity and depolarizing term. But we compute cooperativity just from the cavity contribution. 
Given that \(J_x\rho J_x\) error comes with the probability of \((2N\alpha+8p_d/3)\), where \(\alpha=\pi/(4d_N\sqrt{C})\) and \(p_d\) is the depolarizing error probability coming from imperfect \(Y\)-rotations, we can write the ratio \(p_{\mathrm{cavity}}/p_2\) as \(m\), and rearrange this equation to determine the cooperativity, denoted as \(C_{th}\), at \(p_2 = p_{\mathrm{th}}\). For GHZ state preparation, we have \(\theta = \pi/2\) and \(d_N = 1/\sqrt{2(1 + 2^{-N})}\) \cite{jandura2023nonlocal}, where \(N\) is the weight of the GHZ state. Thus, the \(C_{th}\) is given by 

\begin{align}
    C_{th} = \left( \frac{N\pi}{mp_{\mathrm{th}}\sqrt{2(1+2^{-N})}} \right)^2.
\end{align}
Using this relation, we can determine \(C_{th}\) for a given \(p_{\mathrm{cavity}} = mp_{\mathrm{th}}\). This establishes a crucial link between code performance, characterized by \(p_{\mathrm{th}}\), and the critical experimental parameter \(C_{th}\), which represents the minimum cooperativity required to achieve \(p_{\mathrm{th}}\). In other words, a system can be scaled if the cooperativity satisfies \(C \geq C_{th}\). Cooperativity serves as a key metric for experimentalists, as it quantifies the coupling quality between bosonic modes and Rydberg atoms.

\section{Spread of Pauli errors} \label{App:spread of Pauli errors}

\begin{figure*}[htbp]
    \centering
    \subfloat[]{
        \begin{quantikz}
            \gate[style={draw=red}]{X} & \qw & \ctrl{1} & \qw\\
            \gate{\mathbb{I}} & \qw & \targ{} & \qw
        \end{quantikz}
        $=$
        \begin{quantikz}
            \qw & \ctrl{1} & \gate[style={draw=red}]{X} & \qw\\
            \qw & \targ{} & \gate[style={draw=red}]{X} & \qw
        \end{quantikz}
    }
    \hspace{0.05\textwidth}
    \subfloat[]{
        \begin{quantikz}
            \gate[style={draw=blue}]{Z} & \qw & \ctrl{1} & \qw\\
            \gate{\mathbb{I}} & \qw & \targ{} & \qw
        \end{quantikz}
        $=$
        \begin{quantikz}
            \qw & \ctrl{1} & \gate[style={draw=blue}]{Z} & \qw\\
            \qw & \targ{} & \gate{\mathbb{I}} & \qw
        \end{quantikz} 
    }

    \vspace{0.05\textwidth}

    \subfloat[]{
        \begin{quantikz}
            \gate{\mathbb{I}} & \qw & \ctrl{1} & \qw \\
            \gate[style={draw=blue}]{X} & \qw & \targ{} & \qw
        \end{quantikz}
        $=$
        \begin{quantikz}
            \qw & \ctrl{1} & \gate{\mathbb{I}} & \qw \\
            \qw & \targ{} & \gate[style={draw=blue}]{X} & \qw
        \end{quantikz}
    }
    \hspace{0.05\textwidth}
    \subfloat[]{
        \begin{quantikz}
            \gate{\mathbb{I}} & \qw & \ctrl{1} & \qw\\
            \gate[style={draw=red}]{Z} & \qw & \targ{} & \qw
        \end{quantikz}
        $=$
        \begin{quantikz}
            \qw & \ctrl{1} & \qw & \gate[style={draw=red}]{Z}\\
            \qw & \targ{} & \qw & \gate[style={draw=red}]{Z}
        \end{quantikz}
    }
    \caption{Error propagation through the CNOT gate. (a) Spread of \(X\) error from control qubit to target qubit after CNOT gate. (b) \(Z\) error on the control qubit remains unchanged after the CNOT gate. (c) An \(X\) error on the target qubit commutes with the CNOT gate. (d) \(Z\) error on the target qubit propagates to the control qubit after the CNOT gate.}
    \label{app:fig.error_prop}
\end{figure*}
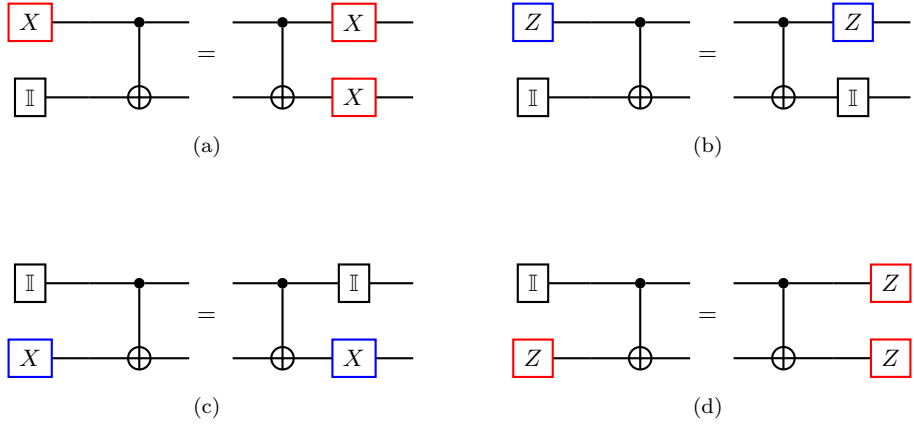

During the execution of circuit operations, errors will propagate and build up over time. Let's take the example of a CNOT gate applied to two qubits. As shown in Fig.~\ref{app:fig.error_prop}, we can track the evolution of a set of 2-qubit Pauli operators $\{XI, ZI, IX, IZ\}$ under an ideal CNOT operation. In the worst-case scenario, the size of the error doubles. \(X\) errors ``flow" down a CNOT and \(Z\) errors ``flow" up. Similarly, CZ gates also propagate errors. However, since these gates are diagonal in the computational basis, they do not impact products of $Z$ operators—$ X\otimes I$ transforms into $ X \otimes Z$, and $I \otimes X$ becomes $Z \otimes X$.

\section{Details of Code Construction} \label{App:Details of Code Construction}

There are a variety of ways to obtain a sparse parity check matrix. One specific way which we use in this work is through a polynomial. We can define a $n \times n$ circulant matrix $\mathbb{X}$ with entries belonging to field $F_q$ expressed as,
\begin{equation}
\label{eq:X_matrix}
   \mathbb{X}=
    \begin{pmatrix}
    a_0 & a_1 & a_2 &\cdots&a_{n-1}\\
    a_{n-1}& a_0 & a_1&\cdots&a_{n-2}\\
    a_{n-2}& a_{n-1}&a_0&\cdots&a_{n-3}\\
    \vdots&\vdots&\vdots&\ddots&\vdots\\
    a_1&a_2&a_3&\cdots&a_0
   \end{pmatrix}.
\end{equation}
Where $a_0,a_1,a_2,...,a_{n-1}\in F_q$. A code $C$ is cyclic if $(a_0,a_1,...a_{n-1}) \in C$ implies $(a_{n-1},a_0,...,a_{n-2}) \in C$. Codes that are linear and cyclic can be written in terms of a polynomial $a(x)=a_0+a_1x+a_2x^2+...+a_{n-1}x^{n-1}$. It is possible to show that any such code consists of a polynomial which is a multiple of a single generator polynomial $g(x)$, which must divide $x^n-1$. The quotient defines the check polynomial $h(x)$, given by $h(x)=g(x)/x^n-1$, which is the generator polynomial of the dual code. The degree of the generator polynomial is deg $g(x)=n-k$, which gives the number of stabilizer generators in the stabilizer group. While the degree of check polynomial $h(x)$ is $k$, which gives us idea of the number of logical qubits. The classical code corresponding to check polynomial $h(x)$ will have $k$ logical bits. Starting with a generator polynomial $h(x)=a_0 x^0+a_1 x^1+..+a_{n-1} x^{n-1}$ of a cyclic code, we can collect the coefficients $a_0,a_1,..,a_{n-1}$ and arrange them in matrix form like Eq.~\ref{eq:X_matrix}, and continue to permute the entries into rows below. Now, we can decide the length of the starting vector (or columns of the matrix $\mathbb{X}$) which we call lift. We provide the following examples for in depth code construction. 

\subsection{Surface Code from repetition Code}\label{App:surface code from repetition code}

Suppose $h(x) = 1 + x$, so $a_0 = 1$ and $a_1 = 1$. Let the lift be 5. We start with the vector $a_0 = 1, a_1 = 1, a_2 = 0, a_3 = 0, a_4 = 0$, represented as $11000$, and continue permuting it. Note that after ``lift+1" steps, we return to the original vector. Let's denote this matrix as $H$, which is:
\begin{equation}
 H=
 \begin{pmatrix}
  1&1&0&0&0\\
  0&1&1&0&0\\
  0&0&1&1&0\\
  0&0&0&1&1\\
  1&0&0&0&1\\
  \end{pmatrix}
\end{equation}

The matrix $H$ defines a classical code with $n = \text{lift} = 5$ bits and $k = n - \text{rank}(H) = 0$ logical bits. When there are no logical bits, the code's distance is denoted as $\infty$, resulting in a $[5,0,\infty]$ code. Since $H$ is of full rank, there are 0 logical bits. An intuitive way to understand this is by considering that we initially have $n$ physical degrees of freedom. The rank of $H$, which represents the number of stabilizers, imposes constraints. When these constraints equal the number of physical degrees of freedom (as with a full-rank $H$), there is no room left for encoding logical information, leading to $0$ logical bits. To obtain a logical bit, we can relax some of the constraints by deleting rows from $H$. For instance, if we remove the last row from $H$ we get the following parity check matrix, 

\begin{equation}
\label{app:eqn_starting matrix for surface code}
\tilde{H}=
\begin{pmatrix}
  1&1&0&0&0\\
  0&1&1&0&0\\
  0&0&1&1&0\\
  0&0&0&1&1\\
\end{pmatrix}.
\end{equation}

\begin{figure}[htbp]
    \centering
    \includegraphics[width = 0.5\columnwidth]{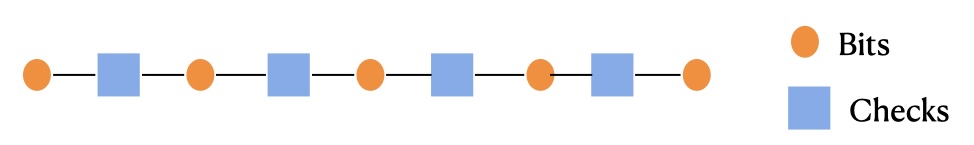}
    \caption{Factor graph of the distance-5 repetition code: Orange circular nodes represent bits, and purple square nodes represent checks. The corresponding parity check matrix is given in Eq. 6}
    \label{app:fig_rep_code_d5}
\end{figure}

\begin{figure}[htbp]
    \centering
    \includegraphics[width=0.5\linewidth]{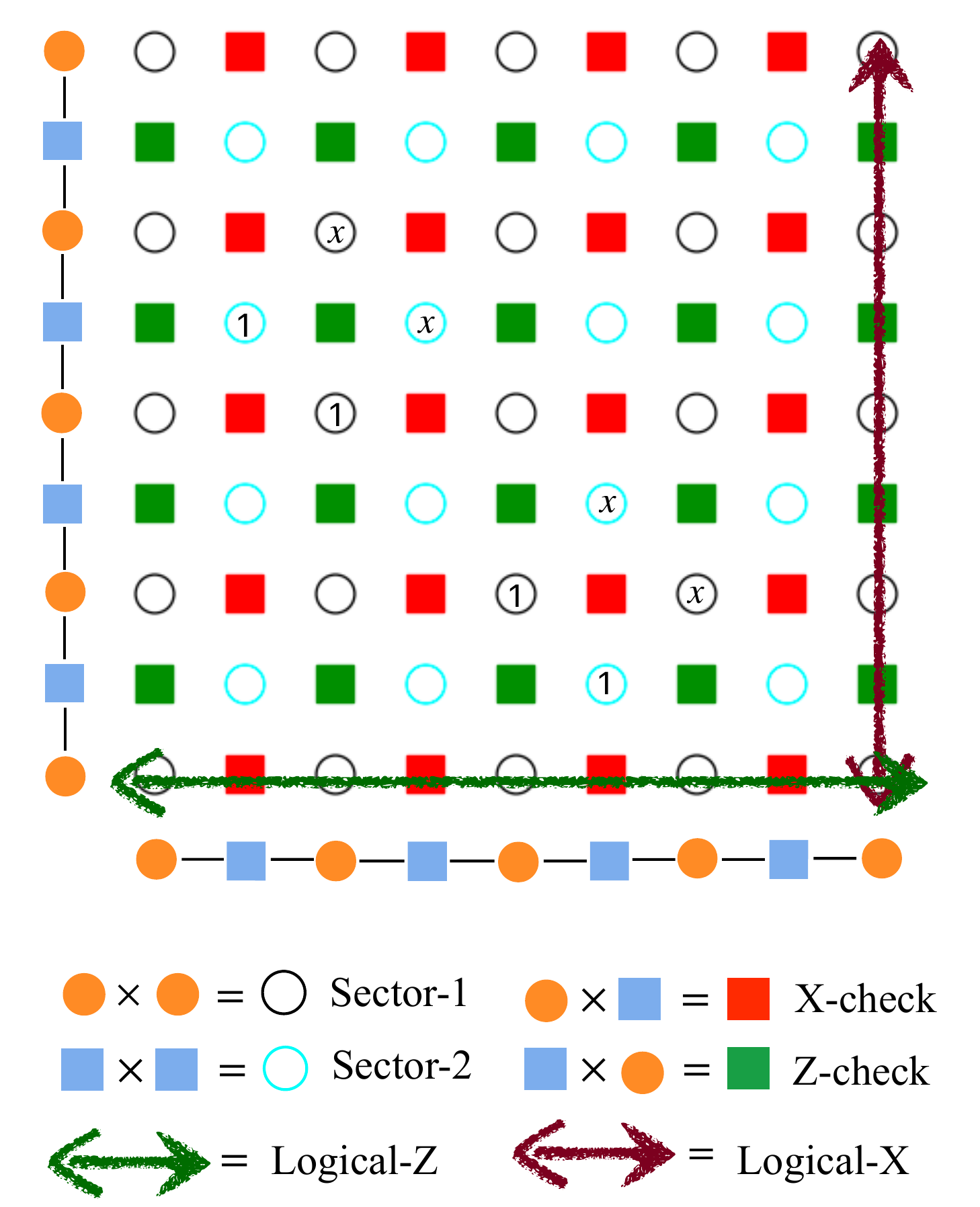}
    \caption{A pictorial depiction of hypergraph product of distance $=5$ repetition code with itself yielding a $d=5$ surface code.}
    \label{app:fig_surface_code}
\end{figure}

The rank of $\tilde{H}$ is $4$, which gives $1$ logical bit, resulting in $[[5,1,5]]$ code. Similarly, by varying the rank of the matrix $H$ and the lift, we can generate codes with different numbers of physical and logical bits. Next, let's take the hypergraph product of the repetition code shown in Fig.~\ref{app:fig_rep_code_d5} with itself. In terms of factor graphs, this hypergraph product corresponds to the process described in Fig.~\ref{app:fig_surface_code}. The resulting quantum code is a $[41,1,5]$ surface code.

As previously mentioned, we have $n_1 n_2$ sector-1 qubits and $(n_1-k_1)(n_2-k_2)$ sector-2 qubits. Sector-1 qubits are formed by the product of two classical bits, while sector-2 qubits are formed by the product of two classical checks. The product of a classical bit with a check creates an X-stabilizer (red), and the product of a check with a bit creates a Z-stabilizer (green). For the surface code, there is 1 logical qubit, with both its logical-X and logical-Z observables fully supported on sector-1 qubits \ref{app:fig_surface_code}. This arises from the symmetry in the hypergraph product relative to the classical codes. If we had started with a symmetric parity check matrix, we would obtain a toric code with 2 logical qubits and 4 logical observables. In that case, 2 logical observables would be supported by sector-1 qubits, and the remaining 2 by sector-2 qubits. However, for the surface code, the initial parity check matrix as shown in Eq.~\ref{app:eqn_starting matrix for surface code} lacks this symmetry, leading to an unequal distribution of logical observables between sector-1 and sector-2 qubits \cite{quintavalle2023partitioning}.

Recall that we started with the generator polynomial $h(x) = 1 + x$. The influence of this generator polynomial is evident in the shape of the stabilizers. Note the polynomial labeling of qubits shown in Fig.~\ref{app:fig_surface_code}. Each green square represents a Z-stabilizer, while each red square represents an X-stabilizer. The support for each green or red square extends to its nearest neighbors on the left, right, top, and bottom.

\subsection{Construction of HGP codes from \texorpdfstring{$h(x)=1+x+x^3+x^7$}{h(x)=1+x+x3+x7}}

Let's consider the check matrix defined by $h(x) = 1 + x + x^3 + x^7$. We perform the hypergraph product of this check matrix with itself. Given that the degree of $h(x)$ is 7, the resulting quantum code with periodic boundary conditions will have a total of $2 \times 7^2 = 98$ logical qubits. Table \ref{app:tab_big codes (appendix)} lists the specifications of the code for different lifts.

We will use the codes in Table \ref{app:tab_big codes (appendix)} for our numerical studies. These codes were chosen because they provide $98$ logical qubits with sufficient distance and require only a few thousand physical qubits. This makes them a promising option for near-term implementation.

Classical code corresponding to $h(x)$ with lift=15 gives us a $[15,7,5]$ code. Let us consider the first code from Table \ref{app:tab_big codes (appendix)}. This was obtained using hypergraph product of the parity check matrix $H$ with itself. The classical code specified by $H$ is: $C_1 = C_2 = [[15, 7, 5]]$, and by $H^T$ is $C_1^T = C_2^T = [[15, 7, 5]]$. The resulting quantum code is: $[[15^2 + 15^2 = 450, 7^2 +7^2 = 98, 5]]$. As mentioned above, $15^2 = 225$ qubits belong to sector-1 and $15^2 = 225$ qubits belong to sector-2. The next step involves creating a $n_1 \times n_2$ lattice, which is $n_1=n_2=15$ for our case thus, $15 \times 15$ two-dimensional lattice with each number corresponding to a qubit from sector-1. Then, place the qubits from sector-2 in between, as illustrated in Fig.~\ref{app/fig/[[450,98,5]] layout}. 
\begin{figure}[htbp]
\centering
\includegraphics[width=0.5\linewidth]{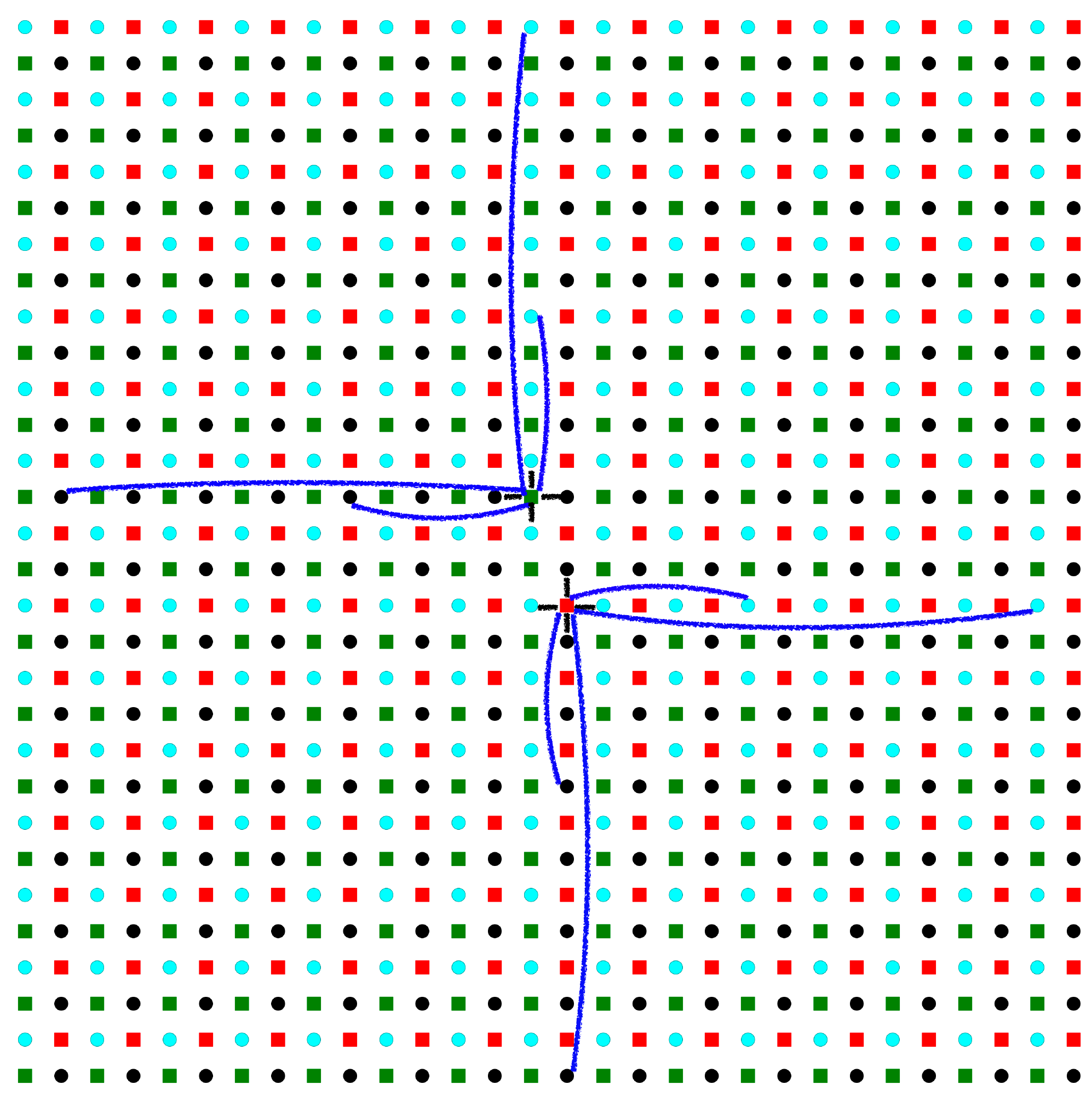}
\caption{2-dimensional layout of $[[450,98,5]]$ code achieved via hypergraph product of check matrices $h(x)=1+x+x^3+x^7$. Black circles indicate sector-1 qubits, while cyan circles denote sector-2 qubits. Additionally, red and green squares represent \(X\) and \(Z\) stabilizers, respectively. The layout displays one-to-one symmetry between sector-1 and sector-2 qubits.}
\label{app/fig/[[450,98,5]] layout}
\end{figure}

We employ a mapping similar to the initial example to maintain manageability. We isolate the sector-1 and sector-2 qubits as before and observe symmetry between \(X\) and \(Z\) stabilizers. This characteristic is not unique to this particular code but is inherent in any hypergraph product code. Attempting to display all the stabilizers of codes listed in Table \ref{app:tab_big codes (appendix)} is impractical. Instead, we show the shape of a $X$ and $Z$ stabilizers in Fig.~\ref{app/fig/[[450,98,5]] layout}, and the remaining stabilizers can be derived by merely shifting the entire pattern horizontally and vertically with periodic boundary condition. Non-local gates for stabilizer measurements are shown in blue, while local gates are shown in black for both \(X\) and \(Z\) stabilizers.

In Fig.~\ref{app/fig/[[450,98,5]] layout}, black squares indicate sector-1 qubits, while cyan squares denote sector-2 qubits. Red and green rectangles represent \(X\) and \(Z\) stabilizers, respectively. We apply the same product method described in Fig.~\ref{app:fig_surface_code} to derive qubits and stabilizers from bits and checks, as in the surface code example above. Figure \ref{app/fig/[[450,98,5]] layout} depicts the shape of \(X\) and \(Z\) stabilizers. The red and green squares represent the X-type and Z-type stabilizers, respectively. The influence of the check polynomial $h(x)$ on the shape of the stabilizers is evident. Additionally, we can see that the placement of sector-1 and sector-2 qubits is dual to each other, as are the shapes of the \(X\) and \(Z\) stabilizers, which is as expected. Understanding the shape and connectivity of stabilizers is crucial, as we will use cavities to do non-local gates. The stabilizer shape determines the optimal placement of these cavities.

With periodic boundary conditions, there is a one-to-one symmetry in the layout between sector-1 and sector-2 qubits, as well as between \(X\) and \(Z\) stabilizers, including the logical operators. Half of the logical operators are supported on sector-1 qubits, and the other half on sector-2 qubits. However, under open boundary conditions, gaps appear in the layout due to the asymmetric distribution of qubits between sectors, resulting from the unequal number of qubits in each sector. This disparity, caused by the open boundary condition, leads to all logical operators being fully supported on sector-1 qubits. We do not go into the details of the codes with open boundary conditions.

\begin{table}[htbp]
\centering
\fbox{
\begin{tabular}{|c|c|}
\hline
\textbf{lift} & \textbf{Codes:Periodic Boundaries} \\ \hline
$15$ & $[[450,98,5]]$ \\ \hline
$30$ & $[[1800,98,10]]$ \\ \hline
$45$ & $[[4050,98,15]]$
\end{tabular}}
\caption{Quantum codes generated via hypergraph product of $h(x)=1+x+x^3+x^7$ as the classical check matrix with itself with different lifts.}
\label{app:tab_big codes (appendix)}
\end{table}

\subsection{Construction of LP codes}

We selected the LP codes described in \cite{xu2024constant} for our simulations. We list down the seed protographs behind \([[544,80,\leq 12]]\) and \([[714,100,\leq16]]\) code. Denoting \(B^l_d\) as a base matrix with a lift size \(l\) and a classical code distance \(d\) after the lift, the base matrices are

\begin{equation}
B_{12}^{16} =
\begin{bmatrix}
1 & 1 & 1 & 1 & 1 \\
1 & x^2 & x^4 & x^7 & x^{11} \\
1 & x^3 & x^{10} & x^{14} & x^{15}
\end{bmatrix}
,\quad
B_{16}^{21} =
\begin{bmatrix}
1 & 1 & 1 & 1 & 1 \\
1 & x^4 & x^7 & x^9 & x^{17} \\
1 & x^6 & x^{14} & x^{18} & x^{12}
\end{bmatrix}.
\end{equation}

\section{Decoding} \label{App:Decoding}

The techniques for decoding quantum error-correcting codes are inspired by classical decoding algorithms. Notable examples include Belief Propagation \cite{poulin2008iterative}, the Union-Find decoder \cite{delfosse2021almost}, Minimum-Weight Perfect Matching \cite{dennis2002topological}, and Belief Propagation with Ordered Statistics Decoding \cite{Panteleev2021degeneratequantum}, among others. The decoding problem can be formalized as follows:
 
\begin{align}
    H  e =s
\end{align}

Here, $H$ is the parity check matrix associated with the code, $e$ is the error that occurred, and $s$ is the syndrome obtained by measuring the stabilizers. Given $s$, the decoder determines the most likely correction operator $c$ that produces the same syndrome $s$, $H \cdot c=s$, such that applying $c$ together with $e$ nullifies the syndrome, i.e.,
\begin{align}
H \cdot (c + e) =& 0, \\
(e+c) \in & \ \mathrm{rowspace}(H). \nonumber
\end{align}

All algebraic operations are performed over the binary field $F_2^N = \{0,1\}^N$ and are taken modulo 2. The condition \((e+c) \in \text{rowspace}(H)\) implies that the combined operator \(w = (e+c)\) must belong to the stabilizer group \(S\). If \(w \notin S\), a logical error has occurred, indicating that the decoding attempt has failed. 

Our review of BPOSD decoder is guided by a comprehensive review paper by iOlius et al. \cite{demartireview} that provides an in-depth examination of various decoding algorithms.
There are two key differences between decoding classical and quantum codes:
\begin{enumerate}
    \item \textbf{Types of Errors:} Classical codes only address bit flip errors, while quantum codes must handle both bit flip and phase flip errors.
    \item \textbf{Degeneracy:} In quantum codes, a single syndrome can correspond to multiple correction operators. This degeneracy means that several correction operators may satisfy the syndrome equation, but not all will correctly fix the error.
\end{enumerate}
These differences make decoding quantum codes more complex than decoding classical codes. Let's define the two decoding problems:

\begin{itemize}
    \item \textbf{Maximum likelihood decoding (MLD)}: This method seeks to determine the most probable error pattern corresponding to the observed error syndrome. Specifically, it solves the following optimization problem:

\begin{align}
    \hat{E} = \text{argmax}_{E \in \Pi^N} P(E | s)
\end{align}

where \( P \) represents the probability distribution function of the error vector \( \hat{E} \), \( \Pi^N \) denotes the \( N \)-qubit Pauli group, and \( s \) is the given syndrome. It is important to note that this method performs an exhaustive search over the \( \Pi^N \) group, disregarding the presence of degeneracy, and is thus referred to as non-degenerate decoding.

\item \textbf{Degenerate Maximum likelihood decoding (DMLD)}: 
An operator in the \( N \)-qubit Pauli group \( \Pi^N \) can be decomposed into three components: a pure error term forming the centralizer coset, a logical operator term forming the stabilizer coset, and a stabilizer component. An error operator \( E \) can be expressed in these three components to determine the exact stabilizer coset. Once the correct stabilizer coset is identified, any operator from this coset can be applied as a correction, as all elements in the stabilizer coset are equivalent up to a stabilizer, which acts trivially on the codespace.

The process of identifying the specific coset begins with a given syndrome, followed by the identification of the pure error component of the probable error operator, and finally, the logical component of the error. Mathematically, this is expressed as:

\begin{align}
    \hat{Q} = \text{argmax}_{Q \in \mathcal{Q}} P(Q | s)
\end{align}

where \( \mathcal{Q} \) represents the coset partitioning of the \( N \)-qubit Pauli group \( \Pi^N \), and \( Q \) is the coset from this partition. \(\hat{Q}\) denotes the correct stabilizer coset, and once identified, any element of this coset can be used for correction, as they all have the same effect on the logical codewords. For a detailed description of coset partitioning, we refer to the work by Fuentes et al. \cite{fuentes2021degeneracy}.

\end{itemize}

The MLD problem has been proven to be \textsf{NP}-complete, while the Degenerate Maximum Likelihood Decoding (DMLD) problem falls into the $\#$\textsf{P} complexity class \cite{iyer2015hardness}. Problems in the $\#$\textsf{P} class are computationally even more challenging than those in \textsf{NP}, presenting a significant obstacle to achieving the fast decoding necessary for effective quantum error correction. 
According to \cite{fuentes2021degeneracy}, degeneracy should theoretically enhance the performance of quantum codes by allowing multiple errors to be corrected using the same recovery operation. In practice, degeneracy has indeed been shown to improve the performance of certain quantum codes. Moreover, decoders have been proposed that specifically address the challenges associated with degeneracy \cite{cao2023qecgpt}.

\subsection{Belief Propagation (BP)}
Belief Propagation (BP), also known as the Sum-Product Algorithm (SPA), is a message-passing algorithm used for inference on probabilistic graphical models. In this discussion, we will refer to this algorithm as BP. Given a syndrome \( s \), BP aims to find the minimum-weight (MW) error pattern \( \hat{e} \) that satisfies \( H\hat{e} = s \).

A classical or quantum error-correcting code can be succinctly represented using a Tanner graph or factor graph. For instance, Figs.~\ref{app:fig_rep_code_d5} illustrate the factor graphs for a classical repetition code. In these graphs, derived from the parity-check matrix of a code, the columns correspond to bits/qubits, while the independent rows correspond to checks/stabilizer checks. A classical code includes a single type of check, which addresses only bit-flip errors. In contrast, a quantum code includes two types of checks: one for addressing \(X\) errors and another for addressing \(Z\) errors. Essentially, a quantum code can be seen as comprising two classical codes—one for protecting against \(X\) errors and another for \(Z\) errors.

The factor graph of a quantum code is a bipartite graph $G=(V \cup C, E)$, where  $V,C$ represent variable and check nodes, respectively, and $E$ is the set of edges between nodes $V$ and $C$. For example, Fig.~\ref{app:fig_surface_code} represents factor graph of surface code. We can see two types of checks, \(X\) (red) and \(Z\) (green) for correcting bit-flips and phase-flips. The edges correspond to the support of checks on physical qubits. Belief Propagation (BP) is employed as a message-passing algorithm between the nodes $V$ and $C$. 

For quantum codes, a modified version of BP, derived from its classical counterpart, is employed. In classical codes, BP identifies the most likely error pattern given a syndrome, achieving a global optimum that resolves the syndrome equation. In contrast, for quantum codes, BP seeks the qubit-wise most likely error pattern, targeting a marginal optimum. This approach aims to identify an error configuration that maximizes the marginal probability of individual qubit flips. This can be mathematically represented as,

\begin{align} \label{app:eq_marginals}
    P_i(E_i) =\text{argmax} \sum_{\text{all configurations}} P(E_1,..E_i=1,..,E_n|s) 
.
\end{align}

The summation is performed over all possible configurations where $E_i=1$ that satisfy the syndrome equation. Similarly, the bit-wise marginal probability is computed for each qubit. $P_i(E_i)$ represents the soft-decision for qubit $i$. The final decision is made using a hard-decision for each bit according to,
\begin{align}
    (E_{MW})_i=\left\{\begin{array}{ll}
1 & \text { if } P_1\left(E_i\right) \geq 0.5 \\
0 & \text { if } P_1\left(E_i\right)<0.5     
\end{array}\right. 
.
\end{align}

Here, \((E_{MW})_i\) represent the minimum weight configuration for qubit \(i\) among all possible configurations that satisfy Eq.~\ref{app:eq_marginals}. By determining the marginal for each qubit in this manner, we derive an overall minimum weight configuration that satisfies the given syndrome. The detailed description of each step of BP can be found in Appendix-C of \cite{roffe2020decoding}. BP is an effective decoding algorithm for classical codes with nearly loop-free factor graphs. Some classical codes have been shown to approach the Shannon limit capacity when decoded using BP \cite{mackay1999good, chung2001design}. However, the factor graphs of quantum LDPC codes exhibit high degeneracy, as previously discussed. This increased degeneracy leads to factor graphs with numerous short loops, causing BP to become stuck and preventing it from converging to a solution—a phenomenon known as quantum trapping sets \cite{raveendran2021trapping}. Consequently, BP encounters significant challenges when applied to quantum LDPC codes, failing to achieve a decoding threshold \cite{roffe2020decoding}.

\subsection{Post processing of BP: Ordered Statistics Decoding (OSD)}

We previously discussed that BP struggles in the presence of short cycles in a factor/Tanner graph, failing to converge to a solution and resulting in the absence of a threshold and an error floor \cite{richardson2003error}. This issue becomes particularly evident at low physical error rates, where the hard decisions based on soft decisions start to introduce errors. To address this challenge, a post-processing technique known as Ordered Statistics Decoding (OSD) was introduced after BP, collectively referred to as BPOSD. This approach was first implemented to quantum LDPC codes by Panteleev and Kalachev \cite{panteleev2021quantum}. BPOSD initially runs BP and then uses its output as the input for the OSD post-processing step. This approach has demonstrated strong performance across a range of random quantum LDPC codes, as evidenced in this work \cite{panteleev2021quantum}. Their method performs remarkably well for any random quantum LDPC codes. 

When BP fails to converge to a solution, the Ordered Statistics Decoding (OSD) post-processing step is invoked. Despite BP's inability to converge, it provides marginal probabilities for each qubit as shown in Eq.~\ref{app:eq_marginals} and an estimate of the error pattern $\hat{E}$. However, not all estimates of BP are necessarily incorrect. OSD utilizes the marginal information from BP to find a valid solution. OSD comes in various complexities known as OSD-w, where $w\in [0,...,H]$ and $H \in N$. OSD-0 when $w=0$ is the least complex case which we use in our simulations. The steps of OSD-0 goes as follows: 
\begin{enumerate}
    \item Utilize the marginals from Eq.~\ref{app:eq_marginals} or the soft-outputs from BP, and rank them from most likely to least likely to have been flipped. Store this list of bit indices as [BP].
    \item Reorder the columns of the parity-check matrix \( H \) according to the ranking of bit indices [BP], and denote the reordered matrix as \( \Xi \).
    \item Select the $\text{row-rank}(H)$ columns of the reordered matrix \( \Xi \), denoted as \( \Xi_{[BP]} \). Ensure that these selected columns are linearly independent, as the new matrix must have full rank.
    \item Invert the matrix \( \Xi_{[BP]} \) and solve the equation \( \hat{E}_{[K]} = \Xi_{[BP]}^{-1}s \).
    \item The final solution across all bits/qubits is given by \( \hat{E} = [\hat{E}_{[K]}, \hat{E}_{[\Bar{K}]}] = [\hat{E}_{[K]}, 0] \), where \( \Bar{K} \) represents the most reliable set, which can be assumed to be zero. The OSD-0 will always satisfy the syndrome equation.
\end{enumerate}

\subsubsection{Higher order OSD}
The motivation behind higher-order OSD is to find a solution $\hat{E}_{\Xi_w}$ with a lower Hamming weight than the solution from OSD-0, denoted as $\hat{E}_{\Xi_0}$. OSD-w follows a process similar to OSD-0 up to the fourth step; the distinction lies in the fifth step. In OSD-0, the first four steps yield the vector $\hat{E}_{[K]}$.  In OSD-w, we seek solutions $\hat{E}_{\Xi_w}=[\hat{E}_{[K]},\hat{E}_{[\Bar{K}]}]$, where $\hat{E}_{[\Bar{K}]}\neq 0$ by solving: 
\begin{align}
    \mathbf{\hat{E}}_{\Xi_w}=\left[\hat{E}_{[\Bar{K}]}, \hat{E}_{[\bar{K}]}\right]=\left[\hat{E}_{[K]}^{w=0}+\Xi_{[K]}^{-1} \Xi_{[\bar{K}]} \hat{E}_{[\bar{K}]}, \hat{E}_{[\bar{K}]}\right],
\end{align}

The vector \( \hat{E}_{[\bar{K}]} \neq 0 \) has a dimension of \( n - \text{row-rank}(H) \), suggesting that, in theory, one could attempt to find a solution with minimal Hamming weight by exploring all possible chains of lengths up to \( n - \text{row-rank}(H) \). However, this approach is computationally intensive, making it practical only for short chain lengths. To mitigate this challenge, the authors of \cite{roffe2020decoding} proposed the combination sweep strategy, a greedy search method that simplifies the identification of \( \hat{E}_{[\bar{K}]} \). For more details, see Appendix B of \cite{roffe2020decoding}.

OSD-\( w \) extends the search for a minimum Hamming weight solution beyond OSD-0. As the order \( w \) increases, the likelihood of finding a solution with minimal Hamming weight improves. However, this advantage comes with the cost of higher computational complexity. For instance, we compared the performance of OSD-$w$ and OSD-0 and found no significant difference. However, varying the parameter $w$ (OSD-4,5,6,7) resulted in a considerable increase in computation time compared to OSD-0. Therefore, we opted to use OSD-0.

\section{Simulating errors in \texttt{STIM}} \label{App:correlated and else correlated function}

We initiate the ancilla state, $\rho$, in the all-zero state and attempt to encode $\rho$ into a GHZ state using the cavity. However, as mentioned in the main text, in the presence of losses, the effect of the cavity can be described by map:

\begin{eqnarray}\label{app:eq_cavity_error_equation}
\mathcal{E}_{D^{-1}}(\rho) = \tau + \frac{2\theta}{\sqrt{C}d_N}\hat{J}_x \tau \hat{J}_x,
\end{eqnarray}

where $\tau$ is the perfect GHZ state. The cavity error introduces a bit-flip on any of the participating qubits with equal probability $p$ given by $2\theta/\sqrt{C}d_N$. Due to the limitations of \texttt{STIM}, we cannot directly apply this map. To simulate the errors, we use the \texttt{CORRELATED\_ERROR} and \texttt{ELSE\_CORRELATED\_ERROR} functions. These two functions always appear in pairs, with the correlated term always preceding the else correlated term. We modify the probability arguments within these functions so that they apply a bit-flip to all participating qubits with the same probability. 

Consider the example of a case where six qubits participate in a cavity operation, say for weight-6 GHZ state preparation. The final state will resemble Eq.~\ref{app:eq_cavity_error_equation}. With some probability $p$, we may have a one-bit-flipped state. To model this scenario, we use the \texttt{CORRELATED\_ERROR} and \texttt{ELSE\_CORRELATED\_ERROR} functions. Please refer to the code snippet below:

\begin{lstlisting}

for kk in np.arange(len(sx_list)):
    if kk==0:
        hgc_circuit.append_operation("CORRELATED_ERROR",stim.target_x(sx_list[kk]+2*n),p_encoding/(len(sx_list)-kk*p_encoding))
    else:
        hgc_circuit.append_operation("ELSE_CORRELATED_ERROR",stim.target_x(sx_list[kk]+2*n),p_encoding/(len(sx_list)-kk*p_encoding))

\end{lstlisting}

where \texttt{len(sx\_list)} represents the weight of the X-type stabilizer, and \texttt{p\_encoding} denotes the cavity error probability. The following example illustrates the function of this code. All ancilla qubits (indexed as $106$, $112$, $118$, $140$, $141$, and $142$) have an equal probability of experiencing an \(X\) error, but in any given sampling, only one qubit will actually incur the error.

\begin{lstlisting}
E(0.000166667) X106
ELSE_CORRELATED_ERROR(0.000166694) X112
ELSE_CORRELATED_ERROR(0.000166722) X118
ELSE_CORRELATED_ERROR(0.00016675) X140
ELSE_CORRELATED_ERROR(0.000166778) X141
ELSE_CORRELATED_ERROR(0.000166806) X142
\end{lstlisting}

\section{Relation between Cooperativity and \texorpdfstring{\(\frac{p_{\text{cavity}}}{p_2}\)}{p\_cavity/p}}
\label{App:relation cooperativity and p}

The $\hat{J}_x$ terms occur with a probability of \(2\alpha\) where \(\alpha=N\theta / \sqrt{C}d_N\) \ref{app/final cavity error term}. 
\begin{align}
    p_{cavity} & = \frac{2N\theta}{\sqrt{C}d_N}
\end{align}
The idea is to compute cooperativity say \(C_{th}\) at \(p_2=p_{th}\), where $p_{th}$ denotes threshold. Suppose the ratio $p_{cavity}/p_{th}=m$, then the above relation becomes
\begin{align}
    mp_{th} &= \frac{2N\theta }{\sqrt{C_{th}}d_N},\\
    C_{th} &= \left(\frac{2N\theta}{md_N p_{th}}\right)^2
\end{align}
For GHZ state preparation, $\theta = \pi/2$ and $d_N = \sqrt{2(1 + 2^{-N})}$ \cite{jandura2023nonlocal}, where $N$ is the number of qubits involved in the non-local gates, which is 6 for the codes with periodic boundaries listed in Table~\ref{tab:Codes h(x)=1+x+x^2}. For periodic boundaries, \(C_{th}\) becomes:

\begin{align}
    C_{th}=&\left(\frac{N\pi}{mp_{th}\sqrt{2(1+2^{-N})}}\right)^2
\end{align}

Using this relation, we can compute the Cooperativity \(C\) for a given value of \(p_{\text{cavity}} = mp_{\text{th}}\). This provides a key relationship between the code performance, denoted by \(p_{\text{th}}\), and the critical experimental quantity, cooperativity. The Cooperativity \(C\) is a significant metric for experimentalists, as it determines the quality of coupling between bosonic modes and Rydberg atoms.

\section{Projecting to combined GHZ state} \label{app:projecting into combined GHZ state}

 As illustrated in green in Ancilla-1 in Fig.~\ref{Fig:tril-layer stack}, a non-local resource is used to prepare a $\ket{GHZ}_3$ state on qubits arranged horizontally and vertically. Let's label the horizontal qubits as $h_1, h_2, h_3$ and the vertical qubits as $v_1, v_2, v_3$. We first use the non-local resource to prepare the GHZ state on the horizontal qubits: $\ket{GHZ}_{h_1 h_2 h_3}$. Then, another non-local resource is used to prepare the GHZ state on the vertical qubits: $\ket{GHZ}_{v_1 v_2 v_3}$. We can measure the parity between any two horizontal and vertical qubits, such as $Z_{h_3} Z_{v_1}$, to project the system into a combined GHZ state.

\begin{align}
\ket{GHZ}_{h_1 h_2 h_3}=&\left(\ket{000}_{h_1 h_2 h_3}+\ket{111}_{h_1 h_2 h_3}\right) \\
\ket{GHZ}_{v_1 v_2 v_3}=&\left(\ket{000}_{v_1 v_2 v_3}+\ket{111}_{v_1 v_2 v_3}\right)
\end{align}

Upon measuring $Z_{h_3} Z_{v_1}$, with the measurement outcome $m$, the combined state $\ket{GHZ}_{h_1 h_2 h_3} \otimes \ket{GHZ}_{v_1 v_2 v_3}$ is projected into the $\ket{GHZ}_6 = \left(\ket{000,000} + \ket{111,111}\right)/\sqrt{2}$ state if $m=0$. If $m=1$, apply the correction $X_{h_1} X_{h_2} X_{h_3}$ or $X_{v_1} X_{v_2} X_{v_3}$ to return to the $\ket{GHZ}_6$ state. The correction term after measurement can be written as $\left(X_1 X_2 X_3\right)^m$.

\section{Detailed simulation results}

\subsection{Results for codes generated from \texorpdfstring{$h(x)=1+x+x^2$}{h(x)=1+x+x2}}
\begin{figure*}[htbp]
    \centering
    \begin{minipage}{0.5\textwidth}
        \centering
        \rotatebox{0}{
        \includegraphics[width=0.7\columnwidth]{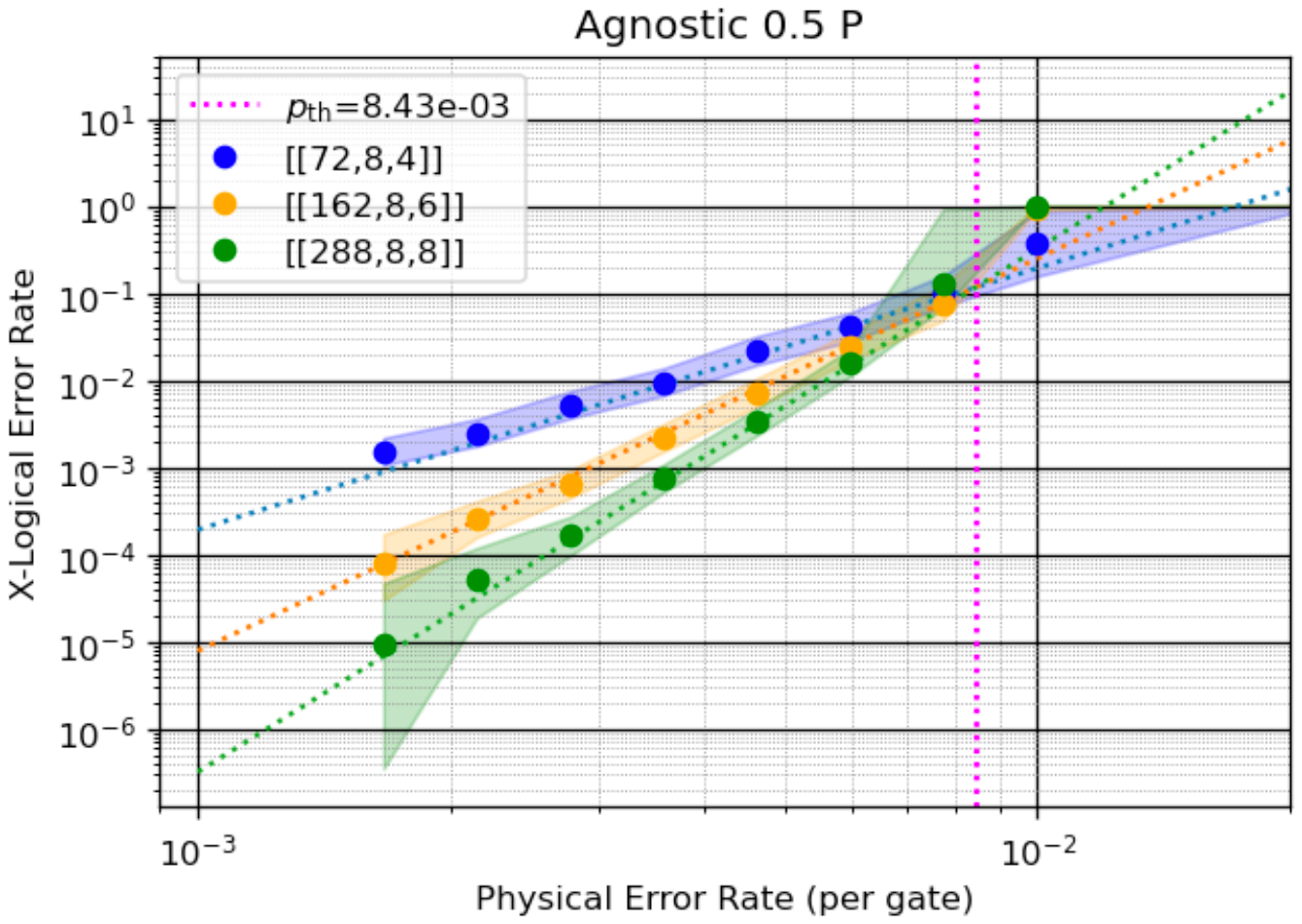}
        }
    \end{minipage}%
    \hspace{0.01\columnwidth}
    \begin{minipage}{0.45\textwidth}
        \centering
        \rotatebox{0}{
        \includegraphics[width=0.7\columnwidth]{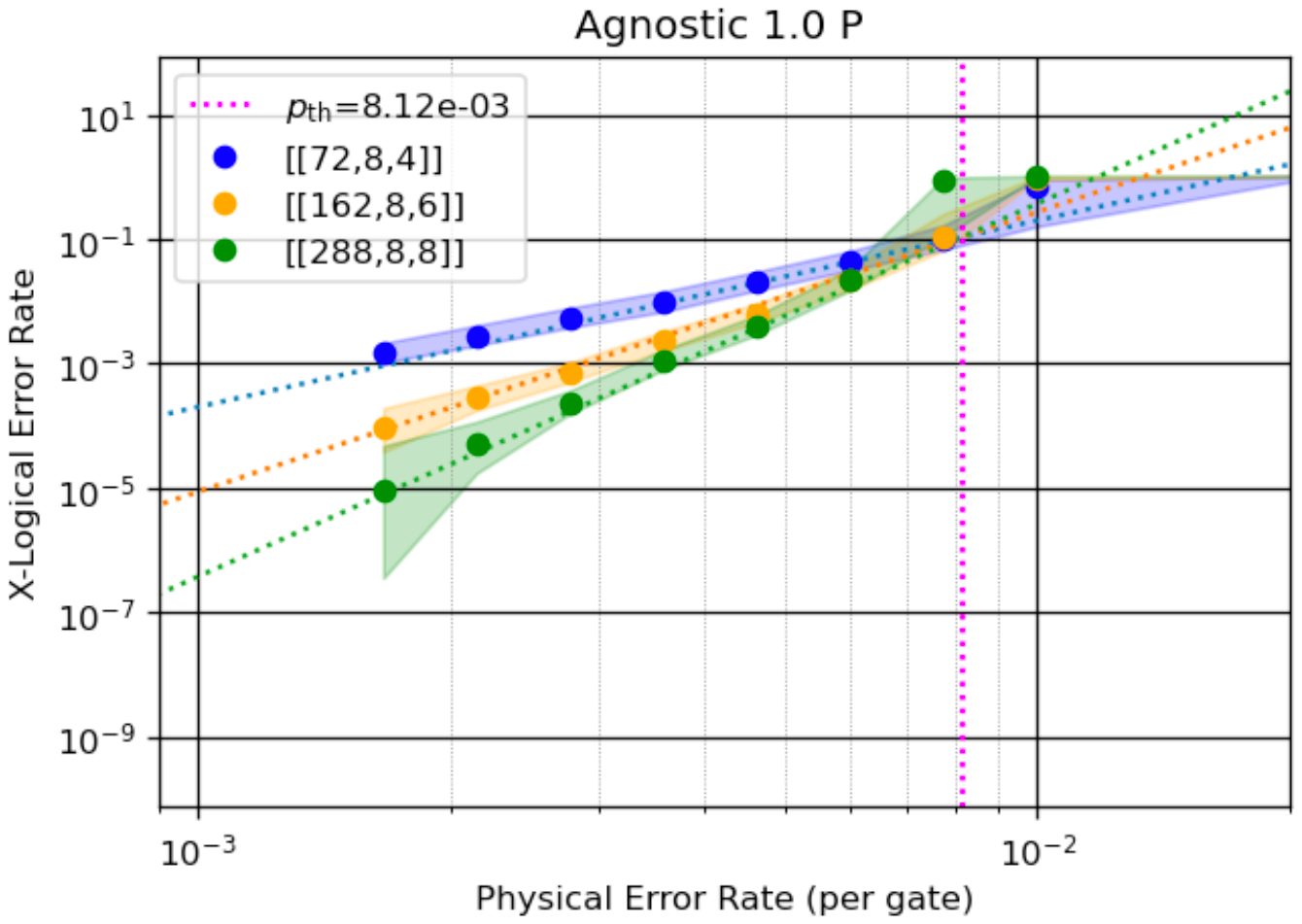}
        }
    \end{minipage}
    \vspace{0.01cm} 
    \begin{minipage}{0.45\textwidth}
        \centering
        \rotatebox{0}{
        \includegraphics[width=0.7\columnwidth]{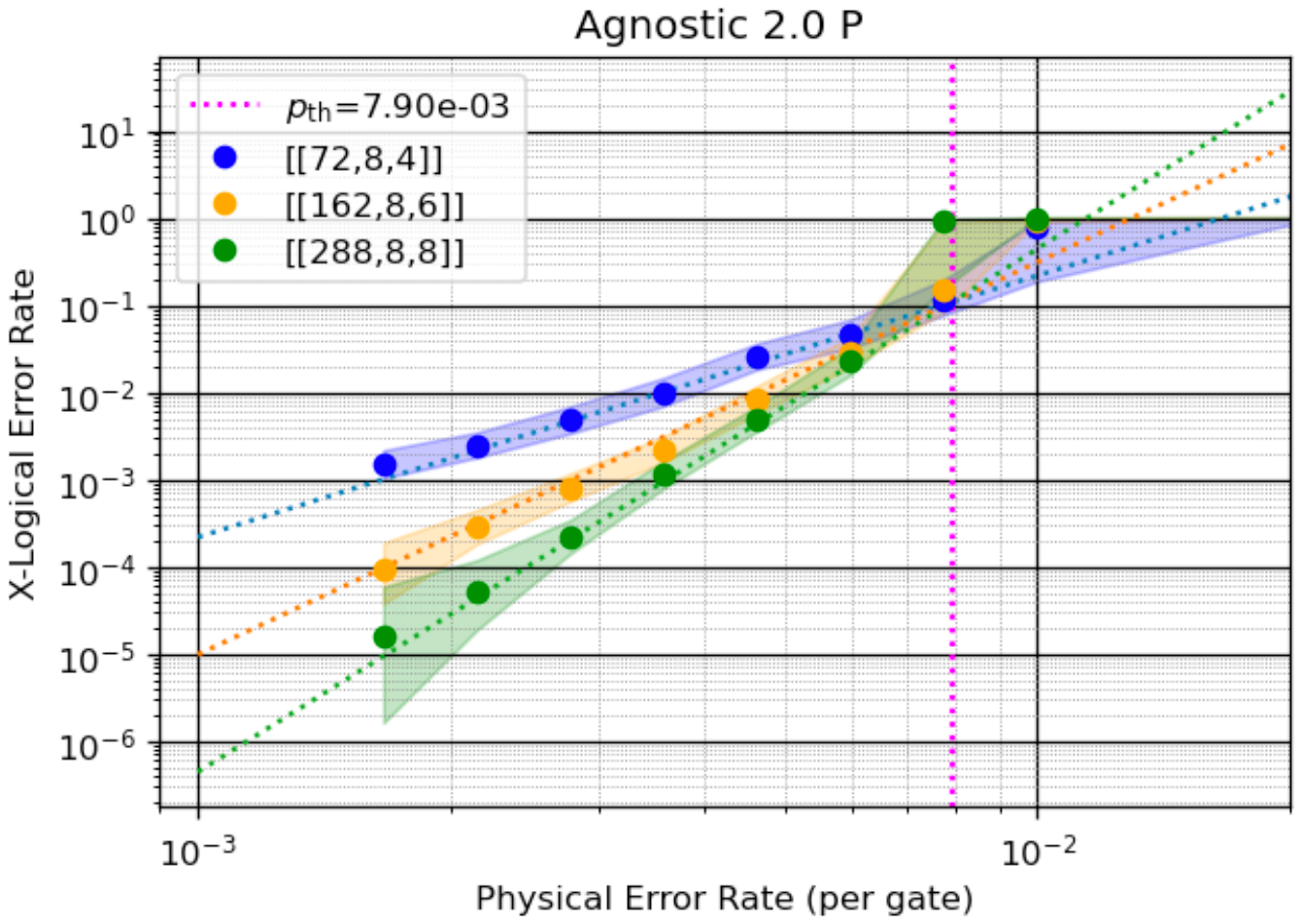}
        }
    \end{minipage}%
    \hspace{0.01\columnwidth}
    \begin{minipage}{0.45\textwidth}
        \centering
        \rotatebox{0}{
        \includegraphics[width=0.7\columnwidth]{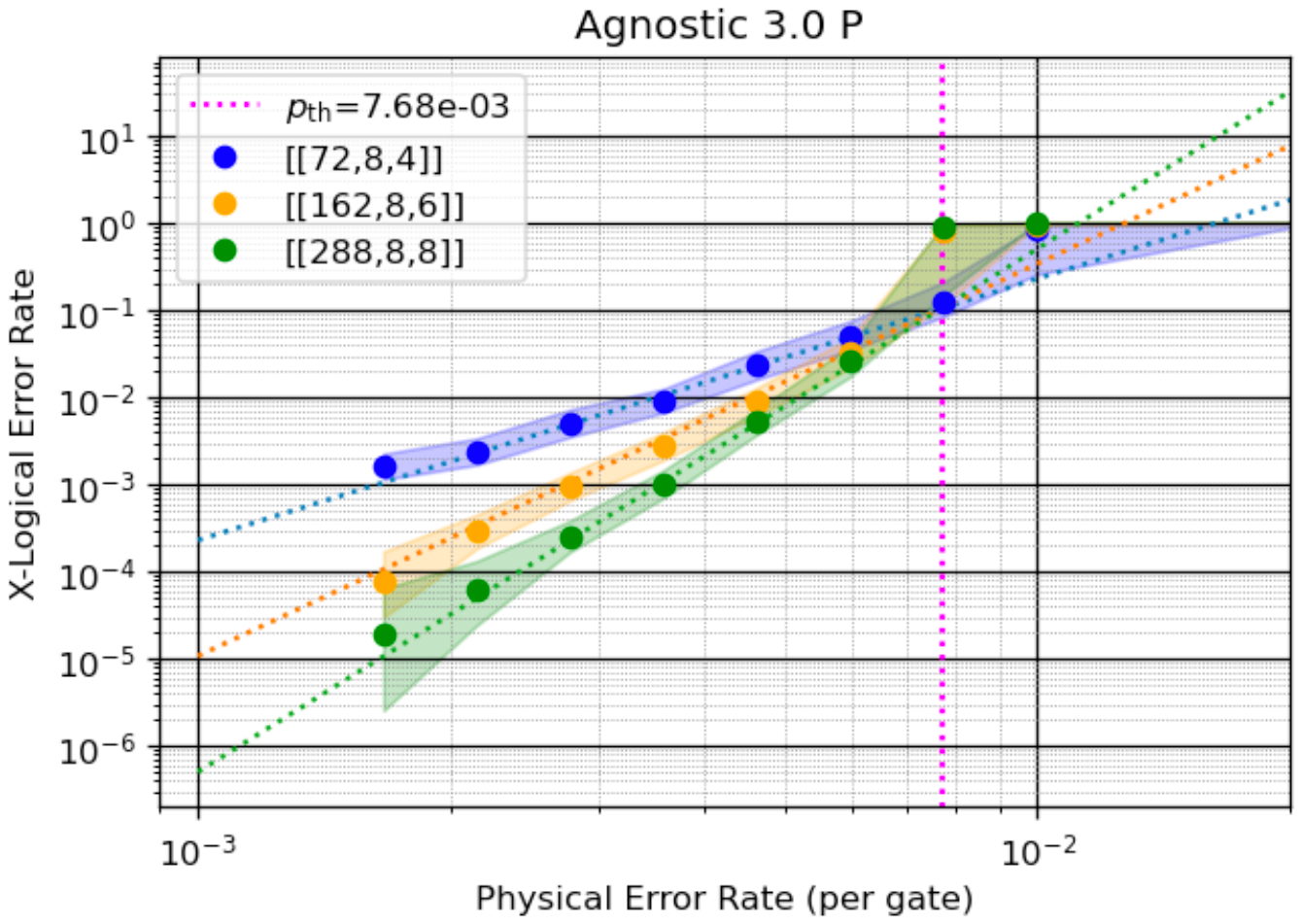}
        }
    \end{minipage}
    \vspace{0.01cm}

    \begin{minipage}{0.45\textwidth}
        \centering
        \rotatebox{0}{
        \includegraphics[width=0.7\columnwidth]{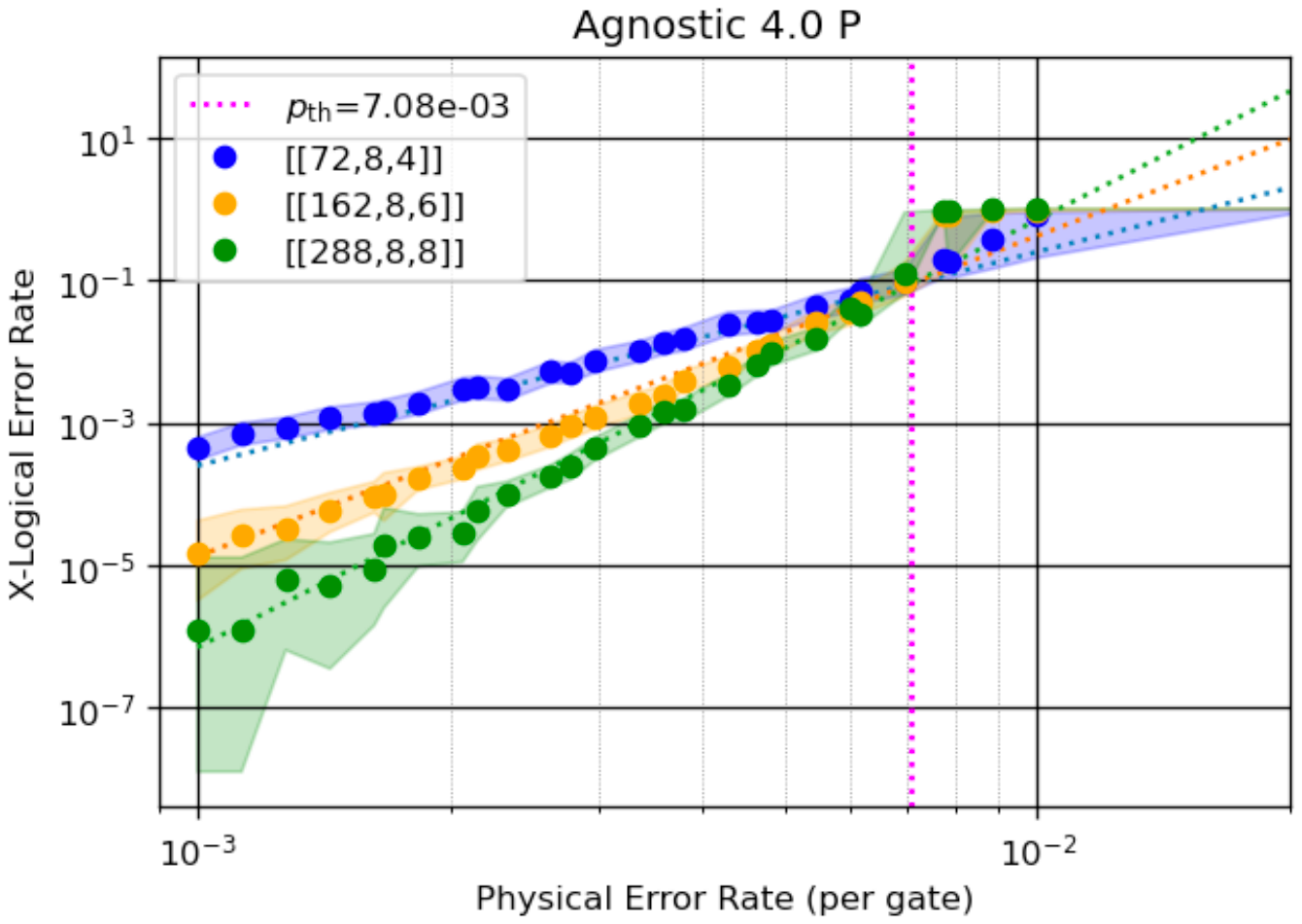}
        }
    \end{minipage}%
    \hspace{0.01\columnwidth}
    \begin{minipage}{0.45\textwidth}
        \centering
        \rotatebox{0}{
        \includegraphics[width=0.7\columnwidth]{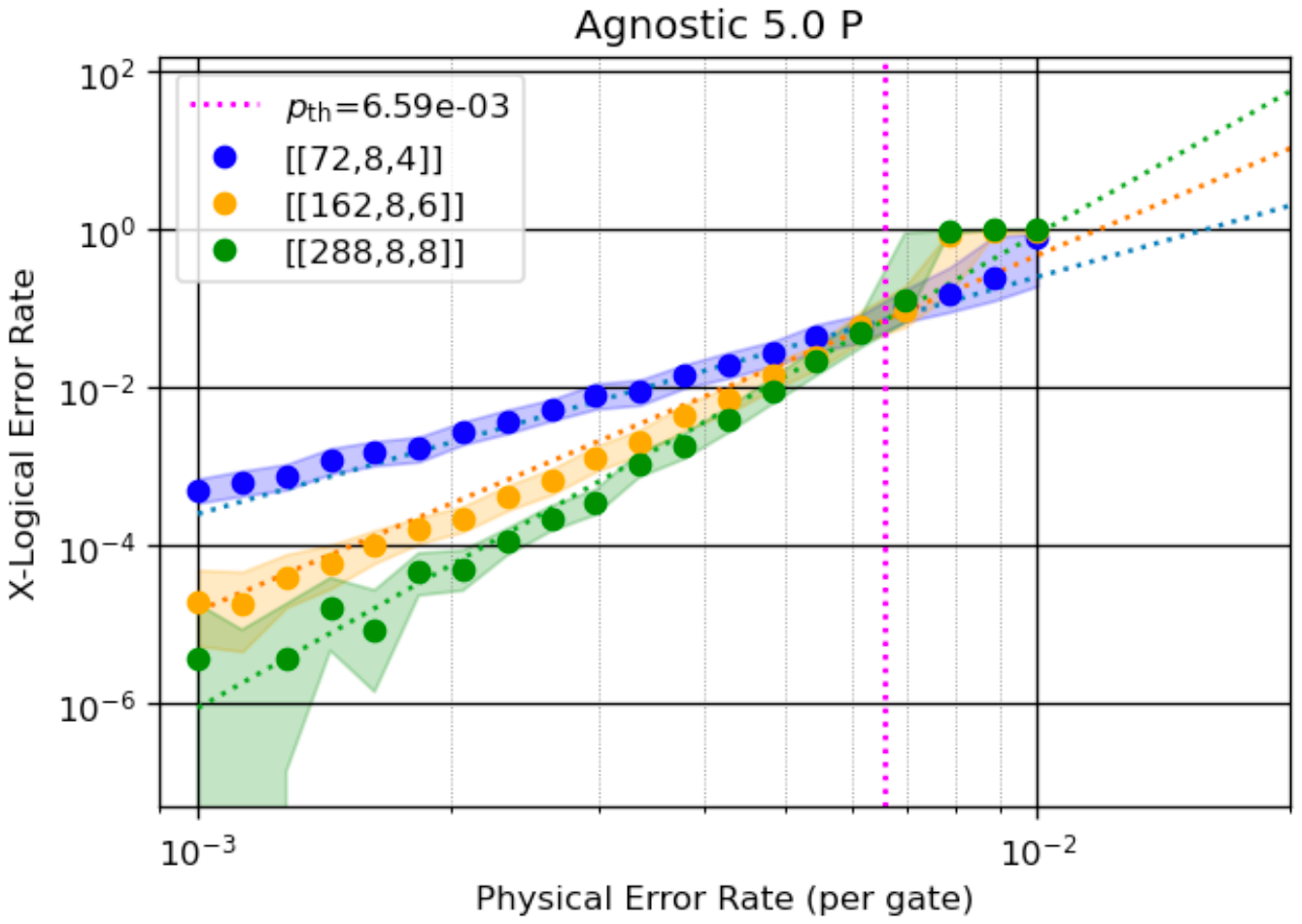}
        }
    \end{minipage}    
    \vspace{0.01cm}    
    \begin{minipage}{0.45\textwidth}
    \centering
    \rotatebox{0}{
        \includegraphics[width=0.7\columnwidth]{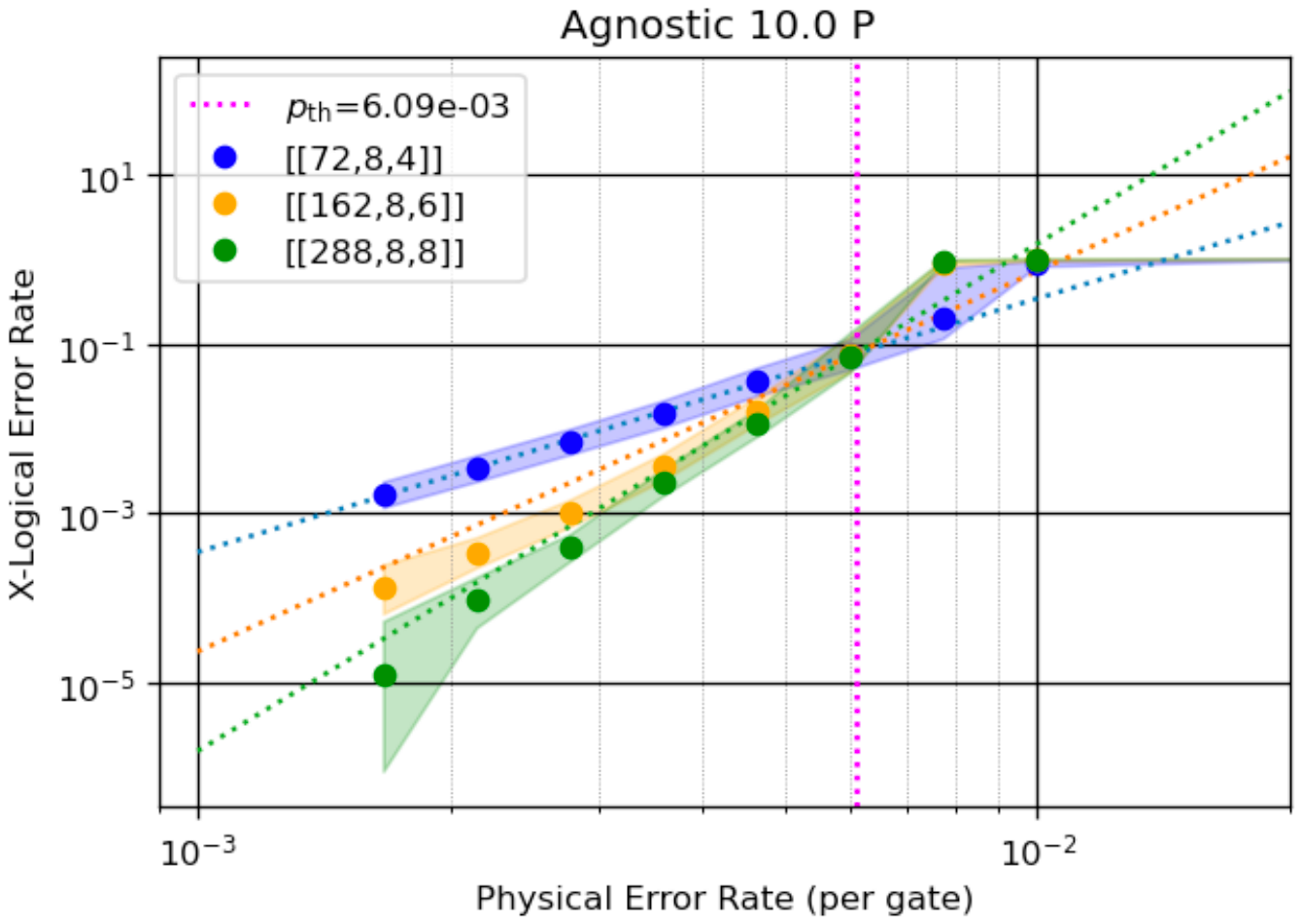}
        }
    \end{minipage}
    \caption{Agnostic circuit-level noise simulation results for the codes listed in Table~\ref{tab:Codes h(x)=1+x+x^2}. The plots are presented on a log-log scale, where the Y-axis represents the logical error rate and the X-axis represents the physical error rate (per gate). The simulations were carried out using \texttt{STIM}, with each data point based on $10^5$ Monte Carlo samplings. The fitting lines were obtained using fitting Eq.~\ref{eq:fitting equation} with \(a\approx 1/2\). }
    \label{App:fig_agnostic}
\end{figure*}

\begin{figure*}[htbp]
    \centering
    \begin{minipage}{0.45\textwidth}
        \centering
        \rotatebox{0}{
        \includegraphics[width=0.7\columnwidth]{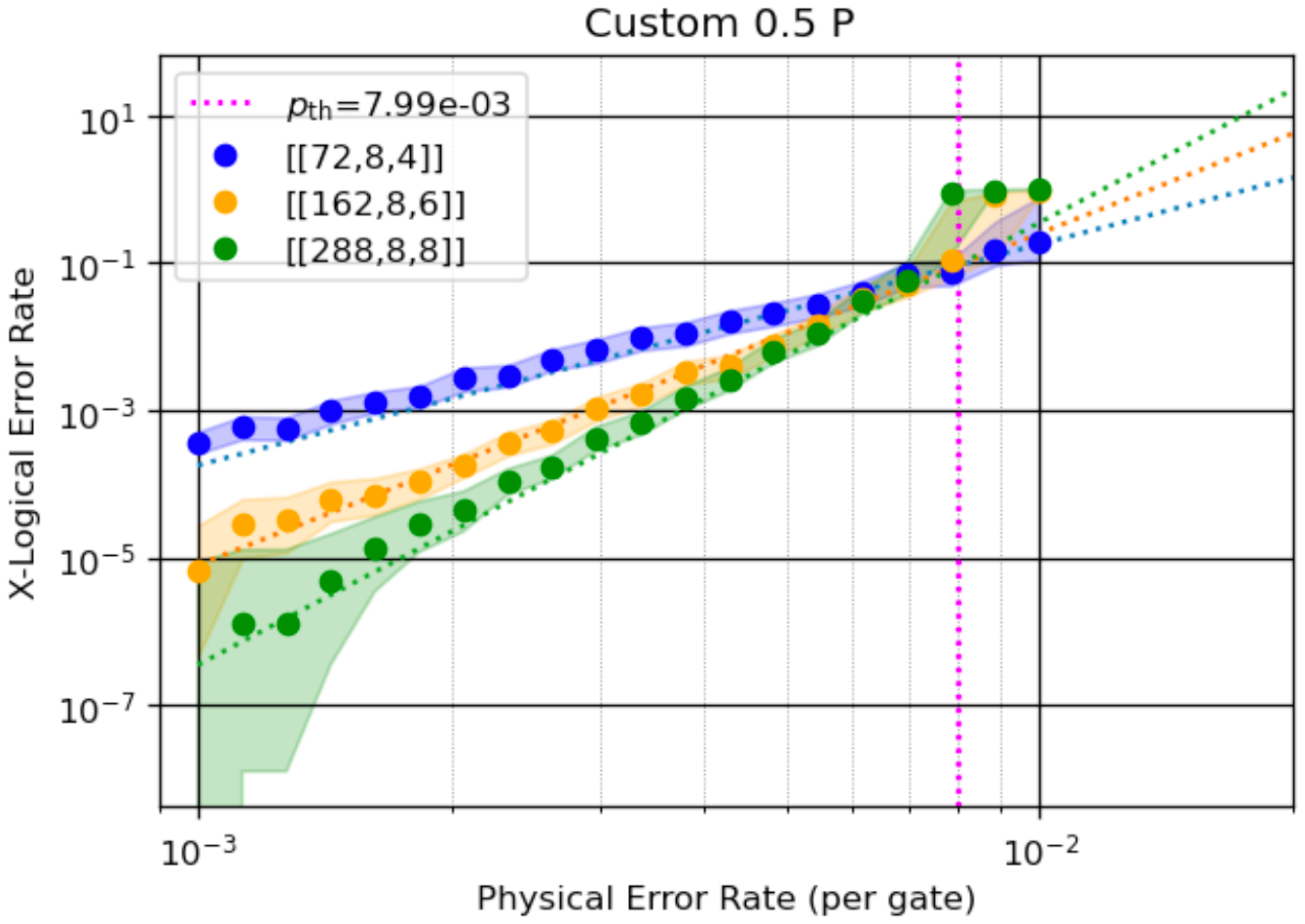}
        }
    \end{minipage}%
    \hspace{0.01\columnwidth} 
    \begin{minipage}{0.45\textwidth}
        \centering
        \rotatebox{0}{
        \includegraphics[width=0.7\columnwidth]{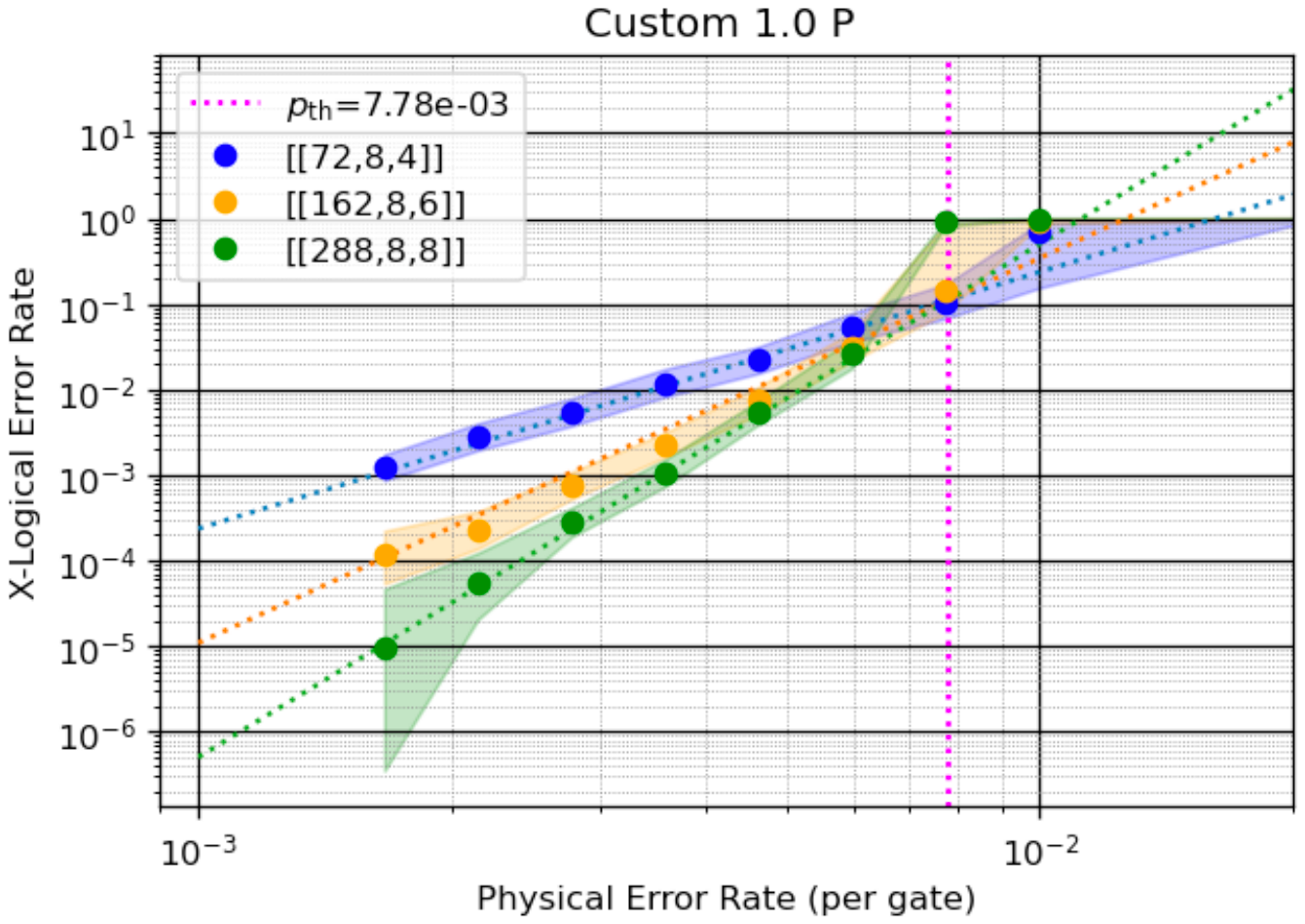}
        }
    \end{minipage}
    \vspace{0.01cm} 
    
    \begin{minipage}{0.45\textwidth}
        \centering
        \rotatebox{0}{
        \includegraphics[width=0.7\columnwidth]{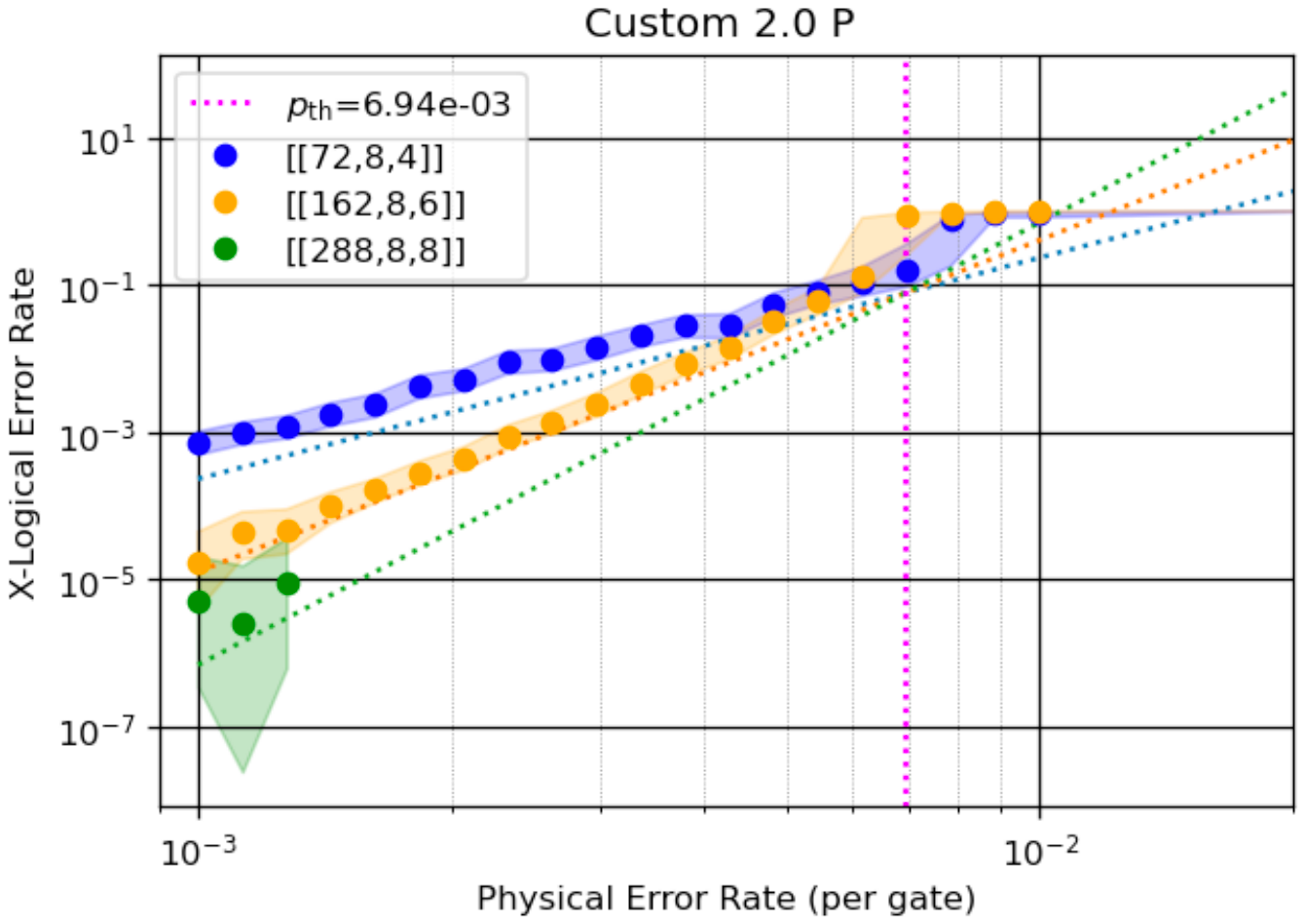}
        }
    \end{minipage}%
    \hspace{0.01\columnwidth}
    \begin{minipage}{0.45\textwidth}
        \centering
        \rotatebox{0}{
        \includegraphics[width=0.7\columnwidth]{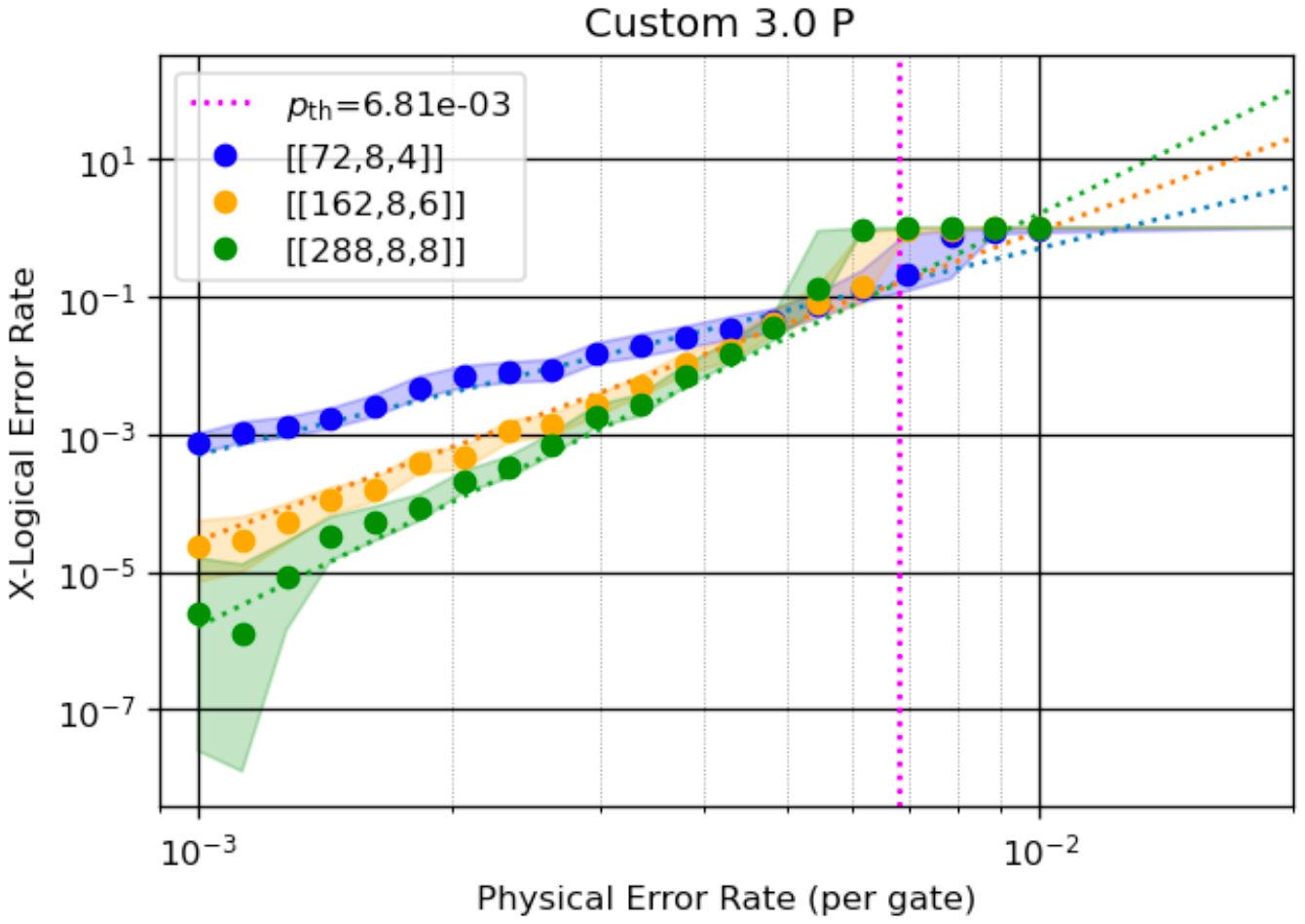}
        }
    \end{minipage}

    \vspace{0.01cm}
    \begin{minipage}{0.45\textwidth}
        \centering
        \rotatebox{0}{
        \includegraphics[width=0.7\columnwidth]{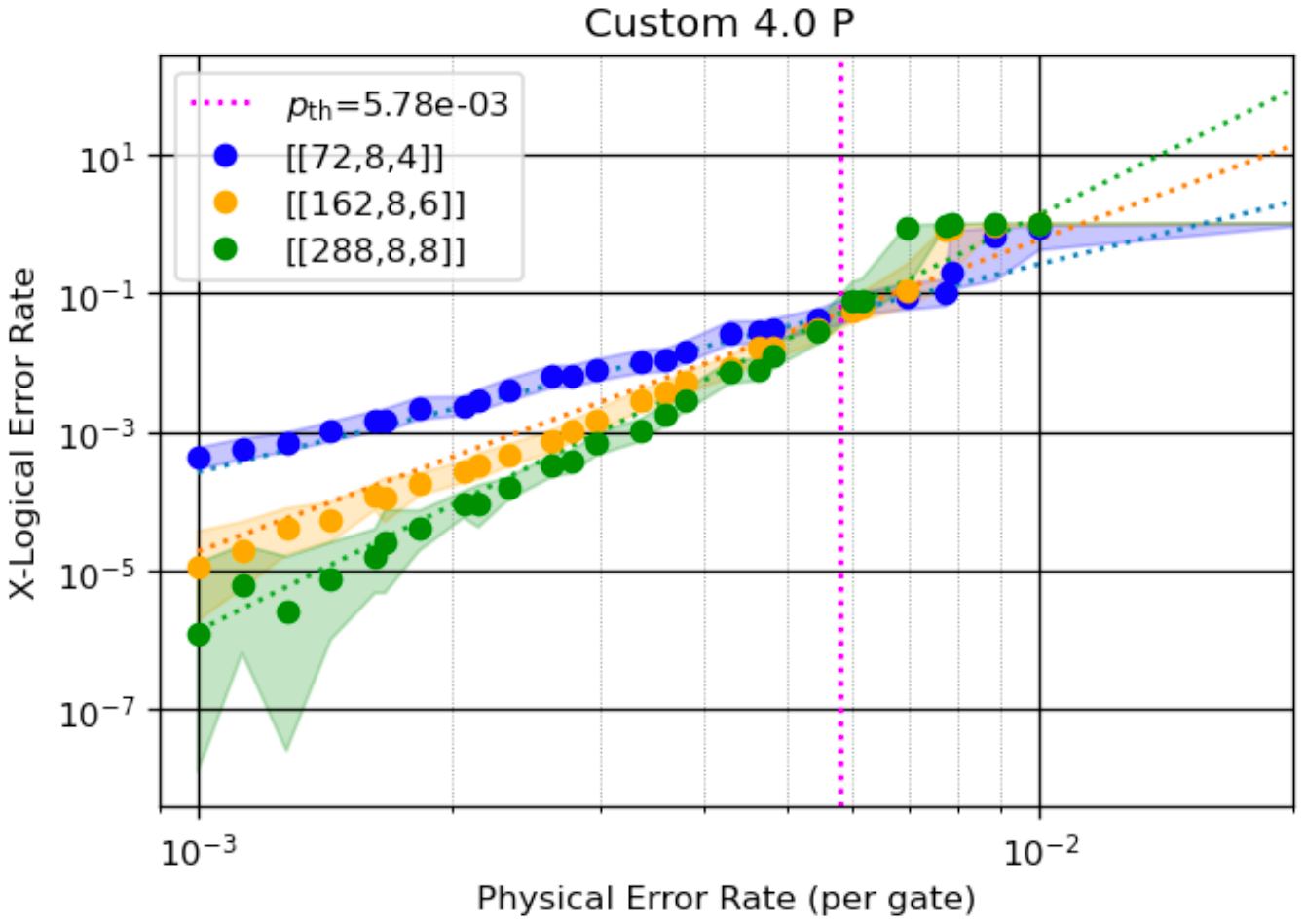}
        }
    \end{minipage}%
    \hspace{0.01\columnwidth}
    \begin{minipage}{0.45\textwidth}
        \centering
        \rotatebox{0}{
        \includegraphics[width=0.7\columnwidth]{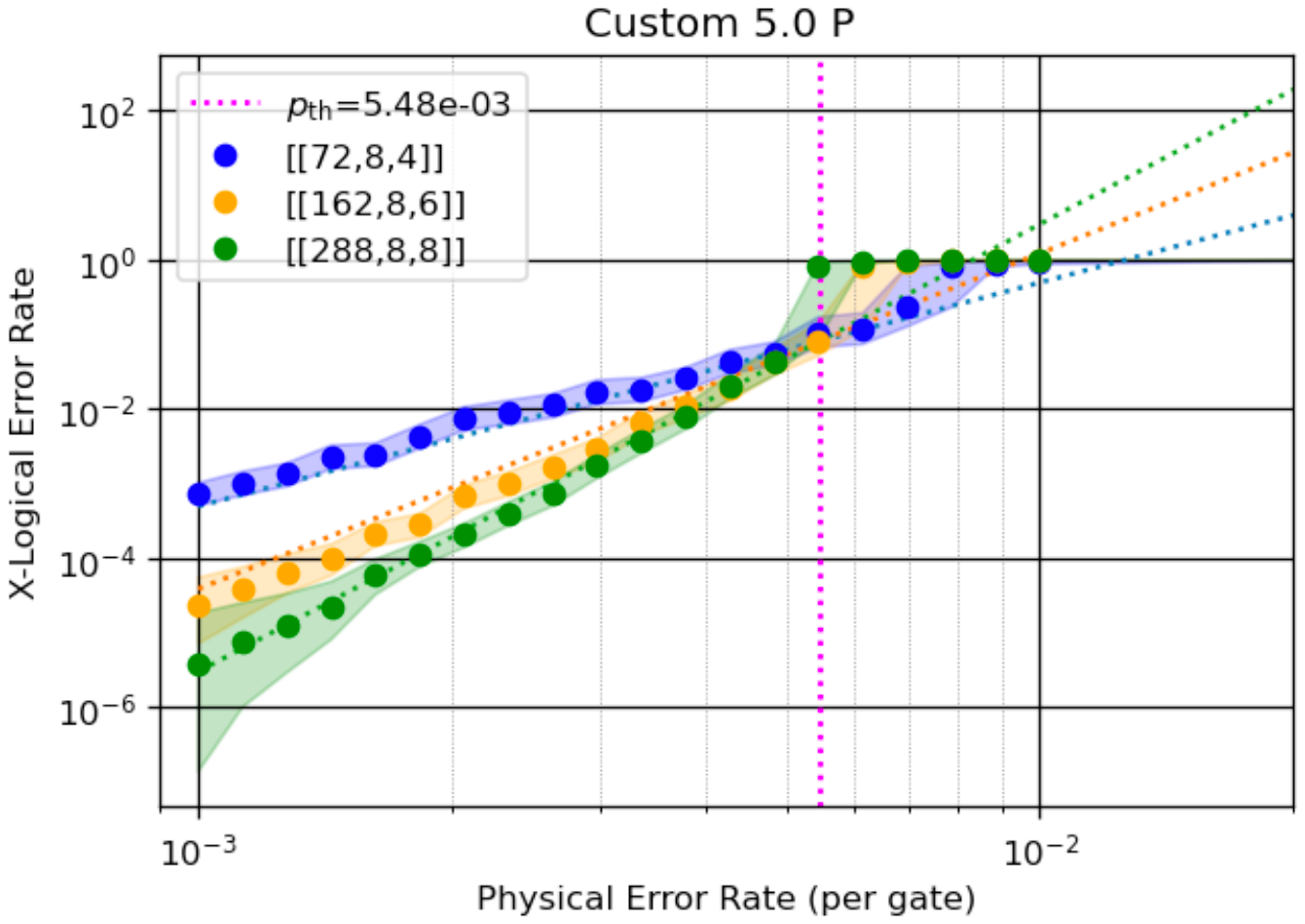}
        }
    \end{minipage}%
    \hspace{0.01\columnwidth}
    \begin{minipage}{0.45\textwidth}
        \centering
        \rotatebox{0}{
        \includegraphics[width=0.7\columnwidth]{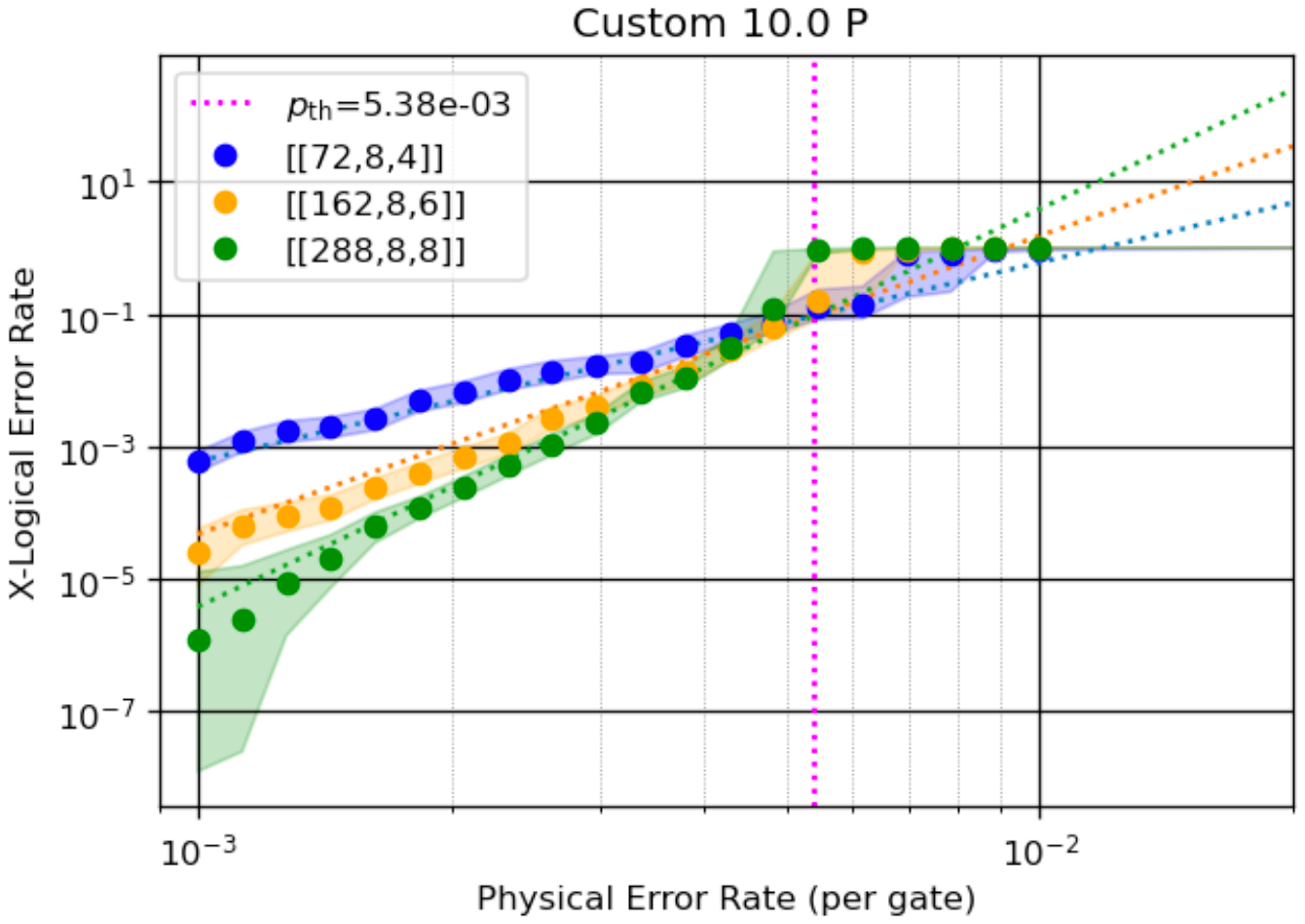}
        }
    \end{minipage}
    \caption{Custom circuit-level noise simulation results for the codes listed in Table~\ref{tab:Codes h(x)=1+x+x^2}. The plots are presented on a log-log scale, where the Y-axis represents the logical error rate and the X-axis represents the physical error rate (per gate). The simulations were carried out using \texttt{STIM}, with each data point based on $10^5$ Monte Carlo samplings. The fitting lines were obtained using fitting Eq.~\ref{eq:fitting equation} with \(a \approx 1/2\).}
    \label{App:fig_custom}    
\end{figure*}

Here we present all the simulation results. Figure~\ref{App:fig_agnostic} shows the results for Hardware-agnostic error model. See Fig.~\ref{App:fig_custom} for the results for custom error model. In both cases the ratio \(p_{\mathrm{cavity}}/p_2\) is varied. The title of each subfigures shows the ratio. We fit all the plots for a given ratio to the equation,
\begin{align} 
    P_L(p) & = A\left(\frac{p}{p_{th}}\right)^{a d},
\end{align}

The threshold is calculated using this fit function and is presented in Table~\ref{tab:threshold_agnostic_custom_periodic}.

\end{document}